\documentclass[12pt,a4paper,oneside]{article}
\usepackage{graphicx,amsmath,amssymb}
\usepackage{mdwlist}
\usepackage{stmaryrd}
\usepackage{dsfont}
\usepackage[utf8]{inputenc}
\usepackage[hang]{subfigure}
\usepackage{mdwlist}
\usepackage{float}
\usepackage{graphicx}
\usepackage{array}
\usepackage{algorithmic}

\usepackage{pgf}
\usepackage{tikz}
\usetikzlibrary{positioning,shapes,shadows,arrows,calc,snakes}

\usepackage{chngcntr} 

\usepackage{verbatim} 
\usepackage{enumerate}
\usepackage{enumitem}
\usepackage{listings}
\usepackage{siunitx}
\usepackage[normalem]{ulem}

\usepackage{url}

\usepackage{color}

\newcommand{\comR}[1]{{\leavevmode\color{red}{\emph{#1}}}}

\usepackage{todonotes}

\usepackage{xspace}
\newcommand{\Snoopy} {\textit{Snoopy}\xspace}
\newcommand{\Charlie}{\textit{Charlie}\xspace}
\newcommand{\Marcie} {\textit{Marcie}\xspace}
\newcommand{\Spike} {\textit{Spike}\xspace}
\newcommand{\Patty} {\textit{Patty}\xspace}

\newcommand{\ie}{i.e.,\xspace}

\newcommand{\SM}{Supplementary Material\xspace}

\newcommand{\CPN}{\ensuremath{\mathcal{CPN}}\xspace}

\newcommand{\QPN}{\ensuremath{\mathcal{QPN}}\xspace}
\newcommand{\HPN}{\ensuremath{\mathcal{HPN}}\xspace}

\newcommand{\SPN}{\ensuremath{\mathcal{SPN}}\xspace}

\newcommand{\node}[1]{\textit{#1}}
\newcommand{\rnumber}{\ensuremath{\mathcal{R}}}

\newcommand{\CPNC}{\ensuremath{\mathcal{CPN^C}}\xspace}
\newcommand{\QPNC}{\ensuremath{\mathcal{QPN^C}}\xspace}
\newcommand{\HPNC}{\ensuremath{\mathcal{HPN^C}}\xspace}

\newcommand{\SPNC}{\ensuremath{\mathcal{SPN^C}}\xspace}

\newcommand{\CTMC}{\text{CTMC}\xspace}

\usepackage{xr}
\usepackage{xr-hyper}
\usepackage{hyperref}

\usepackage{cite}
\usepackage[T1]{fontenc}

\DeclareFontFamily{OT1}{pzc}{}
\DeclareFontShape{OT1}{pzc}{m}{it}{<-> s * [1.10] pzcmi7t}{}
\DeclareMathAlphabet{\mathpzc}{OT1}{pzc}{m}{it}

\def\R{{\mathbb{R}}}
\def\R+{{\mathbb{R^+}}}

\def\C{{\mathbb{C}}}

\def\N{{\mathbb{N}}}

\def\BibTeX{{\rm B\kern-.05em{\sc i\kern-.025em b}\kern-.08em
    T\kern-.1667em\lower.7ex\hbox{E}\kern-.125emX}}

\def\blankpage{%
      \clearpage%
      \thispagestyle{empty}%
      \addtocounter{page}{-1}%
      \null%
      \clearpage}

\newcounter{petrinet}
\renewcommand{\labelitemi}{$\bullet$}

\usepackage{xspace}

\newenvironment{myitemize}
{ \begin{itemize}
    \setlength{\itemsep}{1pt}
    \setlength{\parskip}{1pt}
    \setlength{\parsep}{1pt}     }
{ \end{itemize}      
}

\usepackage{fancyhdr}
\begin{document}

\lhead[{\it From Epidemic to Pandemic Modelling}]{{\it From Epidemic to Pandemic Modelling}}
\rhead[ ]{ }
\lfoot[\today]{30 March 2021}
\cfoot[ ]{ }
\rfoot[\thepage]{\thepage}

\bibliographystyle{alpha}


\title{Technical Report
}

\author{Brunel University London /\\Brandenburg Technical University Cottbus-Senftenberg
\\
\\
\\
{\bf\LARGE From Epidemic to Pandemic Modelling}\\
\\
Shannon Connolly$^{(1)}$, David Gilbert$^{(1)}$,
 Monika Heiner$^{(1,2)}$\\
 \\
$^{(1)}$ Brunel University London, UK\\
shannonfayeconnolly@hotmail.com\\ david.gilbert@brunel.ac.uk\\
\\
$^{(2)}$ Brandenburg Technical University, Cottbus, Germany\\
   monika.heiner@b-tu.de\\
  \\
  30 March 2021
}

\date{}
\maketitle
\thispagestyle{empty}

\begin{abstract}

We present a methodology for systematically extending epidemic models to multilevel and multiscale spatio-temporal pandemic ones.
Our approach builds on the use of coloured stochastic and continuous
Petri nets facilitating the sound component-based extension of basic SIR models to include
population stratification and also spatio-geographic information and travel connections, represented as graphs,
resulting in robust stratified pandemic metapopulation models.
This method is inherently easy to use, producing scalable and reusable models with a high degree of clarity
and accessibility which can be read either in
a deterministic or stochastic paradigm.
Our method is supported by a publicly available platform {\it PetriNuts}; it
enables the visual construction and
editing of models; deterministic, stochastic and hybrid simulation as well as structural and behavioural analysis.  
All the models are available as supplementary material, ensuring reproducibility.

\end{abstract}

\newpage
\thispagestyle{empty}
\vspace*{\fill}
\noindent
cite as

\noindent
\small{
\begin{verbatim}
@techreport{CGH21.report,
  author = {S Connolly and D Gilbert and M Heiner},
  title = {{From Epidemic to Pandemic Modelling}},
  institution = {Brunel University London / 
                 Brandenburg University of Technology Cottbus-Senftenberg},
  year = {2021},
  month = {30 March},
  note = {Technical Report},
  pdf = {},
  url = {},
}
\end{verbatim}
}
\normalsize

\newpage
\pagenumbering{roman}
\tableofcontents

\newpage
\listoffigures
\listoftables
\blankpage

\newpage
\pagenumbering{arabic}
\setcounter{page}{1}
\section{Introduction}

We present a methodology for modelling pandemics, the development of which was motivated by the current 
COVID19 outbreak.
The key point of our approach is the introduction of geography and travel connections
in a well founded manner, facilitating intuitive
understanding by modellers, clinicians, epidemiologists, public health strategists, politicians, and other key
stakeholders.

Our modelling approach builds on the use of coloured stochastic and continuous
Petri nets which facilitates the sound component-based extension of 
basic SIR models to include
population stratification and also spatio-geographic information, resulting in robust 
stratified pandemic metapopulation models.
This method is inherently
easy to use, producing scalable and reusable models with a high degree of clarity and accessibility which 
can be read either in
a stochastic or  
deterministic paradigm, the latter mapping to Ordinary Differential Equations (ODEs). 
The ODEs are uniquely defined by the Petri net models and are automatically generated by our simulators.
In turn this conforms with sound model engineering practice,
with the aim of  minimising encoding errors.  

The models which we obtain by our method are multilevel, the lower
level comprising homogeneous SIR models, and the upper level being a
network of geographical connections. Thus our methodology enables
the construction and analysis of  metapopulation models.  The models
are multiscale in terms of distances, which are explicit at the
upper geographical level and implicit at the lower homogeneous
level, and multiscale in terms of time in that we assume that
geographical connections (travel) occur at a much lower rate than
infections in lower level SIR components.

Our method is supported by a publicly available platform {\it PetriNuts} which enables visual construction and
editing of models; deterministic, stochastic and hybrid simulation as well as
structural and behavioural analysis.  All
the models presented in this paper are provided as Supplementary Material,
and can be processed using the platform, thus ensuring reproducibility.


\section{Results}
\label{section:results}

\subsection{Basic epidemic SIR model in Petri nets}

\begin{figure}[!hb]
            \centering
            \includegraphics[width=0.8\textwidth]{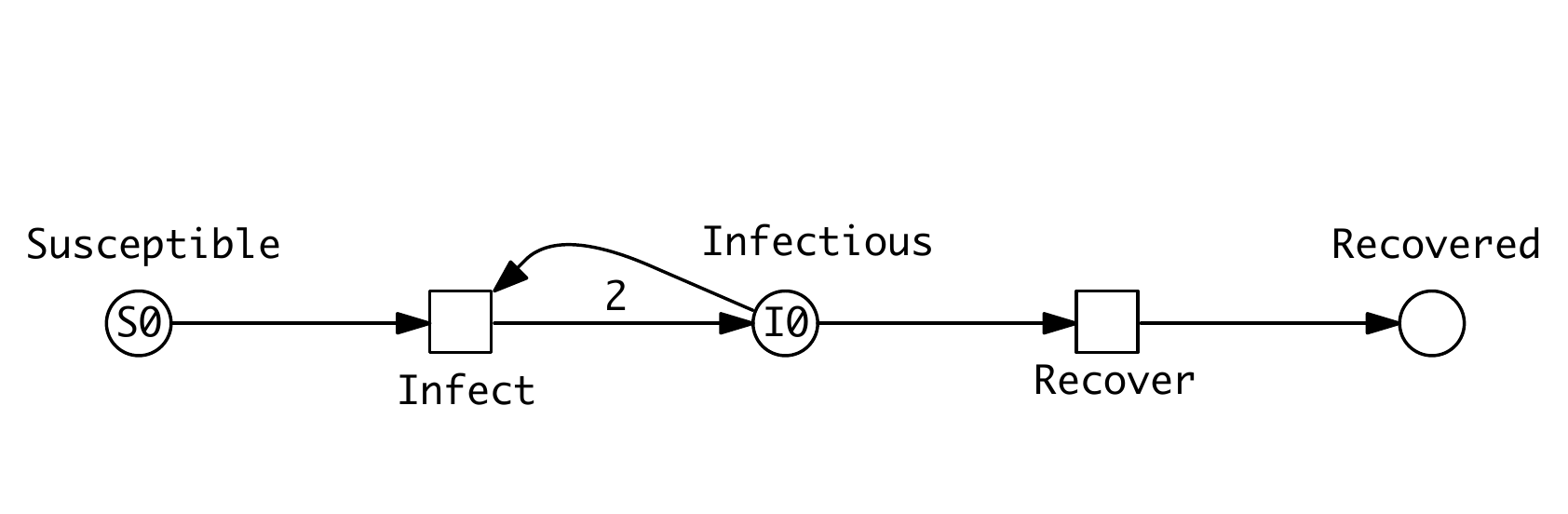}\\
            \includegraphics[width=0.32\textwidth]{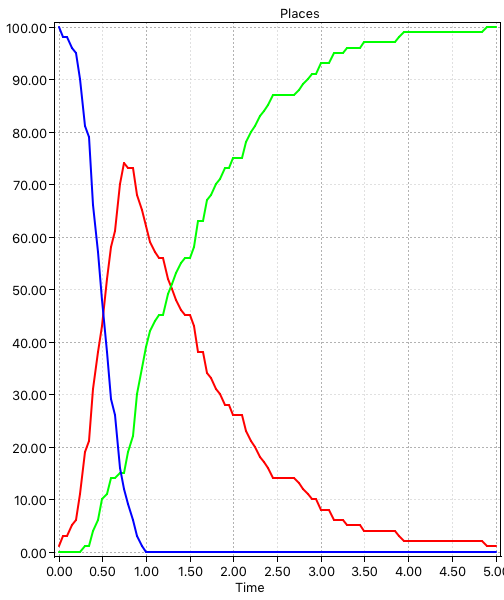}
            \includegraphics[width=0.29\textwidth]{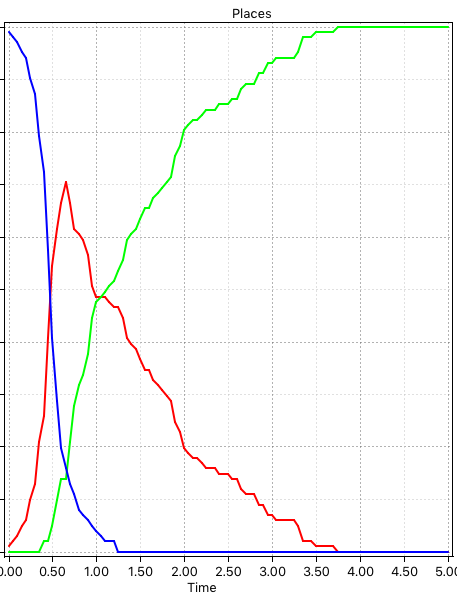}
               \includegraphics[width=0.35\textwidth]{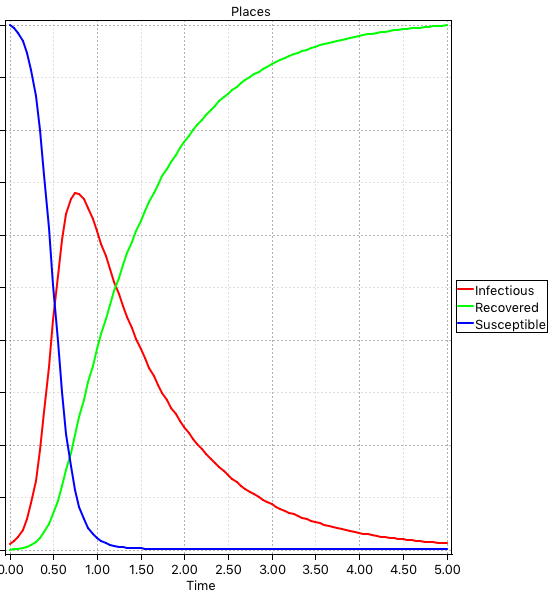}
            \caption[SIR model]{
{\bf SIR model}.
{\bf(Top)} A Petri Net describing the SIR model where Petri net places (circles)
represent the
Susceptible~(S), Infectious~(I) and Recovered~(R) compartments. 
Petri net transitions 
(squares)
represent the actions (events) of becoming infectious and recovering, and arcs connect places with transitions enabling events to occur and describing the effect of the occurrence. 
At the \node{Infect} transition, an infection will only occur when a susceptible and
infectious person meet reducing the number in both compartments by 1. 
This results in two people being added to the \node{Infectious} place which is
represented by an arc weight of 2 (shorthand notation for two arcs). 
Note that the default arc weight of 1 is not given.
Also the number of infectious people can be reduced by 1 via the
\node{Recover}
transition, increasing \node{Recovered} by 1.
Each transition preserves the number of persons (the number of  incoming arcs
equals the number of outgoing arcs), thus the net is conservative and the total population size is constant.
{\bf(Bottom)} From left to right: Two single stochastic runs and the deterministic run, with $S0=100$, $I0=1$, $k_{infect}=0.1$, $k_{recover}=1$.
Note, in the second stochastic run, \node{Infectious} dies out much earlier than in the deterministic simulation.
}
\label{figure:SIR}
\end{figure}

We start off with the standard SIR epidemic model and the usual assumptions (see Methods section); 
its Petri net representation is given in Figure~\ref{figure:SIR}.

The infection rate is effectively the observed increase in
the number of daily infections, more generally the number of infections in a
given time period, typically described by mass action
kinetics~\cite{wilson1945law}.  An alternative approach,
called the standard incidence, 
normalises the infection rate by the total population size $N$, based on the assumption that daily encounters between individuals in the population
are largely independent of community size~\cite{hethcote2000mathematics}.
The {\it rate constant} of infection $k_{infect}$
(the parameter for mass action kinetics) is driven by both the
infectiousness of the disease, as well as any societal prevention
measure in place (isolation etc.).  Likewise, the recovery rate is governed by
the rate constant of recovery $k_{recover}$ which 
reflects the contribution of all the factors
contributing to recovery - health of the individuals in the population,
treatment etc.

These infection and recovery rates correspond to the rate functions
associated with the transitions \node{Infect} and \node{Recover} in the Petri net model.
Rate functions can be interpreted in two different ways: stochastic or deterministic, thus yielding stochastic Petri nets (\SPN) or continuous Petri nets (\CPN).
The semantics of \SPN is given by continuous time Markov chains (\CTMC), and the semantics of \CPN by Ordinary Differential Equations (ODEs). The ODEs generated reading the Petri
net in Figure~\ref{figure:SIR} as a \CPN are given in Equation~\ref{equation:SIR_ODEs}, 
which correspond exactly to the ODEs given in the literature.
We observe that each place gets its own equation (see Methods section for the automatic translation of \CPN into ODEs).
In the equations we have
replaced the long place names by their standard short forms for readability.
The formulae governing the rates according to the standard incidence can be easily adjusted to use the mass action kinetics by setting $N=1$.

\begin{equation}
\label{equation:SIR_ODEs}
\frac{dS}{dt} =  -k_{infect}\cdot \frac{S}{N} \cdot I\\
\end{equation}
\begin{equation*}
\frac{dI}{dt} =  k_{infect}\cdot \frac{S}{N} \cdot I - k_{recover}\cdot I\\
\end{equation*}
\begin{equation*}
\frac{dR}{dt} =  k_{recover}\cdot I\\
\end{equation*}
\smallskip

To obtain a flexible model, we introduce constants 
$S_0$, $I_0$ and $R_0$ to initialise the 
places \node{Susceptible}, \node{Infectious} and \node{Recovered};
thus, $N = S_0 + I_0 + R_0$. 
In general, the initial values of all the places
in a model form its initial state at time $t=0$. We also use, for example, the notation $S_0$
to stand for the value of $S_{t=0}$, or generally $S_t$ for time point $t$.

These models are typically simulated, which for \SPN means making random walks through the \CTMC, applying e.g. Gillespie's stochastic simulation algorithm (SSA)~\cite{Gillespie:1976},
and for \CPN solving the underlying ODEs. Figure~\ref{figure:SIR} gives simulation traces for both cases.
Even without realistic rate constants, this basic model already permits some
interesting computational experiments, see section on parameter fitting.

\vspace{3em}

\newpage
We make some observations about the behaviour of the system.
There is a trivial, i.e. disease free steady state for $I_0=0$ so that
no infections or recovery
occur, otherwise the steady state is reached when $I_t=0$. 
We note some interesting \SPN behaviour of single runs -- 
there is a probability that the infection dies out early, whatever the balance
between the rates of infection and recovery, due to the underlying Markov chain.
In contrast, the ODE behaviour can be thought of as the
average over many stochastic runs, and thus ignores the variability inherent
in a stochastic model.
Besides exploring time series of model variables, we can also define derived measures as observers, such as the
reproduction number \rnumber~\cite{heffernan2005perspectives,vandendriessche2017reproduction}; see Methods section for details.

All our models can be equally read as \SPN or \CPN.


\subsection{Extended epidemic models}

The SIR model can be extended by 
extra compartments such as 
deaths (D), maternally-derived immunity (M), exposed/incubation period (E) etc., 
and extra variables such as total number of infectious, total deaths, 
yielding variations of the basic SIR model including
SIS, SIRS, SIRD, MSIR, SEIR, SEIS, MSEIR, MSEIRS~\cite{hethcote2000mathematics}.
Their representation as a Petri net model is straightforward, and Petri net animation supports the modeller in getting the model structure right.
Some illustrative models from this paper
can be animated in a web browser at \url{https://www-dssz.informatik.tu-cottbus.de/DSSZ/Research/ModellingEpidemics}.

\begin{figure}[!htb]
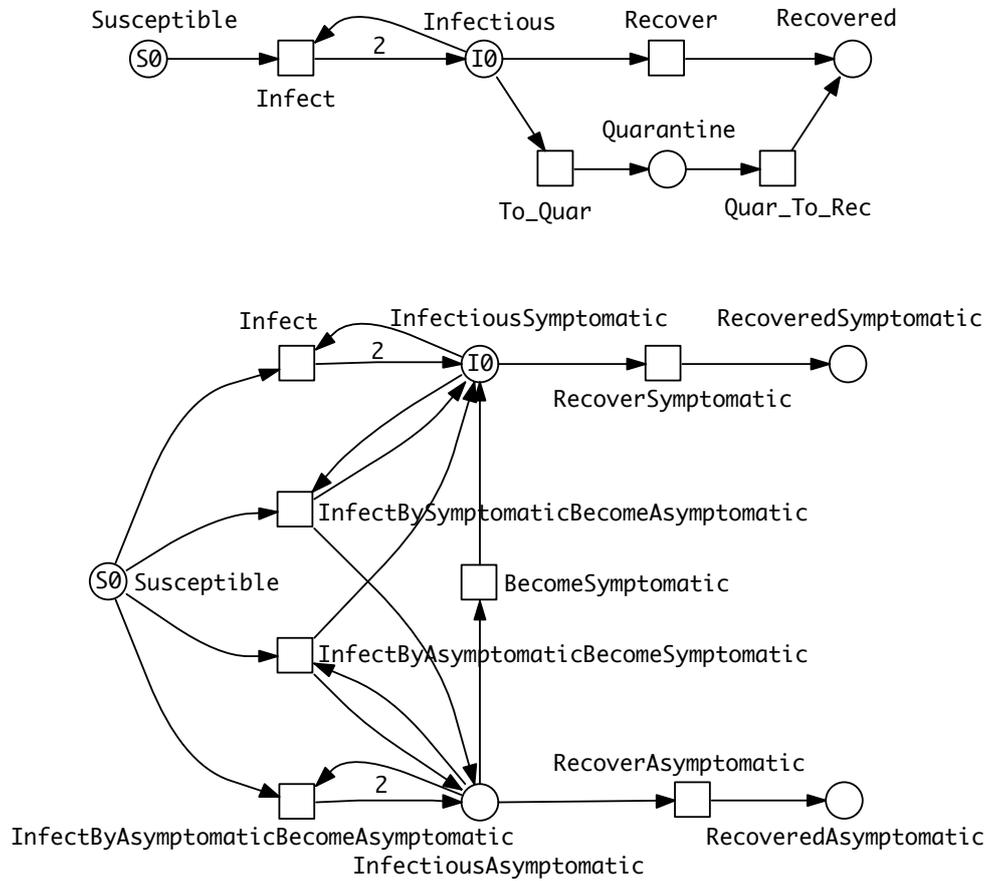

            \centering
            \includegraphics[width=0.75\textwidth]{Figures/SIQR-spn.pdf}
            \includegraphics[width=0.92\textwidth]{Figures/SIAR-spn.pdf}
            \caption[Two extended epidemic models]{{\bf Two extended epidemic
models}. {\bf(Top)} SIQR - an SIR model extended by quarantine.
{\bf(Bottom)} SIAR - an SIR model with symptomatic/asymptomatic compartments.
There are four different cases of
infection reflected by the four transitions connected with the \node{Susceptible} place.  
The two which involve only one
infectious compartment -- \node{InfectiousSymptomatic} or \node{InfectiousAsymptomatic}
-- follow the standard SIR pattern, reflected in the outgoing arc weights of 2.  In the cases of
the two
dual outcome infections -- \node{Susceptible} and \node{InfectiousSymtomatic} 
to \node{InfectiousAsymptomatic} or
{\em vice versa}, there are two outgoing arcs to the two infectious places each of
which has an arc weight of 1 (not shown by default).
The total rate of infection in the whole population is the sum of the rates of the four
infection transitions.
The ODEs of these two models are given in the Supplementary Material.
}
            \label{figure:SIQR-SIAR}
        \end{figure}

Here we consider two examples which are relevant to the current COVID19
epidemic (Figure~\ref{figure:SIQR-SIAR}):
\begin{itemize}
\item
SIQR model which describes,
as an additional route to recovery,
the possibility of
infectious people being subject to quarantine before
recovery~\cite{anand2020predicting}.
\item
SIAR model which differentiates
infectious people into symptomatic and asymptomatic
compartments~\cite{delasen2017new}.
\end{itemize}

\newpage
\subsection{Stratified epidemic models}

In general, modelling the stratification of a population in terms of epidemics means that the overall 
population is partitioned into a number of strata and a suitable model is defined for each stratum
and their appropriate interconnections.
The compartments are equally divided into disjunctive sets, 
such that each partition is internally homogeneous and distinctively differs from all the other partitions, and there are corresponding partitions in all compartments.

\begin{figure}[!htb]
            \centering
            \includegraphics[width=0.81\textwidth]{Figures/SIR_age2-spn.pdf}
            \includegraphics[width=0.76\textwidth]{Figures/SIR_age_enum-colspn.pdf}
            \includegraphics[width=0.34\textwidth]{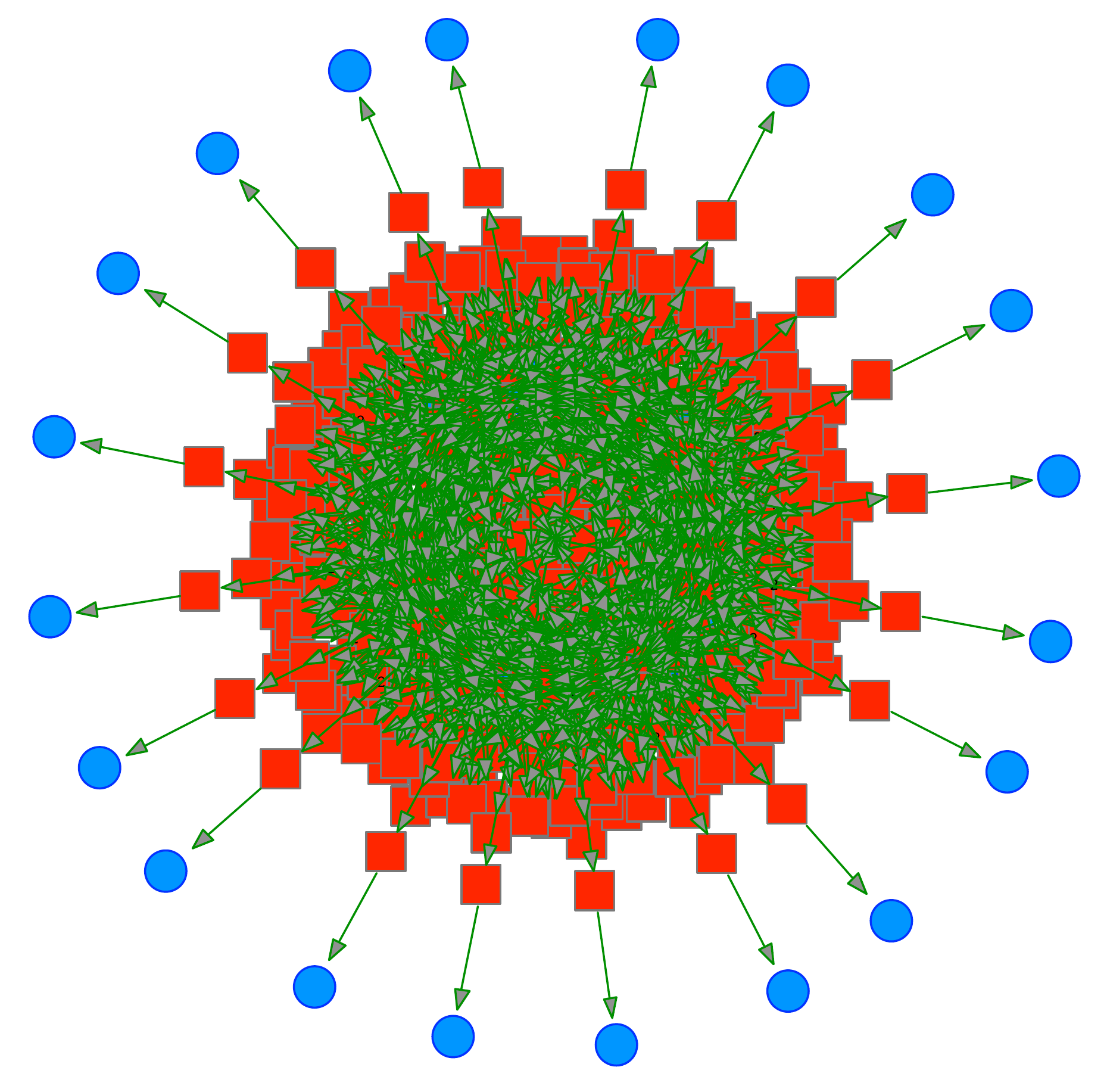}
\caption[SIRage model]{{\bf SIR-$\textrm{S}^\textrm{age}_\textrm{2;20}$ models}. 
Two variations of the SIR model describing 
stratification of a population into two age groups (young and old) or 20 age groups, 
and the cross infections between the groups.
Transmission of infection does not change the age group of the infectious person. 
At the start there are no old infectious people, and only very few young infectious people.
{\bf(Top)}~\CPN for two age strata, each of which is
associated with a separate, but structurally identical SIR model, and linked by cross infections.
{\bf(Middle)}~The same model represented as a coloured \CPN.
The markings of coloured places always show the sum of the markings of the corresponding uncoloured places. 
The colour definitions are given in the Methods section, and
the ODEs in the Supplementary Material.
{\bf(Bottom)}~Unfolded \CPN of a 20-strata SIR model, with coloured graphics.
}
\label{figure:SIRage}
        \end{figure}

In the following 
we assume that the same epidemic model is equally applied to each stratum.
As an example we consider SIR-$\textrm{S}^\textrm{age}_\textrm{2}$, an SIR model with
two age strata~(Figure~\ref{figure:SIRage}~(Top)).
See Methods section for an explanation of our naming convention for models. 

Clearly the stratification in this model is very minimal; the standard practice when
collecting statistics of infections is to use more than two age groups - often in
10 deciles from 0 to 100 years old.  A model should describe all the possible
combinations of cross infections, in this case we obtain
$10\cdot 10=100$ infect transitions.  The modelling effort invites modelling errors;
this will be further compounded if gender is added to age in the stratification.
To cope with this modelling challenge, we use coloured Petri nets.

\paragraph{Coloured Petri nets} provide an elegant solution to the given combinatorial problem by pattern
compression of the repeated parts of a Petri net, while maintaining the connections between those parts.  Each
repeated component is associated with a unique integer (`colour'); see Methods section for details.
This colouring principle can be equally applied to \SPN or \CPN, yielding coloured \SPN (\SPNC) and coloured \CPN (\CPNC).

In summary, Figure~\ref{figure:SIRage} gives the same model in two different but equivalent representations,
uncoloured (Top) and coloured (Bottom), which also means that both generate the same ODEs. Obtaining the coloured model from the uncoloured one is called {\em
folding}, and the reverse operation is called {\em unfolding}.  Folding is typically performed manually
because the degree of folding is subjective, while unfolding can be done automatically as long as all colour
sets are discrete and finite.

Increasing the number of strata while ensuring that all needed
connectivities are present just requires extending the colour set Strata; nothing else
in the structure of the coloured model needs to be touched,
apart from adjusting the initial marking.
In this way, our modelling approach supports scalability in an easy to use manner.
In general the unfolded SIR model for $s$ strata 
comprises $3s$ places, $s^2+s$ transitions and $4s^2+2s$ arcs,
and the ODEs generated consist of $3s$ equations. 

\clearpage

\subsection{Pandemic models}

The distinguishing characteristic of a pandemic model is the concept of multiple
intercommunicating epidemic models in a spatial (geographic) context.  In the
following we prefix standard epidemic model names by `P' for pandemic, thus obtaining
PSIR, PSIQR, etc.

\begin{figure}[!htb]
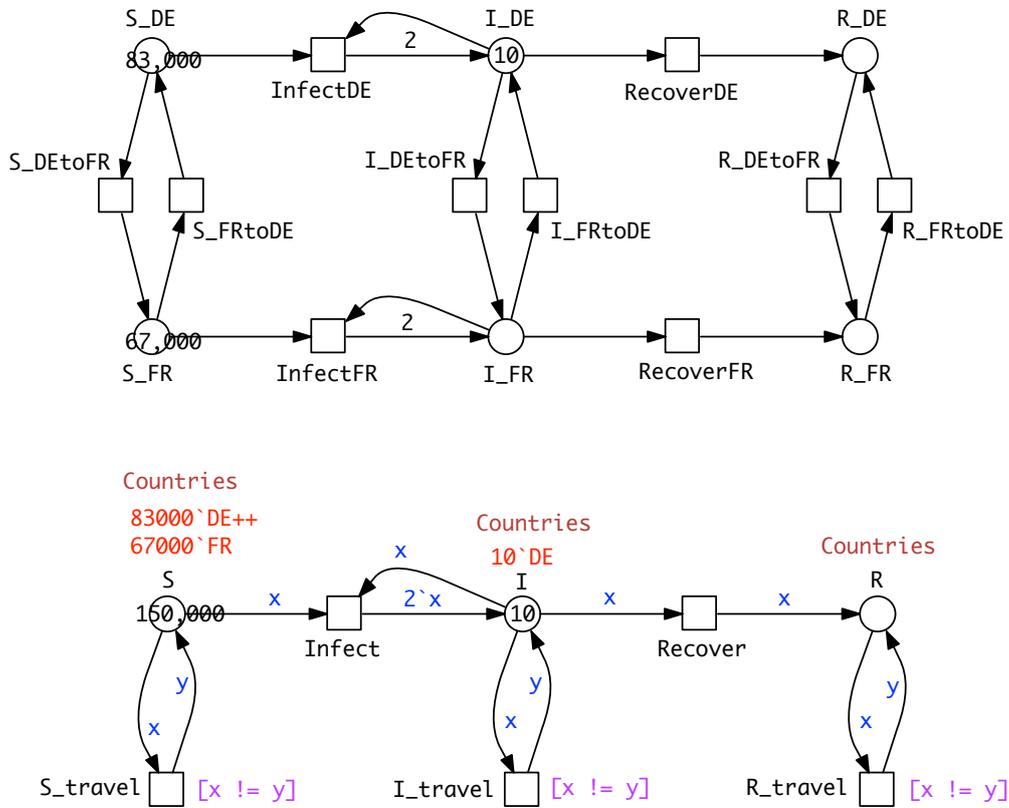

            \centering
            \includegraphics[width=0.95\textwidth]{Figures/SIR_TRAVEL2-spn.pdf}
            \includegraphics[width=0.9\textwidth]{Figures/SIR_Travel2-colspn.pdf}
\caption[PSIR2 model]{{\bf P$_2$SIR model}. 
{\bf(Top)} \CPN for two countries, each of which is
associated with a separate, but structurally identical SIR model, and linked by travel.
{\bf(Bottom)} The same model represented as a coloured \CPN. 
Eeach travel transition is associated with a guard
{\bf [x != y]} (x and y are unequal)
to ensure that travellers do not immediately return to their country of origin.
The uncoloured and coloured model are equivalent, thus generate the same ODEs, which
are given in the \SM.
}
\label{figure:SIRtravel2}
\end{figure}

\paragraph{Two countries, linked by travel.} 
Figure~\ref{figure:SIRtravel2}~(Top) shows a PSIR model comprising
epidemic SIR models in each of two countries, with
population movement permitted between the countries for all compartments.
As for the age strata model, we can fold it into a coloured version
(Figure~\ref{figure:SIRtravel2}~(Bottom)). The process of colouring is very similar to
that described for SIR-$\textrm{S}^\textrm{age}_\textrm{2}$ (see Methods section). 

\begin{figure}[!htb]
            \centering
            \includegraphics[width=0.22\textwidth]{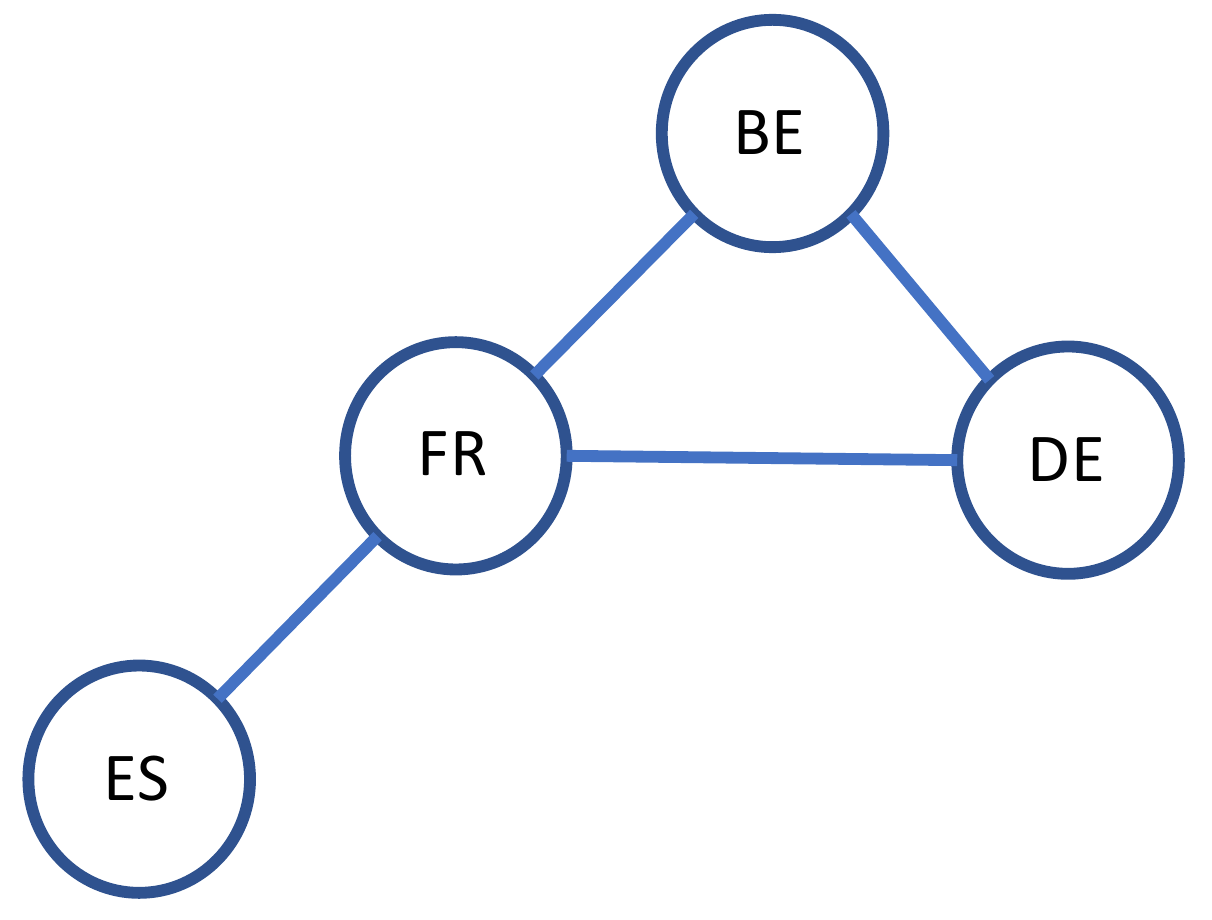}
            \includegraphics[width=0.27\textwidth]{Figures/network_Europe4-spn.pdf}
            \includegraphics[width=0.29\textwidth]{Figures/network_Europe4-colpn.pdf}
            \includegraphics[width=0.82\textwidth]{Figures/P4_SIR-colspn.pdf}
            \includegraphics[width=0.8\textwidth]{Figures/P48_SIR_S10-colspn.pdf}
\caption[SIR Travel]{{\bf P$_n$SIR models}. 
{\bf(Top left)} Connectivity graph for four countries. 
{\bf(Top middle)} Petri net representation of the connectivity graph.
{\bf(Top right)} Coloured Petri net representation of the connectivity graph; see Methods section for details.
{\bf(Middle)} Coloured Petri net representation of the pandemic P$_4$SIR; 
structural identical with P$_2$SIR in Figure~\ref{figure:SIRtravel2}~(Bottom); 
both \CPNC differ by their colour definitions encoding different connectivity graphs. 
The unfolded Petri net of P$_4$SIR is give in 
Figure~\ref{figure:P4-SIR-unfolded}.
{\bf(Bottom)} P$_\textrm{48}$SIR-$\textrm{S}^\textrm{age}_\textrm{10}$ 
coloured Petri net for 48 countries in Europe, each with 10 age strata, see Figure~\ref{figure:P48SIR}
for its unfolded version.
}
\label{figure:Countries4}
        \end{figure}

\paragraph{N countries, linked by travel.} 
Generalising this idea to more than two countries, the travel connectivities between countries need to be
defined.  For example when considering surface travel, not all countries are directly 
connected to each other. 
We assume that the connectivity is described by a graph
(Figure~\ref{figure:Countries4}~(Top left)); this could be directly encoded as a standard Petri net
(Figure~\ref{figure:Countries4}~(Top middle)), which we fold for convenience into a coloured Petri net
(Figure~\ref{figure:Countries4}~(Top right)),
see Methods section for details.
Equally the graph could include air and/or sea connections; even then, connectivity could still be less than fully connected.

\begin{figure}[!htb]
            \centering
            \includegraphics[width=0.45\textwidth]{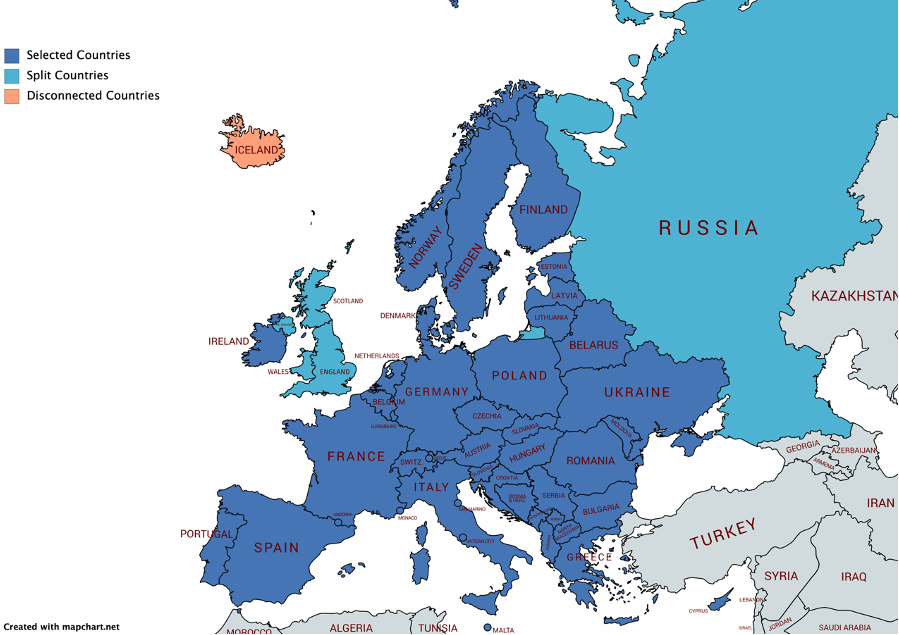}
            \includegraphics[width=0.45\textwidth]{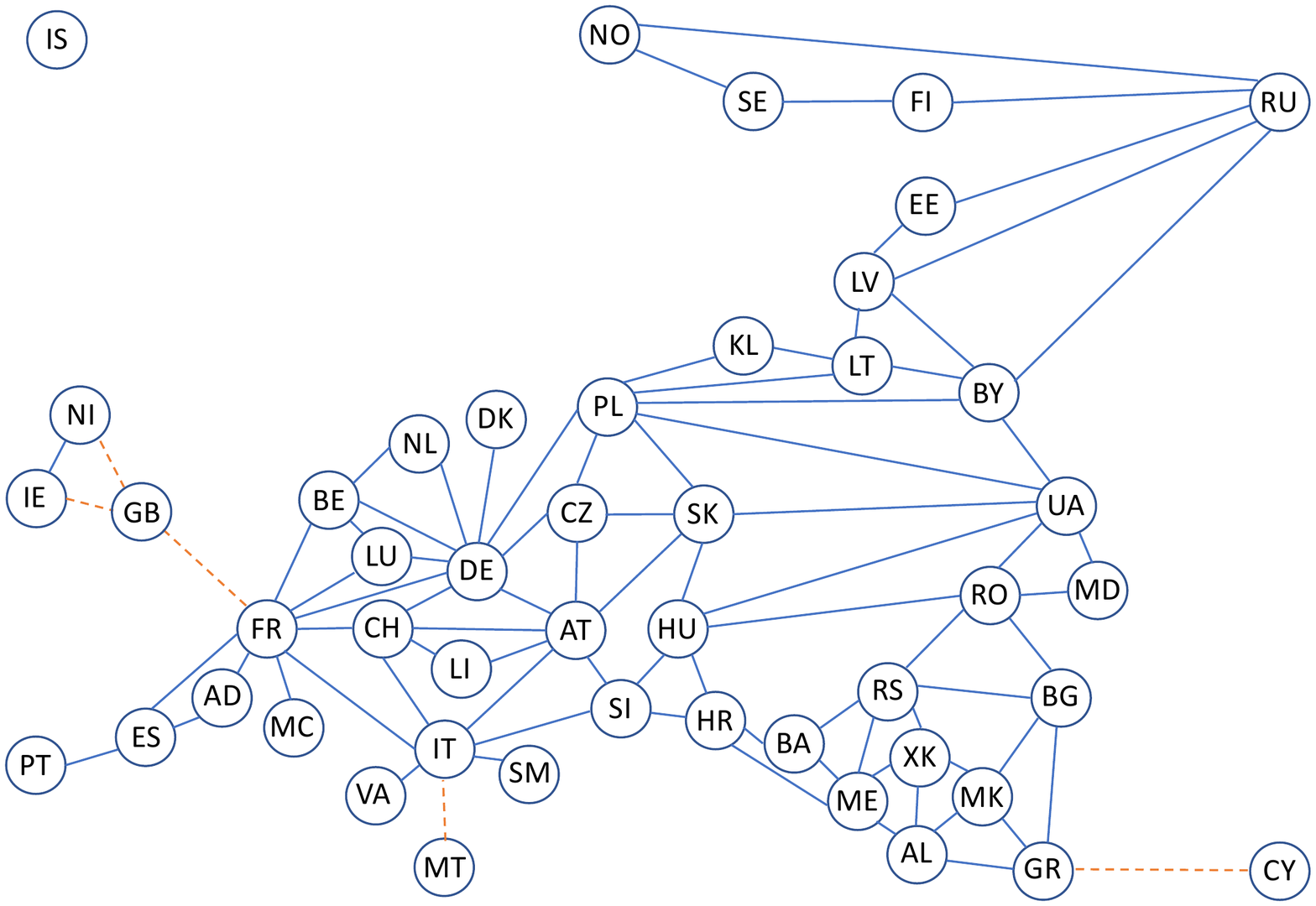}
            \includegraphics[width=0.9\textwidth]{Figures/P48SIR_S10-spn.pdf}
                        
\caption[P48SIR-S10]{{\bf P$_\textrm{48}$SIR-$\textrm{S}^\textrm{age}_\textrm{10}$ model.} {\bf(Top left)} Map of 48 European countries created with \url{https://mapchart.net}.  {\bf(Top right)} Undirected graph
representation of the map where edges represent two-way surface connections. 
Blue edges stand for road connections, orange dashed edges for sea connections.
{\bf(Bottom)} 
Unfolded version of the coloured net in Figure~\ref{figure:Countries4}~(Bottom);
layout automatically generated.  Iceland appears as an
isolated subnet because it is not connected by surface to the rest of Europe, and is thus similar to 
the SIR net with 20 strata in
Figure~\ref{figure:SIRage}~(Bottom). 
Model size:  1,440 places; 10,380 transitions.}
\label{figure:P48SIR}
\end{figure}

The coloured Petri net representation of the PSIR over four countries (Figure~\ref{figure:Countries4}~(Middle)) is derived by 
adjusting the colour definitions in the SIR
for two countries (Figure~\ref{figure:SIRtravel2}~(Bottom)) according to the
connectivity graph, and getting the initial marking right.

The generalisation to connectivity graphs representing different geographical relationships is
straightforward, merely requiring adjusting the two related colour definitions
(\node{Countries}, \node{Connections}) and the initial marking.
This example illustrates how our modelling approach supports reusability in an easy to use manner.
In general the unfolded PSIR model for $n$ countries with overall $m$ mutual connections
comprises $3n$ places, $2n+6m$ transitions and $6n+12m$ arcs,
and the ODEs generated consist of $3n$ equations.

We provide a number of PSIR pandemic models in the Supplementary Material, including Europe (P$_{10}$SIR -- 10 countries, 15 mutual connections, P$_{48}$SIR -- 48
countries, 85 mutual connections), China (P$_\textrm{China}$SIR -- 34 provinces, 71 mutual connections) and the
USA (P$_\textrm{USA}$SIR -- 50 states, 105 connections), and the guidelines how to obtain the corresponding ODEs using our tool platform. 

\subsection{Combined models}
These three extensions -- SIR variations, stratification and space --  are orthogonal, and thus any combination is possible.
For illustration we provide two examples:

\begin{itemize}
\item 
P$_{10}$SIQR combining space (Europe with 10 countries) with an extended epidemic model 
(see Figure~\ref{figure:SC:model7}) 
which yields $4n=40$ ODEs, and

\item
P$_{48}$SIR-S$^\textrm{age}_{10}$
combining space (Europe with 48 countries) and stratification (10 age strata per country; 
coloured version in Figure~\ref{figure:Countries4}~(Bottom), unfolded
version in Figure~\ref{figure:P48SIR}) which yields 1440 ODEs.

In general the unfolding of a P$_{n}$SIR-S$_{s}$ model with $s$ strata and $n$ countries with $m$ mutual connections comprises $3sn$ places, $n(s^2+s) + 6ms$  transitions and $2n(2s^2+s) + 12ms$ arcs. 

\end{itemize}

\subsection{Parameter fitting}

Apart from defining the structure of a model, it is also important to find suitable
rate constants.  This can be naively achieved by scanning over parameter values, or in a more sophisticated
manner heuristically by target driven parameter optimisation.

Even without fitted values derived from real world data, a great deal of
useful analysis can be achieved using comparative rate constant values.
Examples include scanning by
\begin{itemize}
\item varying the ratio between infection rate and recovery rate 
while maintaining the recovery rate constant
resulting in changes in the \rnumber$_t$ value
(SIR), and comparison between geographic regions (PSIR),
\item varying the ratio between direct and quarantine recovery (SIQR), and
varying quarantine regimes 
(Figures~\ref{figure:SC:Figure27+28},~\ref{figure:SC:Figure30}, 
both referring to the model 
P$_{10}$SIQR in 
Figure~\ref{figure:SC:model7} (Top)).
\item varying the initial state of the infectious compartments in the different
countries of a pandemic model (PSIR), 

\item varying the strength of the 
connectivities between countries - road, train, sea and air - in a pandemic model (PSIR).
\end{itemize}

Refining a model automatically involves increasing the number of required kinetic rate
constants. Specifically in the SIR-$\textrm{S}^\textrm{age}_\textrm{2}$ model, the kinetic constants can be different
for both intra-stratum and inter-strata transitions.
This can be achieved in the coloured version 
using colour-dependent rates, see Methods section.

More generally, we can establish equivalence classes for rate constants, for
example
in a PSIR travel model, there can be different rate constants between classes of movements
such as road versus air and/or frequent versus rare, while the rate constants within classes
are the same.

When real world data are available, then we can attempt to fit models to them.
We did this for the pandemic model for ten European countries
(P$_{10}$SIR, see 
Figure~\ref{figure:SC:model6}),  
based on source data for daily new infections 
obtained from World\-Ometers
(\url{https://www.worldometers.info/coronavirus/}) during the period 19th January to
16th August 2020.
The rate constants fitted were for infection, and the inter-country movement for all three
compartments. The values for the \node{Susceptible}, \node{Infectious} and \node{Recovered} compartments for movement in
one direction between a pair of countries were kept the same, thus reflecting the assumption that the rate
constant for travel does not depend on the compartment, but only on the journey.
Initial markings for \node{Infectious} were set at 1\% of the total population for
each country according to Wikipedia, divided by 1000.
Daily new cases were standardised and smoothed using a 7-day rolling 
average 
(Figure~\ref{figure:SC:Figure17}) 
and the magnitude and timing of the peaks were
detected computationally 
(Table~\ref{table:sc:table6}).


The goals of the fitting were 
(Objective~1) to reproduce the order of the time taken from the first 5 deaths to a peak for daily new cases, 
(Objective~2) to reproduce the order of the height of the peaks for daily new cases, and
(Objective~3) combining Objective~1 and Objective~2.
Fitting was performed in all three cases using target driven optimisation employing
Random Restart Hill Climbing (see Methods section for details, and 
Tables~\ref{table:sc:table9}~--~\ref{table:sc:fittedconnectionrates}).

\subsection{Analysis}

For an initial validation of (unfitted) pandemic model
behaviour we performed correlation analyses between real world variables and
model variables for P$_{10}$SIR and P$_{48}$SIR; for details see 
Table~\ref{table:sc:table7} and Figures~\ref{figure:SC:Figure23},~\ref{figure:SC:Figure24}.
These results support our assumption that even the unfitted pandemic 
models behave well, and that the geography incorporated into the model
appropriately influences behaviour.
Thus, an analysis of the unfitted pandemic PSIR model can give useful insights even in
the absence of detailed real data; specifically given the geographical connections
between regions and their populations, we can predict to some extend the
expected peak height of infections and thus inform planners how to cope with
the impact on health services etc.;  
some 
examples are given below.

An initial analysis can be performed by inspection of the visualised time series
plots.  Properties of interest include the values of \node{Susceptible} 
and \node{Recovered} over time (e.g. for herd immunity), and the existence, order and height of
peaks for \node{Infectious}, see 
e.g.~Figure~\ref{figure:SC:Figure20}. 
This can be extended to derived measures such as the \rnumber~value, 
computed over time and plotted for the different geographic regions.

Going beyond visual inspection, the order and height of peaks can be ascertained
computationally from the traces.
More generally, many behavioural properties of interest can be expressed using
linear temporal logic and analysed in practice via simulative model
checking~\cite{donaldson:cmsb:2008,GHGC19}, which is specifically useful for exploring larger models by the use of
property libraries, see Methods section and 
Figure~\ref{figure:MCpeaks}.

In addition, the general relationships between the behaviour of related
compartments or derived measures in different geographical regions can be
determined via cluster analysis.  The results for hierarchical clustering can be
represented for example via dendrograms, 
see Figures~\ref{figure:SC:Figure25},~\ref{figure:SC:Figure26}.

Furthermore, the results of such comparative analyses above could be related to the
genomic data of the causative infectious agent,
for example a virus which exhibits high mutation rates.   
Thus the variants or strains of the virus which evolve over
the time of an epidemic can be related to the behaviour of pandemic models.

Finally, pandemic models can be used to predict the effects of various control
measures.  These include travel restrictions in terms of inter-country connections, and
quarantine restrictions on international travellers.
In addition, the effects of the imposition and lifting of lockdowns in different
regions and at different times can be investigated, but requires 
the dynamic variation of
rate constants during simulation.
In order to facilitate this, we have recently extended the \Spike simulator to include event-driven triggers, which enables the modelling of these
control measures, see for example 
Figure~\ref{figure:JC:lockunlock}.

Given a pandemic PSIR model 
with fitted parameters, as developed in the previous section, 
we can perform a detailed comparative analysis over the geographic regions,
going beyond the analysis based on comparative values;



\section{Methods}

\label{section:methods}

\paragraph{Related work.}
In order to put our work in context, we give a brief overview of related work.
A general review paper on epidemic modelling including
metapopulation models can be found in~\cite{brauer2017mathematical}.
Early basic pandemic approaches include mathematical models based on ODEs for the global 
spread diseases, where travel connectivity is described by adjacency
matrices~\cite{rvachev1985mathematical,longini1988mathematical}, and by
graphs~\cite{sattenspiel1993geographic,sattenspiel1995structured}.

The GLEaMviz computational tool~\cite{vandenbroek2011gleamviz} has been developed to model multiscale mobility
networks and the spatial spreading of infectious diseases building on a series of papers
including~\cite{colizza2008epidemic,balcan2009multiscale,balcan2010modeling} and has been used to model the 2009
H1N1 pandemic~\cite{bajardi2011human}.
Some more recent work by other groups includes~\cite{chen2014transmission} considering  dynamics of a two-city SIR
epidemic model and~\cite{arandiga2020spatial,goel2020mobility} for the current COVID19 pandemic.

The Spatiotemporal Epidemiologic Modeler (STEM) is an Eclipse based monolithic framework to support the development and simulation of geographic disease
models~\cite{edlund2010spatiotemporal}.  It builds on a component-based architecture (plugins); models can be simulatated deterministically or
stochastically.
Models are encoded in Java, and it is not obvious how to export a model if one wants to apply any analysis technique not supported by the framework.

Age-based stratification has been addressed in, for example,~\cite{delvalle2013mathematical,balabdaoui2020age},
and the effects of intervention strategies for COVID19 in~\cite{balabdaoui2020age,prem2020effect,giordano2020modelling}.

Other approaches besides ODE modelling include stochastic models~\cite{nandi2019stochastic,oka2020spatial},
agent based models~\cite{mahmood2020facs} and coloured Petri nets~\cite{castagno2020computational}. 

In the following we describe the methods underlying our technology.

\subsection{Modelling}

\paragraph{SIR modelling assumptions.}
The population in the SIR model population is divided into three compartments -- \node{Susceptible}
(S), \node{Infectious} (I) and \node{Recovered} (R).
These three compartments are considered to be homogeneous, thus there is an equal probability of any event occurring to any member of 
each compartment.
In general, this assumption of homogeneity applies to any compartment in our models.

A susceptible person is infected by an infectious person  so that they are both infectious; 
thus getting infected and becoming infectious coincide.
A susceptible person can become recovered.  Recovered people are resistant and cannot become susceptible or infectious.
This simple model does not consider births, or natural deaths so that the total population size is constant, as can be seen from the structure 
of the SIR model in Figure~\ref{figure:SIR}, which shows a conservative Petri net covered by a single minimal P-invariant (compare paragraph below on Petri net analysis techniques).
It is also generally assumed that $S_0 \gg I_0$, and (for obvious reasons) $I_0>0$ and $R_0=0$.

\paragraph{Petri nets} 
are a family of formal languages which come with a graphical representation as bipartite directed multi-graphs, enjoying an operational (execution) semantics. They can be either untimed and qualitative, or timed and quantitive; for details see~\cite{HGD08}.  
All these net classes share the following graph properties.

\begin{itemize}

\item {\it Bipartite:} 
There are two types of nodes, called places and transitions, which form disjunctive node sets. {\it Places}, graphically represented as circles,  typically model passive objects (here compartments and derived measures, possibly in different locations), while {\it transitions}, graphically represented as squares, model active events (such as infect, recover, travel), changing the state of the passive nodes.

\item {\it Directed:} 
Directed arcs, represented as arrows, connect places with transitions and vice versa, thereby specifying the relationship between the passive and active nodes. The bipartite property precludes arcs between nodes of the same type.

\item {\it Multigraph:} 
Two given nodes may be connected by multiple arcs, typically abbreviated to one weighted arc.
The weight is shown as a natural number next to the arc. The default value is 1, and usually not explicitly given.

\end{itemize}

The execution semantics builds on movable objects, represented as {\it tokens} residing on places. The number zero is the default value, and usually not explicitly given.
The current state (marking $m$) of a model is defined by the token situation on all places, usually given by a place vector, which is a vector with as many entries as we have places, and the entries are indexed by the places. 
The initial state is denoted by $m_0$.

The specific details 
of the execution semantics 
slightly differ for qualitative and quantitative Petri nets, and if the Petri net class belongs to the discrete or continuous modelling paradigm.

We start off with standard Petri nets, which are inherently untimed and discrete.
Their (non-negative integer) token numbers are given by black dots or natural numbers.  The current state is given by the number of tokens on all places, forming a place vector of natural numbers.
The (integer) arc weights specify how many of these tokens on a certain
place are consumed or produced by a transition, causing the movement of
tokens through the net, which happens according to the following rules
(see~\cite{BHM15}):

\begin{itemize}

\item 
{\it Enabledness:} A transition is enabled, if its pre-places host sufficient amounts of tokens according to the weights of the transition's ingoing arcs. An enabled transition may fire (occur), but is not forced to do so.

\item
{\it Firing:} Upon  firing, a transition consumes tokens from its pre-places according to the arc weight of the ingoing arcs, and produces new tokens on its post-places according to the arc weights of the outgoing arcs. The  firing happens atomically (i.e., there are no states in between) and does note consume any time.
Firing generally changes the current distribution of tokens; thus the system reaches a new state.

\item
{\it Behaviour:} We obtain the dynamic behaviour of a Petri net by repeating these steps of looking for enabled
transitions and randomly choosing one single transition among the enabled ones to let it fire.

\end{itemize}

\paragraph{\it Rate functions.} These standard untimed Petri nets, which are entirely qualitative, can be extended by
enriching transitions with firing rate functions. We obtain timed and thus quantitive Petri nets, out of which we use in this paper stochastic Petri nets (\SPN) generating continuous time Markov chains, and continuous Petri nets (\CPN) generating Ordinary Differential Equations (ODEs).
Both net classes merely differ by the interpretation of the rate functions; thus \SPN and \CPN, sharing the same net structure, can be easily converted into each other.

Rate functions are, technically speaking, arbitrary mathematical functions, typically depend on the given state, and are often governed by popular kinetics, such as mass action kinetics~\cite{wilson1945law}.

{\it Stochastic rate functions} specify how often a transition occurs per time unit, technically achieved by stochastic waiting times before an enabled transition actually fires. 
\SPN with exponentially distributed waiting times for all transitions fulfil the Markov property; thus their semantics are described by continuous time Markov chains (CTMC), and the current state of an SPN is still defined by a place vector of natural numbers. The CTMC for a given \SPN is basically isomorphic to the state transition system of its corresponding untimed Petri net. Thus, all Petri net analysis techniques see paragraph on Petri net analysis below) can still be applied, and all behavioural properties that hold for the untimed Petri net are still valid for the corresponding \SPN.

In contrast, {\it deterministic rate functions} define the strength of a continuous flow, 
turning the traditionally discrete modelling paradigm of Petri nets into a continuous one: the discrete number of tokens on each place is replaced by a (non-negative) real number, called token value. 
Thus, the current state of a \CPN is defined by a place vector of real numbers.
A \CPN transition is enabled if the token value of all its pre-places is larger than zero. This coincides for mass action kinetics with transition rates larger than zero. Due to the influence of time, a continuous transition is forced to fire as soon as possible. Altogether, the semantics of a \CPN is defined by ODEs.

\paragraph{How to transform a \CPN into ODEs.}
 Each place subject to changes gets its own equation, describing the continuous change over time of its token value by the continuous increase of its pre-transitions' flow and the continuous decrease by its post-transitions'  flow. 
Thus, in the generated ODEs, places are interpreted as (nonnegative) real variables.
A transition that is pre- and post-transition for a given place yields two terms, which can be reduced by algebraically transforming the right-hand side of the equation. Thus, each equation corresponds basically to a line in the incidence matrix~\cite{HGD08}.

To formalise the transformation of \CPN into ODEs, we need to introduce a few technical notations. Let ${\scriptstyle^\bullet}p$ denote the set of pre-transitions of the place $p$, and $p\,{\scriptstyle^\bullet}$ the set of post-transitions, $v\left(t\right)$ gives the rate function for transition $t$, 
$f\left(t,p\right)$ the weight of the arc going from $t$ to $p$, and 
$f\left(p,t\right)$ the weight of the arc going from $p$ to $t$.
Then the ODE generated for the place $p$ is given by Equation~\ref{equation:cpn2ode}.

\begin{equation}
\frac{d\left(p\right)}{dt}=\sum_{t\in{\scriptstyle^\bullet}p}f\left(t,p\right)v\left(t\right)-
\sum_{t\in p \, {\scriptstyle^\bullet}}f\left(p,t\right)v\left(t\right)
\label{equation:cpn2ode}
\end{equation}

\smallskip

In words: the pre-transitions $t\in{\scriptstyle^\bullet}p$ increase the token value of  $p$, thus their weighted rate functions occur as plus terms, while the post-transitions $t\in p \, {\scriptstyle^\bullet}$ decrease the token value of $p$, thus their weighted rate functions occur as minus terms of the ODE's right hand side.

The models presented in this paper always generate as many ODEs of this style as we have places. This is automatically triggered when simulating a \CPN, which can be done in our platform either with \Snoopy or \Spike.
In summary, \CPN can be seen as a structured approach to write ODEs~\cite{BGHO08}; they uniquely define ODEs, but generally not vice versa, compare~\cite{SH10,HG11}.

\smallskip
\paragraph{Coloured Petri nets}  
combine Petri nets with the powerful concept of data types as known from programming languages~\cite{Genrich:Lautenbach:1981,Jensen:book:2009}. Tokens are distinguished via colours;  there are no limits for their interpretation. For example,
colour allows for the discrimination of compartments by, e.g., age strata or gender, or to distinguish compartments in different geographical locations. In short, colours permit us to describe similar network structures in a compressed, but still readable way. 
A group of similar model components (subnets) is represented by one component coloured with an appropriate colour set, and the individual components become distinguishable by a specific colour in this colour set. 
 See~\cite{LHG19} for a review of the widespread use of coloured Petri nets for multilevel, multiscale, and multidimensional modelling of biological systems.

Coloured Petri nets consist, as uncoloured Petri nets, of places, transitions, and arcs. Additionally, a coloured Petri net is characterised by a set of discrete data types, called colour sets, and related net inscriptions.

\begin{itemize}

\item {\it Places} 
get assigned a colour set and may contain a multiset of distinguishable tokens coloured with colours of this colour set. Our {\it PetriNuts} platform supports a rich choice of data types for colour set definitions, including simple types (dot, integer, string, boolean, enumeration) and 
 product types.


\item {\it Transitions} 
get assigned a guard, which is a Boolean expression over variables, constants and colour functions. The guard must be evaluated to true for the enabling of the transition. The trivial guard true is usually not explicitly given.
 
\item {\it Arcs} get assigned an expression; the result type of this expression is a multiset over the colour set of the connected place.

\item {\it Rate functions} may incorporate predicates, which are again Boolean expressions. They permit us to assign different rate functions for different colours (colour-dependent rates). Otherwise, the trivial predicate true is used; for examples see \SM. 

\end{itemize}

Elaborating the behaviour of a coloured Petri net involves the following
notions (see~\cite{BHM15}).

\begin{itemize}

\item {\it Transition instance:} 
The variables associated with a transition consist of the variables in the guard of the transition and in the expressions of adjacent arcs. For the evaluations of guards and expressions, values of the associated colour sets have to be assigned to all transition variables, which is called {\it binding}. A specific binding of the transition variables corresponds to a transition instance. Transition instances may have individual rate functions.
 
\item {\it Enabling:} Enabling and  firing of a transition instance are based on the evaluation of its guard and arc expressions. If the guard is evaluated to true and the pre-places have sufficient and appropriately coloured tokens, the transition instance is enabled and may  fire. In the case of quantitative nets, the rate function belonging to the transition instance is determined by evaluating the predicates in the rate function definition.
 
\item {\it Firing:} 
When a transition instance  fires, it consumes coloured tokens from its pre-places and produces coloured tokens on its post-places, both according to the arc expressions. which generally results into a new state. 

\end{itemize}

\paragraph{\it Colouring example.} Let us illustrate the use of colour by means of the stratified age SIR example (SIR-$\textrm{S}^\textrm{age}_\textrm{2}$), see Figure~\ref{figure:SIRage}.
The common pattern is a single basic SIR model.  
In order to design the coloured model, shown in Figure~\ref{figure:SIRage}~(Bottom), we introduce an enumerated 
colour set {\bf Strata = \{young, old\}} (where each element is internally mapped onto an integer) and
two variables of this colour set {\bf x} and {\bf y}, each of which can hold either {\bf young} or {\bf old}.
Arcs in this model are now labelled with expressions over these variables.  The bindings to the variables are
considered only in the local environment (incoming and outgoing arcs) of each transition.
The incoming and outgoing arcs of the coloured transition \node{Recover} are both labelled with the same
variable {\bf x} because that transition only operates within one stratum.
However the coloured transition \node{Infect} has connections both within one stratum and between strata; this
is accomplished by using the two variables {\bf x} and {\bf y} on the two incoming arcs from
\node{Susceptible} and \node{Infectious},
each of which independently hold as bindings either {\bf young} or {\bf old}, giving in total four
combinations, two of which standing for the inter-strata cross infections and the other two for the intra-stratum infections.
The {\bf x} represents the person who becomes infectious, and {\bf y} the infecting person.
Thus the outgoing arc from the transition to the place \node{Infectious} is labelled
with the multiset {\bf \{x,y\}} written as {\bf x++y}.  
If {\bf x} and {\bf y} hold the same colour, {\bf x++y} corresponds to an arc weight of 2 in the standard Petri net model.

\paragraph{\it Folding/Unfolding.} 
Any (standard) Petri net can be folded into a coloured Petri net comprising just a single (coloured) place
and a single (coloured) transition. Then, the entire structure is hidden in the colour definitions, which
can always be revealed by automatic unfolding. Generally, the decision regarding
how much structure to fold into colours is a matter of taste. We follow the rule that a model should be still completely comprehensible
by reading the structure and its colour annotations.  In general, folding of epidemic models should
preserve the pattern of the SIR (or related) model.  For illustration, we provide a coloured version of
the SIAR symptomatic/asymptomatic model in the Supplementary Material.

In principle, coloured Petri nets could be simulated directly on the coloured level, thus avoiding blowing up the model structure which often comes with unfolding. However, the corresponding algorithms are much more complicated, often involve an implicit partial unfolding, and only a few of them are actually available in practice~\cite{beccuti2015symmetric}.

In contrast, unfolding to the corresponding uncoloured Petri net can be efficiently achieved~\cite{SRL+20} and permits to re-use the rich choice of analysis and simulation techniques, which have been developed over the years for uncoloured Petri nets, including \SPN and \CPN, many of them are now supported by highly efficient libraries and reliable software tools.

This is why we exclusively adopt the second approach. All our coloured models are automatically unfolded when it comes to analysing and/or simulating them.

\paragraph{Encoding space and locality.}
The use of colour enables us to encode matrices, where colours (natural numbers) represent indices;
multidimensional matrices can be represented by appropriate colour tuples.  Previously we
have used this approach to encode one, two and three dimensional 
Cartesian space where coordinates were represented as tuples over the entire
matrix~\cite{GHLS13},
for example modelling diffusion and cell movement in biological
systems: planar cell polarity in the Drosophila wing~\cite{GGH+13}, phase variation in
bacterial colonies~\cite{PGH+15}, bacterial quorum sensing~\cite{GHGC19} and intra-cellular
calcium dynamics~\cite{IHAH20}.

In this research we use tuples to index sub-matrices, permitting us to represent graphs as
adjacency matrices and hence to encode geographical spatial relationships.
This overcomes the limitations of the previous Cartesian coordinate approach for representing
pandemics, because whole matrices represent travel diffusion-wise in a regular grid, rather than specific
travel connections.

For illustration we consider the encoding of the connectivity graph for four European countries, see Figure~\ref{figure:Countries4}~(Top). In order to achieve this we define a colour set of enumeration type comprising all countries and next a
matrix as a product over this colour set.

\newpage
\small{
\begin{verbatim}
colorsets:
  enum Countries = {BE,DE,ES,FR};
  Matrix = PROD(Countries,Countries);
\end{verbatim}
}
\normalsize

A subset of the matrix is defined by a Boolean expression (given in square brackets) comprising all required connections.

\small{
\begin{verbatim}
colorsets:
  Connections = Matrix[  // subset defined by Boolean expression
    (x=ES & (y=FR)) |
    (x=FR & (y=ES | y=BE | y=DE)) |
    (x=BE & (y=DE | y=FR)) |
    (x=DE & (y=FR | y=BE ))];
\end{verbatim}
}
\normalsize

Finally we define a colour function \node{connected} to be used as guard for the \node{Travel} transitions, which
constrains travel according to the connectivity graph. 

\small{
\begin{verbatim}
colorfunctions:
bool connected(Countries p, Countries q) { (p,q) elemOf Connections };
\end{verbatim}
}
\normalsize

This acts in the same way as the more simple constraint {\bf [x != y]} used in Figure~\ref{figure:SIRtravel2}.
Now, colour-dependent rates
may model different disease control policies in different subpopulations or in different locations.

Besides this, we need two variables of the colour set {\it Countries} to be used in the arc inscriptions and thus occurring as parameters in the transition guard.

\small{
\begin{verbatim}
variables:
  Countries : x;
  Countries : y;
\end{verbatim}
}
\normalsize

\paragraph{Framework.} These colouring principles can be equally applied to qualitative and quantitative Petri nets, yielding, among others, coloured \SPN (\SPNC) and coloured \CPN (\CPNC):
Our {\it PetriNuts} platform supports the handling of all of them, which includes the reading of a quantitative Petri net as either \SPN or \CPN  (supported by \Snoopy and \Spike) and the automatic unfolding of \SPNC and \CPNC to their uncoloured counterparts (supported by \Snoopy, \Marcie and \Spike).  

See 
Figure~\ref{fig:framework}
for an overview of the modelling paradigms integrated 
in the unifying framework of the {\em PetriNuts} platform and the relations among the supported net classes~\cite{HH+12}.

\paragraph{Petri net analysis techniques.} Petri net theory comes with a wealth of analysis techniques~\cite{Murata:1989}, some of them are particularly useful in the given context due to the specificity of the net structures of our \SPN/\CPN models, which we obtain by design.

\begin{itemize}

\item {\it Conservative:} A Petri net is called conservative, if the number of tokens is preserved by any transition firing. This is the case if it holds for all transitions that the sum of the weights of incoming arcs equals the sum of the weights of outgoing arcs. In our models, most of the transitions have exactly one incoming and one outgoing arc, with the exception of \node{Infect} transitions which have two incoming and two outgoing arcs (the latter combined to one arc with a weight of 2).

\item  {\it  P-invariant:}  A P-invariant (as it occurs in the models considered here) corresponds to a set of places, holding in total the same amount of tokens in any reachable state. Generally, P-invariants induce token-preserving subnets. A P-invariant is minimal, if (roughly speaking) it does not contain a smaller P-invariant; see~\cite{H09} for details. The models discussed in this paper have the remarkable property 
that the are covered by non-overlapping minimal P-invariants,
and each minimal P-invariant comprises the places (compartments) of one stratum.

Extending a model by cumulative counters, such as the total number of infectious over time, destroys the conservative net structure and induces  
additional overlapping P-invariants; thus the model is still covered with P-invariants (CPI); 
see Figure~\ref{figure:SC:model7} for an example.

More generally speaking, any epidemic/pandemic model without birth (transition without pre-place) and/or death (transition without post-place) is covered with P-invariants.

\end{itemize}



We use both criteria for model validation to avoid modelling mistakes.
They are independent of the initial marking and can be decided by exploring the net structure only. This is done in our {\it PetriNuts} platform by the tool \Charlie.
Both criteria prove independently the boundedness of a model (the number of tokens on each place is limited by a constant), which ensures in turn a finite state space. 

The finite state space could possibly open the door to analyse \SPN models by numerical methods to determine popular Markov properties~\cite{SH09}, such as the probability to reach a (transient) state where, e.g., half of the population is effected. A closer look on the typical size of the generated CTMC however soon questions this endeavour, as numerical methods rely on a real-numbered square matrix in the size of the state space~\cite{HRSS10}. So we basically have to confine ourselves to simulation and analysis techniques working over simulation traces, such as simulative model checking; see Methods subsection on {\it Analytical methods}.

It is obvious from exploring the model behaviours that they
reach a dead state (there is no enabled transition), as soon as the \node{infectious} compartment becomes empty, because there is no way to add something to the \node{infectious} compartment as soon as it got empty.  
In Petri net terms this corresponds to a 
bad siphon (a siphon not containing a trap); all these are further popular notions of Petri net theory, requiring structural analysis only, while permitting conclusions on behavioural properties; see~\cite{HGD08} for details.

Often, epidemic/pandemic models are by design irreversible, reflecting the (optimistic) assumption that a once gained immunity is never lost, and a model will reach a dead state at the very latest when the entire population is recovered. 
Loosening this assumption by adding a transition from either \node{Infectious} or \node{Recovered} to \node{Susceptible} and refilling \node{Infectious} in each epidemic model component will introduce cycles to a model which now becomes a chance to be reversible and live (which in turn precludes dead states), if it is covered with T-invariants (see~\cite{H09} for details). This obviously requires that the geographical connectivity graph is strongly connected.

\smallskip
\paragraph{Petri net simulation.} 
Besides structural analysis for model validation, we apply simulation techniques generating traces of model behaviour.
Simulation traces are time series reporting the current variable values at the $n+1$ time points 
$t_{i}$ of a specified  output grid $i = 0,\dots n$, typically splitting the simulation time into $n$ equally sized time intervals. 
We consider here two basic types of traces: 
 
\begin{itemize}

\item 
{\em traces of place markings}, i.e., time series of the current marking (state) of the compartments at the specified time points of the simulation run:

\smallskip
\hspace{2em}
$s(t_{i}): P \rightarrow \N$, with $s(t_{0}) = m_0$,
\smallskip

i.e., $s(t_{i})$ are (non-negative) integer place vectors over all place markings, indexed by the set of places $P$. 

\item
{\em traces of transition rates}, i.e., time series of the number of occurrences which
each individual event had in total in the latest time interval, 

\smallskip
\hspace{2em}
$v(t_{i}): T \rightarrow \N$, with  $v(t_{0}) = 0$,
\smallskip

i.e., $v(t_{i})$ are  (non-negative) integer transition vectors over all transition rates, indexed by the set of transitions $T$.

\end{itemize}
Reading a transition rate vector as Parikh vector immediately leads us to the state equation specifying the relation between both traces:

\smallskip
\hspace{2em}
$s(t_{i}) = s(t_{i-1}) + \C \cdot v(t_{i}), i=1\dots n$,
\smallskip

\noindent
where $\C$ is the incidence matrix of a Petri net (see~\cite{HGD08}).
Thus, the place trace can be derived from the transition trace, but generally not vice versa.
In the stochastic setting, the transition trace cannot be uniquely deduced from the place trace due to alternative and parallel transitions, which specifically holds for individual traces. Therefore, we directly 
record the transition traces during simulation. 

Often we consider {\em averaged traces\/}, and the deterministic simulation of \CPN/\CPNC is bound to do so, such that the individual values at each time point are non-negative real numbers ($\R+$) instead
of natural numbers ($\N$):

\smallskip
$s(t_{i}): P \rightarrow \R+$, with $s(t_{0}) = m_0$ {\em (traces of place markings),} 

\smallskip
$v(t_{i}): T \rightarrow \R+$, with  $v(t_{0}) = 0\;\;\;$ {\em (traces of transition rates).} 

\smallskip
Simulation traces may include coloured places/transitions, where a coloured node always gives the sum of the values of the corresponding unfolded nodes. For example, in the SIR-S$_2$ model given in Figure~\ref{figure:SIRage}, the coloured place \node{Susceptible} always shows the sum of its two unfolded nodes \node{Susceptible\_Young} and \node{Susceptible\_Old}; likewise for transitions and their rates.

\paragraph{Derived measures.} 
A measure often used in the daily news is the daily or weekly 
number of new infections, given at time point $t$ by 
$$\Delta^{I1}_t = S_t - S_{t-1} \;\; or \;\; \Delta^{I7}_t = S_t - S_{t-7}\; .$$
These can also be defined by:
$$\Delta^{I1}_t = (I_t - I_{t-1}) + ( R_t-R_{t-1}) \;\; or \;\;
\Delta^{I7}_t = (I_t - I_{t-7}) + ( R_t-R_{t-7})\; .$$ 

These numbers are typically normalised to a specific population size, e.g., 100,000, thus permitting the comparison of the local incidence numbers among regions.

A more scientific  measure for characterising the progress of an epidemic is the
reproduction number \rnumber~\cite{heffernan2005perspectives,vandendriessche2017reproduction} which
is the average number of secondary cases produced by one infectious individual introduced into a population of
susceptible individuals.
It allows modellers to work out the extent of the spread, but not the rate at which the infection grows~\cite{adam2020guide}.
This measure has several interpretations. \rnumber$_0$ assumes that everybody in a population is susceptible to infection; this is only true when a new virus is introduced into a population
which has never experienced it before and is a measure of the potential virulence of the disease.  
\rnumber$_t$ (sometimes called \rnumber$_e$, or `effective \rnumber'), 
is calculated over time as an epidemic progresses, and is a measure of the potential virulence of the disease.
The value of \rnumber$_t$ varies during the course of an epidemic, as people gain immunity after an infection or a vaccination, or have adjusted their
social interactions.
Thus generally it holds that \rnumber$_t$ $\leq$ \rnumber$_0$.

In our SIR model, we calculate \rnumber$_t$ for a time window of 1 by Equation~\ref{equation:reproduction_number}; the generalisation to wider time windows is straightforward.

\begin{equation}
\label{equation:reproduction_number}
\mathcal{R}_t = \Delta^{I1}_t / \Delta^{I1}_{t-1}
\end{equation}
\begin{equation*}
\mathcal{R}_t = (S_t - S_{t-1})/(S_{t-1}-S_{t-2}) 
\end{equation*}
\begin{equation*}
\mathcal{R}_t = ((I_t - I_{t-1}) + ( R_t-R_{t-1})) / ((I_{t-1}-I_{t-2}) + (R_{t-1}-R_{t-2}))
\end{equation*}

\paragraph{Naming convention.} The naming convention of our models basically follows this Extended BNF.
For a summary of the models provided see Table~\ref{table:models}.
\\
\\
\footnotesize{
ModelName ::= [PandemicComponent] EpidemicComponent [`-' StratifiedComponent] \\
PandemicComponent ::= `P' [Parameter] [PandemicRest]\\
PandemicRest ::= `Q'\\
Parameter ::= Number | Country\\
Country ::= `USA' | `China' \\
EpidemicComponent ::= `S' `I' [EpidemicOption] `R' \\
EpidemicOption ::= `A' | `V' | `Q' \\
StratifiedComponent ::= `S' \{StratificationParameter\}$^+$ \\
StratificationParameter ::= Type  Number \\
Types ::= age | gender | $\epsilon$ 
}
\normalsize

\begin{table}[h!]
\caption[Summary table of models]{Summary table of models.
I Basic and extended epidemic models. II Stratified models. III Pandemic models. IV Combined models; CANDL files provided as separate files in \SM.}
\vspace{1em}
\resizebox{\textwidth}{!}{ 
\begin{tabular}{|l|l|p{20em}|l|}
\hline
Type & Id & Long name & Reference\\
\hline 
I  & SIR  & standard SIR model & Figure~\ref{figure:SIR} \\
  & SIVR & SIR with disease variant & Figure~\ref{figure:SIVR}\\ 
 & SIQR & SIR with local quarantine & Figure~\ref{figure:SIQR-SIAR}~(Top)\\
 & SIAR & SIR with asymptomatic infectious & Figure~\ref{figure:SIQR-SIAR}~(Bottom)\\
\hline
II & SIR-S & Stratified SIR & \\
   &  SIR-$\textrm{S}^\textrm{age}_\textrm{2}$ & Stratified SIR, 2 age
strata & Figure~\ref{figure:SIRage}~(Top/Middle) \\
   &  SIR-S$_{10}$ & Stratified SIR, 10 age strata & \\
   &  SIR-$\textrm{S}^\textrm{age}_\textrm{20}$ & Stratified SIR, 20 age
strata & Figure~\ref{figure:SIRage}~(Bottom) \\
   &  SIR-$\textrm{S}^\textrm{age,gender}_\textrm{10x2}$ & Stratified SIR, 10 age and 2 gender strata & \\
\hline
III & PSIR & Pandemic SIR & \\
  & P$_n$SIR & Pandemic SIR, $n$ locations &
Figure~\ref{figure:SIRtravel2}:(n=2), Figure~\ref{figure:Countries4}:(n=4)\\
  & P$_{10}$SIR & Pandemic SIR with CummulativeInfectious, Western Europe
(10 countries, 15 mutual connections) & Figure~\ref{figure:SC:model6}, P10SIR.candl\\ 
  & P$_{48}$SIR & Pandemic SIR for Europe (48 countries, 85 mutual connections) & P48SIR.candl \\
  & P$_\textrm{China}$SIR & Pandemic SIR for China (34 provinces, 71 mutual connections) & PchinaSIR.candl \\
  & P$_\textrm{USA}$SIR & Pandemic SIR for USA (50 states, 105 connections) & PusaSIR.candl  \\
\hline
IV  & P$_{48}$SIR-S$_{10}^{age}$ & Pandemic SIR, Europe, 10 age strata &
Figure~\ref{figure:Countries4} (Bottom), Figure~\ref{figure:P48SIR}\\
  & P$_{10}$Q$_\textrm{I}$SIR &  Pandemic SIR with may-quarantine on
arrival for \node{Infectious}, Western Europe  & Figure~\ref{figure:SC:model7} (Top)\\
  & P$_{10}$QSIR &  Pandemic SIR with must-quarantine on arrival for all
compartments, Western Europe  & Figure~\ref{figure:SC:model7} (Bottom)\\
\hline
\end{tabular}}
\label{table:models}
\end{table}
\normalsize


\subsection{Parameter fitting}


\paragraph{Random Restart Hill Climbing (RRHC).}
There are two types of optimisation which are systematic and local search~\cite{russell2013artificial}. 
Systematic approaches store information about the path to the target,
whereas local searches do not retain such information. Local searches have two key advantages:
(1) they use very little memory due to not storing the path, and (2): they can often find
reasonable solutions in large state spaces~\cite{russell2013artificial}. 

One type of local search
algorithm is the Hill-Climbing algorithm (HCA). This algorithm generates a random solution and
makes small changes to the solution to increase the fit to target data. The HCA is greedy as
it only accepts a solution better than its current state. As a result, the HCA can get stuck
firstly, in a local maximum,
which is a peak higher than each of its neighbours but lower than the global maximum;
secondly, in a flat local maximum, which is a local maximum that has many solutions and is
difficult for the algorithm to navigate; and finally, at a shoulder, which is a flat region of
space that is neither a local or global maximum~\cite{russell2013artificial}. To overcome such
sticking points the HCA can be modified into a stochastic Random-Restart Hill-Climbing
algorithm (RRHCA) by defining a number of times to re-run the whole program. The final
solution is the best of all runs. 

The goals of the fitting were 
(Objective~1) to reproduce the order of the time taken from the first 5 deaths to a peak for daily new cases, 
(Objective~2) to reproduce the order of the height of the peaks for daily new cases, and
(Objective~3) combining Objective~1 and Objective~2.
Fitting was performed in all three cases using target driven optimisation employing
Random Restart Hill Climbing.
RRHC was used for parameter optimisation
with code developed in
Python which called \Spike.
This
Python program automatically optimises input parameters to better represent the target data. Using
this approach, the three objectives were investigated for West Europe10. 

First, optimisation was
performed for the order of peak new daily infections (PNDI). Each country was given a unique
letter and a string produced to represent the order of peaks, for example: France=A, Spain=B, 
if Spain peaked first the string was ``BA'', else it was ``AB''. 
The fitness function for Objective~1 was achieved by encoding the required country order
for peak time as a ten character string, and minimising
the Levenshtein distance between the target string and the corresponding string representing the
order of peaks obtained from the model.  Results are given in 
Table~\ref{table:sc:table9}.
A smaller distance symbolises model
predictions to be closer to the observed data, and the RRHC algorithm searched for country specific and
inter-country infection parameters that reduced the distance measure for this objective and the
two below.

Secondly, optimisation was performed for the magnitude of PNDI by minimising the result of
Equation~\ref{equation:sc:equation3}. The
measure produces a normalised value between 0 and 1, where 0 represented a perfect fit
between target and modelled data, and 1 the worst possible fit, 
see Table~\ref{table:sc:table10}. 
Finally,
for Objective~3, optimisation was performed for both the order and magnitude of PNDI. The distance measure used
was a combination of the LD, and the measure in Objective~2, see Equation~\ref{equation:sc:equation4}.
The distance measure produces a
normalised value between 0 and 1, where 0 represented a perfect fit between modelled and
predicted data, and 1 the worst fit. Order and magnitude were given equal weight; see infection rates in 
Table~\ref{table:sc:table11},
and travel rates in 
Table~\ref{table:sc:fittedconnectionrates}.

\begin{equation}
\small{
\textrm{\small distance}_{\textrm{\small M}} = \frac{
|\frac{\textrm{Target}_1 - \textrm{Model}_1}{\textrm{Max}(\textrm{Target}_1,\textrm{Model}_1)}| +
|\frac{\textrm{Target}_2 - \textrm{Model}_2}{\textrm{Max}(\textrm{Target}_2,\textrm{Model}_2)}| + \ldots +
|\frac{\textrm{Target}_n - \textrm{Model}_n}{\textrm{Max}(\textrm{Target}_n,\textrm{Model}_n)} |
}
{\small n}}
\label{equation:sc:equation3}
\end{equation}

\begin{equation}
\textrm{\small distance}_{\textrm{\small M+O}} =
\frac{\textrm{\small distance}_{\textrm{\small M}} + \frac{LD}{n}}{2} 
\label{equation:sc:equation4}
\end{equation}

\smallskip

\paragraph{Pseudo code for RRHCA:}
\small{
\begin{verbatim}
Input: Randomly generated solution 
Output: Best solution of all random starting solutions

1  for x = 1 to STARTS do: 
2    S(R) = Randomly generated solution  
3    F(R) = Fitness of S(R)
4    for i = 1 to ITERATIONS do:
5       S(C) = Small change to S(R)
6       F(C) = Fitness of S(C)
7       if F(C) better than F(R) then:
8           S(R) = S(C)
9       end if
10   end for
11   return S(R)
12  end for 
\end{verbatim}
}

\smallskip
\paragraph{Levenshtein distance.} 
  The  Levenshtein distance~\cite{levenshtein1966binary} between two strings a,b (of length $\mid a\mid$ and $\mid b\mid$ respectively) is given by $lev_{a,b}(\mid a \mid , \mid b \mid)$, where
      $$
      lev_{a,b}(i,j)=
      \begin{cases}
       max(i,j) & \text{\hspace{15.5em}If min$(i,j)$=0}\\
       min & \begin{cases}
                lev_{a,b}=(i-1,j) + 1\\
                lev_{a,b}=(i,j-1) + 1 ,\hspace{1.9em}& \text{otherwise}\\
                lev_{a,b}= (i-1,j-1) + 1_{(a_{i}\neq b_{j})}
             \end{cases}
       \end{cases}
       $$
    \\ where $1_{(a_{i}\neq b_{j})}$ is the indicator function equal to 0 when $a_{i}=b_{j}$ and equal to 1 otherwise.  A {\it normalised} edit distance between two strings can be computed by
    $$lev\_norm_{a,b} = \dfrac{lev_{a,b}}{max(\mid a \mid , \mid b \mid)}$$

\subsection{Analytical methods}


\paragraph{Linear Temporal Logic and Simulative Model checking.}

Model checking permits us to determine if a model fulfils given properties 
specified in temporal logics, e.g. probabilistic linear-time temporal logic
PLTL~\cite{donaldson:cmsb:2008}.
Here we employ the simulative model checker {\it MC2}~\cite{donaldson:cmsb:2008} 
over time series traces of behaviours generated by simulation 
of epidemic and pandemic models.
Our simulation platforms (\Snoopy, \Marcie, \Spike) not only support the checking of time series behaviours of (uncoloured/coloured) places/compartments but
also of  (uncoloured/coloured) transitions/events, for example \node{Infect}, \node{Recover} and \node{Travel}, or derived measures over both (observers).
All of these can be subsumed under the general term {\em observables}.

The basic element of any temporal logic is the Atomic Proposition (AP) which is part of a property without
temporal operators. 
The syntax of PLTL is: 

\medskip
$\phi  ::= X \phi \;|\;G\phi \;|\;F \phi \;|\;\phi U \phi \;|\;\phi R\phi \;|\;\phi  \lor  \phi \;|\;\phi  \land \phi \;|\;\neg\phi \;|\;\phi  \rightarrow \phi \;|\;AP$


\medskip

$AP ::= AP \lor AP \;|\;AP \land AP \;|\;\neg AP \;|\;AP \rightarrow AP \;|\;$

\hspace{3em}
$z = z\;|\;z \neq z\;|\;z > z\;|\;z \geq z\;|\;z < z\;|\;z \leq z\;|\;true\;|\;false$

\medskip

$z ::= z + z\;|\;z - z\;|\;z * z\;|\;z/z\;|\;\text{<<}x\text{>>}\;|\;d(\text{<<}x\text{>>})\;|\;max(\text{<<}x\text{>>})\;|\;Int\;|\;Real$

\smallskip
where $\text{<<}x\text{>>}$ is the name of any observable in the simulation output, 
Int is any integer number, Real is any real number,
and $d$ is the differential operator.

The temporal operators are:
\begin{itemize}
\item Next (X) - The property must hold true in the next time point.
\item Globally (G) - The property must hold true always in the future.
\item Finally (F) - The property must hold true sometime in the future.
\item Until (U) - The first property must hold true until the second property holds true.
\item Release (R) - The second property can only ever not hold true if the first property becomes true.
\end{itemize}


When checking small models with few variables, it is possible to check the properties of specific observables,
for example that of \node{infectious} in a simple SIR model.  However, when checking larger models with many
observables, for example P$_{48}$SIR-S$_{10}$, we run model checking for a set of observables against a library
of PLTL properties, and employ {\em meta-variables} indicated by double angle brackets in our queries, to be substituted one by one by all  
observables in the set:

\smallskip
$P_{\ge1}[\,G(\text{<<}x\text{>>}>10)\,]$  

\smallskip

\noindent
which ``Holds for any observable whose value is always greater than 10'', which in turn could be generalised to a general
pattern ``$x$ is always greater than some threshold''.

A library of appropriate property patterns, provided in the \SM, allows us to categorise all model observables
into (not necessarily disjunctive) sets fulfilling  the individual property patterns. 
Example patterns include

\smallskip
\noindent
(1) {\em Always zero}.  When applied to places, this indicates that a place has never been initialised to a
value greater than zero, and that it never participates in the progress of the model.  For transitions it
indicates those which are never active, indicating that they are part of the network that is always dead.

\smallskip
$P_{\ge1}[\,G(\text{<<}x\text{>>}=0\,)\,]$,
which is equivalent to $P_{\ge1}[\,\neg F(\text{<<}x\text{>>}\neq 0\,)\,]$


\medskip
\noindent
(2) {\em At some point greater than zero, and eventually zero.}  Places which have some marking, and then
hold a zero value at then end of the simulation; transitions with changing activity and finally a steady
state of zero activity. 

\smallskip

$P_{\ge1}[\, F(d(\text{<<}x\text{>>}) \neq 0) \wedge F(G(\text{<<}x\text{>>}=0\, \wedge\, d(\text{<<}x\text{>>})=0)) \,]$ 

\medskip
\noindent
(3) {\em Activity peaks and falls:}
\smallskip

$P_{\ge1}[\,F(d(\text{<<}x\text{>>})>0) \,\wedge\, (d(\text{<<}x\text{>>})>0 \;U\; G(d(\text{<<}x\text{>>})<0))\,]$

\medskip
\noindent
(4) {\em Activity peaks and falls then steady state:}

\smallskip

$P_{\ge1}[\,F(d(\text{<<}x\text{>>})>0) \,\wedge\, (d(\text{<<}x\text{>>})>0 \;U\;$ \\
\hspace*{3.3em}$(F(d(\text{<<}x\text{>>})<0 \;\wedge\; G(d(\text{<<}x\text{>>})<0 \;U\;
G(d(\text{<<}x\text{>>}))=0)))\,]$ 

\medskip
\noindent
(5) {\em At least one peak:}

\smallskip

$P_{\ge1} [\,F((d(\text{<<}x\text{>>})>0 ) \,\wedge\, F (( d(\text{<<}x\text{>>})<0)))\,]$

\medskip
\noindent
(6) {\em At least two peaks:}

\smallskip
$P_{\ge1} [\,F((d(\text{<<}x\text{>>})>0 ) \,\wedge\, F (( d\text{<<}x\text{>>})<0) \,\wedge\,
F((d(\text{<<}x\text{>>})>0 ) \,\wedge\, F (( d\text{<<}x\text{>>})<0)))))\,]$

\bigskip
We have applied this to P$_\textrm{China}$SIR, and discovered some provinces with multiple peaks in \node{Infectious} when the infection was initiated
in Hubei province (the capital of which is Wuhan), due to their geographical placement, see 
Figure~\ref{figure:MCpeaks}.


\paragraph{Cluster analysis} is a method of creating groups where objects in one group are very similar and
distinct from other groups~\cite{gan2020data}. In particular, agglomerative hierarchical cluster
analysis was performed, which sequentially combines individual elements into larger clusters
until all elements are in the same cluster. 
Time-series cluster analysis was performed on the \node{Infectious} compartment in non-fitted
Europe10 
(Figure~\ref{figure:SC:Figure25}) 
and  non-fitted Europe48 
(Figure~\ref{figure:SC:Figure26}), 
where the infection was started in all locations. 
The distance between time-series traces of Infectious
compartments for each country was computed using Euclidean distance and a distance matrix was
constructed. Clustering was performed using the complete link method, which takes the largest
distance between clusters to construct a dendrogram.

\paragraph{Bar charts.} 
Exploratory data analysis involves using graphs and summary statistics to explore data;
this was exploited largely by plotting model outputs
for visualisation to ascertain whether or not the models were behaving as expected. The bar
chart in 
Figure~\ref{figure:SC:Figure30} 
was produced using simulation data exported
as CSV files, and generated in Excel.

\paragraph{Correlation matrices} 
were computed to assess model behaviour in more depth and was used to validate both 
unfitted and fitted models. 
Non-fitted models were simulated several times with the infection
starting in a different location, and then with all locations infected. The
magnitude and timing of peak infections were detected for all simulations and combined into a
single data frame. 
For the P$_{10}$SIR fitted model, the best result from the RRHCA
optimisation was taken, and peak and timing of peaks were detected. Data were loaded into R
and real-world data on country population density, country size, and the number of borders a
country had were joined. Correlations between variables were computed to determine the
parameters that were closely linked to higher numbers of infections. The correlation matrix
was then plotted to visualise and statistical significance determined at $p < 0.05$.
Non-significant correlations were displayed with a black cross.
For details see 
Table~\ref{table:sc:table7}
and
Figures~\ref{figure:SC:Figure23}~--\ref{figure:SC:Figure24}.

\subsection{Hardware and software requirements}

\paragraph{Hardware.} A standard computer with a Windows, Mac or Linux operating system (preferably 64-bit). The minimum RAM required is 8GB; depending on the size of the constructed model, more memory may be needed. More heavy simulation experiments take advantage of multiple cores.

\paragraph{Software.} The research reported in this paper has been achieved by help of the {\it PetriNuts} platform, which is also required to reproduce the results. 
The platform  comprises several tools, which can be downloaded from http://www-dssz.informatik.tu-cottbus.de/DSSZ/Software. All these tools can be installed separately, are free for non-commercial use, and run on Windows, Linux, and Mac OS.

\begin{itemize}

\item \Snoopy~\cite{HH+12} is
Petri net editor and simulator supporting various types of hierarchical (coloured) Petri nets, including all Petri net classes used in this paper, with an automatic conversion between them. Snoopy supports several data exchange formats, among them the Systems Biology Markup Language (SBML, level 1 and 2). For communication within the {\it PetriNuts} platform, Snoopy reads and writes ANDL (Abstract Net Description Language) and CANDL (Coloured ANDL) files, and writes simulation traces (places, transitions, observers) as CSV files. 
The ODEs induced by a \CPN or \CPNC can be exported to LaTeX, as plain text, or be written to be read by Octave, Matlab, or  ERODE~\cite{cardelli2017erode}.
Snoopy's proprietary file format uses XML technology; default file extensions indicate the net class (i.e., pn, spn, cpn, colpn, colspn, colcpn). A Snoopy2LATEX generator supports documentation. 

\item \Patty is a JavaScript to support Petri net animation (token flow) in a web browser; it does not require any installation on the user's site. \Patty reads (uncoloured) Petri nets in Snoopy's proprietary format. All uncoloured Petri nets given in this paper can be animated 
in a purely qualitative manner at \url{https://www-dssz.informatik.tu-cottbus.de/DSSZ/Research/ModellingEpidemics}.

\item \Charlie~\cite{HSW15} is an
analysis tool of (uncoloured) Petri net models applying standard techniques of Petri net theory (including structural analysis, such as conservativeness, and invariant analysis), complemented by explicit CTL and LTL model checking. Charlie reads ANDL files, and writes some analysis results (invariants, siphons, traps) into files, to be read by Snoopy for visualisation; for details see~\cite{BHM15}.

\item \Marcie~\cite{HRS13} is
Model checker for qualitative Petri nets and \SPN, and their coloured counterparts; it combines exact analysis techniques gaining their efficiency by symbolic data structures (IDD)~\cite{HRST16} with approximative analysis techniques building on fast adaptive uniformization (FAU) and parallelized stochastic simulation (Gillespie, tau leaping, delta leaping). It supports CTL, CSL and PLTLc model checking. Marcie reads ANDL and CANDL files, and writes stochastic simulation traces (places, transitions, observers) as CSV files. 

\item \Spike~\cite{CH19} is a
command line tool for reproducible stochastic, continuous and hybrid simulation experiments of large-scale (coloured) Petri nets. 
Spike reads a couple of file formats, among them are SBML, ANDL and CANDL, and writes simulation traces (places, transitions, observers) as CSV files.

\Snoopy, \Marcie and \Spike share a library of simulation algorithms, comprising 
four stochastic simulation algorithms (besides Gillespie's exact SSA, three approximative algorithms) and a  couple of stiff/unstiff solvers for 
continuous simulation (some of them use the external library SUNDIAL CVODE~\cite{Hindmarsh05});
see~\cite{CH19} for a quick reference.

\item {\it MC2}~\cite{donaldson:cmsb:2008} is
a Monte Carlo Model Checker for LTLc and PLTLc, operating on stochastic, deterministic, and hybrid simulation traces or even wetlab data, given as CSV files. We use {\it MC2} to analyse deterministic and stochastic traces generated by \Snoopy or \Spike. 

\end{itemize}

Additionally, we recommend the use of several third-party tools for post-processing of simulation traces.

\begin{itemize}
\item 
R - 3.6.2 and R Studio - Version 1.1.463, with packages
dplyr 0.8.5,
ggplot2 3.3.0,
openxlsx 4.1.4,
corrplot 0.84,
RColorBrewer 1.1-2,
dtwclust 5.5.6,
reshape2 1.4.3.
\item Python - Version 3.6.5  and  Visual Studio Code - 1.42.1, used for
\begin{itemize}
\item
RRHCA (Random Restart Hill Climbing) 
using packages pandas 1.1.0, Levenshtein 0.12.0, and packages pre-installed with Python: os, time, random, datetime.

\item
Python Web Scraper, used to to collect COVID19 Data, 
using packages urllib3 1.25.10, requests 2.24.0, lxml 4.5.2, bs4 0.0.1.
\end{itemize}

\item Excel 16.40

\end{itemize}


\section{Discussion }
\label{section:discussion}

\paragraph{Model engineering} is
in general the science of designing, constructing and analysing computational models.  We
have previously discussed this concept in the context of biological models~\cite{HG13,BHM15}, and the
principles are equally applicable to epidemic and pandemic models. 
Major objectives of this approach are the sound reusability of models, and the reproducibility of their simulation
and analysis results, both being facilitated by the use of colour in Petri nets.
In terms of the models presented in this paper, we employ an approach based on orthogonal extensions of the
basic SIR model, enabling robust step-wise model development.  
Examples include the extension of epidemic SIR
models to pandemic models in a reusable manner, where models can be scaled by modifying stratification colour
sets, or adjusted to different geographies by merely replacing the related colour definitions.
More specifically, model engineering in this context comprises three aspects: engineering model structure,
spatial aspects, and rates.

\paragraph{Engineering model structure.}
We exploit the concept of colour as supported in coloured Petri nets, which facilitates parameterised repetition
of model components, as for example required to encode stratified populations.
This enables modelling at a higher level of abstraction akin to principles in high level programming languages.
Abstraction generally reduces design errors in large and complex models, but can introduce some implementation errors due to
misunderstanding of the coloured annotations.

This calls for robust model validation before entering the simulation phase.
With increasing model size, this can
hardly be accomplished by visual inspection only. Petri nets offer
structural analysis techniques for this purpose.
For example, all
the models presented in this paper are by construction 
covered by 
P-invariants, and most models are also conservative. Both properties can
be easily checked using our Petri net analysis tool \Charlie (see Methods section for details). Besides
this, animation of discrete Petri nets offers the option to play
with a net in order to follow the token flow, suggested as an early check
to reveal design faults.  

Our use of colour enables us to construct multilevel models in a sound manner.  Specifically they encode two
levels, the lower one being based on an SIR or its extensions, and the upper level comprising a network of geographical
connections, resulting in a metapopulation model~\cite{brauer2017mathematical}.  
The models are multiscale in terms of distances at the upper level; regarding
time, we assume that geographical connections, i.e. travel, occur at much lower rates than infections in
the lower level SIR components.

\paragraph{Engineering spatial aspects.}
The crucial difference between epidemic and pandemic models is the notion of space and connectivities permitting
population movement
and hence the geographic transmission of disease.
Our approach which is based on colour and associated functions enables the encoding, for example, of
set operators. 
This in turn facilitates model design based directly on abstract representations of spatial relationships as
graphs.  Modification of geographical relationships then merely involves modifying the corresponding graph at
the coloured high level, rather than attempting to change the expanded and more complex unfolded low level.
This all occurs within one smooth coherent framework of coloured Petri nets, implemented seamlessly in our
{\it PetriNuts} platform.

Petri nets can be designed in a graphical or textual way, depending on personal preferences. 
A combination of both is also possible; for example, designing the Petri net structure graphically and 
writing all colour-related definitions using CANDL; see \SM for details.

\paragraph{Engineering rates}


The increase in model size results in an increase both in the number of kinetic parameters that need to be
fitted, as well as an increase in the complexity of their interdependencies.
Our paper is a methodology paper, and parameter fitting has been included as an illustration of what can be done
with our approach.  


In this paper we have discussed models where the rate constants are fixed for the
entirety of a simulation.
Modelling lockdown situations, however, requires that rate constants are changed 
dynamically during a simulation run.
The imposition of lockdown measures which are driven by social disease control policies
effectively diminishes the infection 
rate constant, and likewise lifting of the measures increases this value.
These changes are typically event driven, in response to changing \rnumber$_t$ values
for example.
We have recently extended the \Spike simulator to include event-driven triggers, which enables the modelling of these
control measures, see for example 
Figure~\ref{figure:JC:lockunlock}.
The methodology enables the imposition of changes in an
some incremental manner, in order to better represent the 
transitional nature of the uptake of lockdown measures in practice.
In more general terms modelling features such as the 
dynamic modification of rate constants is a step towards self-adaptable models.



\paragraph{Acknowledgments}
We would like to thank George Assaf, Jacek Chodak, Robin Donaldson, Mostafa Herajy, Ronny Richter, Christian Rohr, Martin Schwarick and many
students at the Brandenburg Technical University, Cottbus, Germany for developing the {\it PetriNuts} platform. 

\subsubsection*{Authors}

{\bf Shannon Connolly} is a Masters graduate in Data Science and Analytics from Brunel University London, UK. She currently works as an analyst helping to model data for healthcare, pharmaceutical and financial clients, as well as working on tools to automate parts of the systematic review process.

{\bf David Gilbert} is a Professor of Computing in the Department of Computer Science, Brunel University London, UK.  His research
interests include systems biology: modelling and analysis of biological systems; synthetic biology: computational design of novel biological systems.

{\bf Monika Heiner} is a Professor of Computing Science in the Department of Computer Science, Brandenburg Technical University, Cottbus, Germany.  Her research interests include modelling and analysis of technical as well as biological systems using qualitative and quantitative Petri nets, model checking and simulation techniques.

\normalsize

\newpage
\appendix
\counterwithin{figure}{section}
\counterwithin{table}{section}

\newpage
\addcontentsline{toc}{section}{References}
\bibstyle{alpha}
\bibliography{epidemics}

\newpage
\section*{Appendix}
\addcontentsline{toc}{section}{Appendix - Supplementary Material}

\section{PetriNuts Platform -- Quick Reference}

For newcomers to the {\em PetriNuts} platform we provide guidelines for some basic use cases.

\subsection{The Framework}

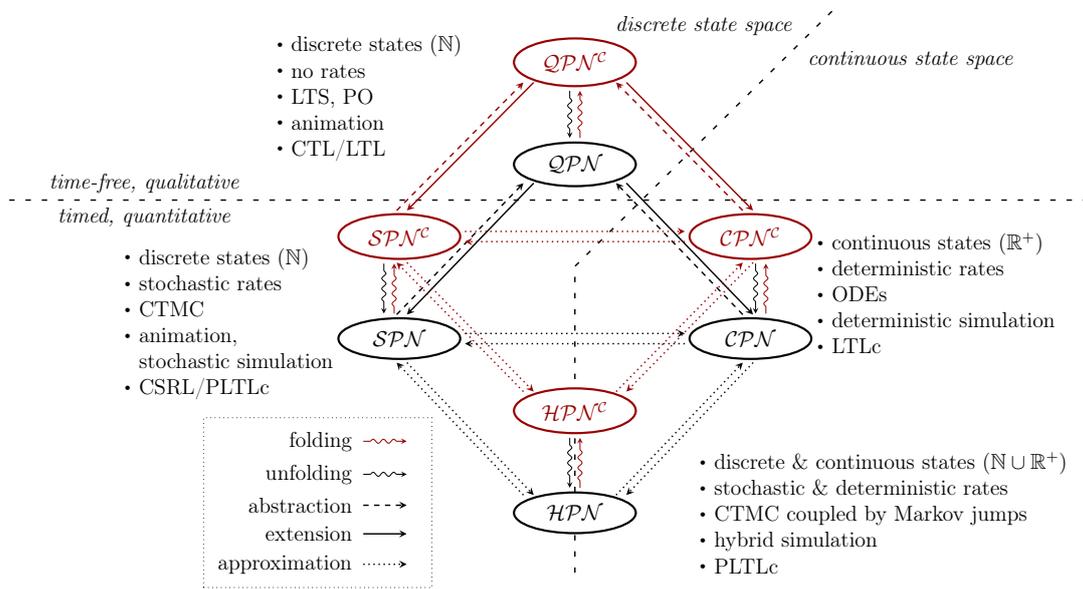
\begin{figure} [!th]
  \resizebox{\columnwidth}{!}{
		\begin{tikzpicture}[node distance=2cm,>=stealth,auto]
	\definecolor{darkred}{rgb}{.6,0,0}
	\tikzstyle{myred}=[color=darkred]
    \tikzstyle{comment}=[align=left]
    \tikzstyle{node}=[draw,ellipse,text width=1.4cm,align=center,fill=white,very thick]
    \tikzstyle{unfolding}=[->, decorate, decoration={snake,amplitude=.4mm,segment length=2mm,pre length=2mm, post length=2mm}, shorten <=2pt, shorten >=2pt]
    \tikzstyle{folding}=[->, decorate, decoration={snake,amplitude=.4mm,segment length=2mm,pre length=2mm, post length=2mm},  myred, shorten <=2pt, shorten >=2pt]
    \tikzstyle{abstraction}=[->, thick, dashed, shorten <=2pt, shorten >=2pt]
    \tikzstyle{extension}=[->, thick, shorten <=2pt, shorten >=2pt]
    \tikzstyle{approximation}=[->, thick, dotted, shorten <=2pt, shorten >=2pt]
    \tikzstyle{line}=[-, thick, loosely dashed]

	\draw[line] 
        (-11,-2.7) --
		node[align=center,above,very near start]{{\it time-free, qualitative}}
		node[align=center,below,very near start]{{\it timed, quantitative}}
	    (10,-2.7);
	\draw[line] 
		(5,1) -- 
		node[align=center,above left,very near start]{{\it discrete state space}}
		node[align=center,below right,very near start]{{\it continuous state space}}
		(0,-4) -- (0,-10);

	\node(QPNc)[myred,node]{\QPNC };
	\node(QPN)[node,below of=QPNc]{ \QPN };

	\node(SPNc)[myred,node, below left of=QPN, xshift=-2cm]{ \SPNC };
	\node(SPN)[node,below of=SPNc]{ \SPN };

	\node(CPNc)[myred,node, below right of=QPN, xshift=2cm]{ \CPNC };
	\node(CPN)[node,below of=CPNc]{ \CPN };

	\node(GHPNc)[myred, node, below right of=SPN, xshift=2cm]{ \HPNC };
	\node(GHPN)[node,below of=GHPNc]{ \HPN };

	\draw[folding] ($(QPN.north)+(.1,0)$) --  ($(QPNc.south)+(.1,0)$);
	\draw[unfolding] ($(QPNc.south)-(.1,0)$) -- ($(QPN.north)-(.1,0)$);
	\draw[folding] ($(SPN.north)-(.1,0)$) --  ($(SPNc.south)-(.1,0)$);
	\draw[unfolding] ($(SPNc.south)-(.3,0)$) -- ($(SPN.north)-(.3,0)$);
	\draw[folding] ($(CPN.north)+(.3,0)$) --  ($(CPNc.south)+(.3,0)$);
	\draw[unfolding] ($(CPNc.south)+(.1,0)$) -- ($(CPN.north)+(.1,0)$);
	\draw[folding] ($(GHPN.north)+(.1,0)$) --  ($(GHPNc.south)+(.1,0)$);
	\draw[unfolding] ($(GHPNc.south)-(.1,0)$) -- ($(GHPN.north)-(.1,0)$);

	\draw[extension, myred] ($(QPNc.south west)+(.1,0)$) --  ($(SPNc.north)+(.1,0)$);
	\draw[abstraction, myred] ($(SPNc.north)-(.1,0)$) -- ($(QPNc.south west)-(.1,0)$);
	\draw[extension, myred] ($(QPNc.south east)+(.1,0)$) --  ($(CPNc.north)+(.1,0)$);
	\draw[abstraction, myred] ($(CPNc.north)-(.1,0)$) -- ($(QPNc.south east)-(.1,0)$);
	
	\draw[extension] ($(QPN.south west)+(.1,0)$) --  ($(SPN.north)+(.1,0)$);
	\draw[abstraction] ($(SPN.north)-(.1,0)$) -- ($(QPN.south west)-(.1,0)$);
	\draw[extension] ($(QPN.south east)+(.1,0)$) --  ($(CPN.north)+(.1,0)$);
	\draw[abstraction] ($(CPN.north)-(.1,0)$) -- ($(QPN.south east)-(.1,0)$);

	\draw[approximation,myred] ($(SPNc.east)+(0,.1)$) --  ($(CPNc.west)+(0,.1)$);
	\draw[approximation,myred] ($(CPNc.west)-(0,.1)$) --  ($(SPNc.east)-(0,.1)$);
	
	\draw[approximation] ($(SPN.east)+(0,.1)$) --  ($(CPN.west)+(0,.1)$);
	\draw[approximation] ($(CPN.west)-(0,.1)$) --  ($(SPN.east)-(0,.1)$);
			
	\draw[approximation, myred] ($(CPNc.south)+(.1,0)$) --  ($(GHPNc.north east)+(.1,0)$);
	\draw[approximation, myred] ($(GHPNc.north east)-(.1,0)$) -- ($(CPNc.south)-(.1,0)$);
	\draw[approximation, myred] ($(SPNc.south)+(.1,0)$) --  ($(GHPNc.north west)+(.1,0)$);
	\draw[approximation, myred] ($(GHPNc.north west)-(.1,0)$) -- ($(SPNc.south)-(.1,0)$);
	
	\draw[approximation] ($(CPN.south)+(.1,0)$) --  ($(GHPN.north east)+(.1,0)$);
	\draw[approximation] ($(GHPN.north east)-(.1,0)$) -- ($(CPN.south)-(.1,0)$);
	\draw[approximation] ($(SPN.south)+(.1,0)$) --  ($(GHPN.north west)+(.1,0)$);
	\draw[approximation] ($(GHPN.north west)-(.1,0)$) -- ($(SPN.south)-(.1,0)$);

	\node[comment,left of=QPNc, xshift=-2cm, yshift=-0.7cm]
	    {
		{\tiny$^\bullet$} discrete states ($\N$)\\
		{\tiny$^\bullet$} no rates\\
		{\tiny$^\bullet$} LTS, PO\\
		{\tiny$^\bullet$} animation\\
		{\tiny$^\bullet$} CTL/LTL
	    };

	\node[comment,below left of=SPNc, xshift=-1.9cm, yshift=-0.3cm]
	    {
		{\tiny$^\bullet$} discrete states ($\N$)\\
		{\tiny$^\bullet$} stochastic rates\\
		{\tiny$^\bullet$} CTMC\\
		{\tiny$^\bullet$} animation,\\
		{\tiny$\;\;$} stochastic simulation\\
		{\tiny$^\bullet$} CSRL/PLTLc
	    };

	\node[comment,below right of=CPNc, xshift=2.2cm, yshift=0.3cm]
	    {
		{\tiny$^\bullet$} continuous states ($\R+$)\\
		{\tiny$^\bullet$} deterministic rates\\
		{\tiny$^\bullet$} ODEs\\
		{\tiny$^\bullet$} deterministic simulation\\
		{\tiny$^\bullet$} LTLc
	    };

	\matrix [draw,dotted,column sep=1cm,left of=GHPN,xshift=-3cm, yshift=0.2cm] 
	{
		\node[left](a) {folding}; & \node (b) {}; \\ 
		\node[left](c) {unfolding};	& \node (d) {}; \\
		\node[left](e) {abstraction};	& \node (f) {}; \\
		\node[left](g) {extension};	& \node (h) {}; \\
		\node[left](i) {approximation};	& \node (j) {}; \\	
	}; 
	\draw [folding] (a) -- (b); 
	\draw [unfolding] (c) -- (d); 
	\draw [abstraction] (e) -- (f); 
	\draw [extension] (g) -- (h); 
	\draw [approximation] (i) -- (j);

	\node[comment,right of=GHPN, xshift=4cm]
	  {
		{\tiny$^\bullet$} discrete \& continuous states ($\N \cup \R+$)\\
		{\tiny$^\bullet$} stochastic \& deterministic rates\\
		{\tiny$^\bullet$} CTMC coupled by Markov jumps\\
		{\tiny$^\bullet$} hybrid simulation\\
		{\tiny$^\bullet$} PLTLc
	  };

\end{tikzpicture}
 }
\caption[Paradigms of the unifying framework]{{\bf Paradigms of the unifying framework.} Paradigms integrated in the unifying framework of the {\em PetriNuts} platform (adapted from~\cite{HH+12}). The arrows describe the relations supported by automatic file transformation (export, unfolding); except {\it folding}, which in most cases has to be done manually.}
  \label{fig:framework}
\end{figure}

\noindent
\small{
{\bf Net classes} given in Figure~\ref{fig:framework}:
\renewcommand\labelitemi{\tiny$^\bullet$}
\begin{itemize}
    \setlength{\itemsep}{0pt}
    \setlength{\parskip}{0pt}
    \setlength{\parsep}{0pt}  
\item \QPN, \QPNC - (coloured) qualitative Petri nets (standard Petri nets)
\item \SPN, \SPNC - (coloured) stochastic Petri nets
\item \CPN, \CPNC - (coloured) continuous Petri nets
\item \HPN, \HPNC - (coloured) hybrid Petri nets
\end{itemize}
}

{\it Remark:} There are also fuzzy extensions of all six quantitative Petri nets~\cite{AHL19}.

\bigskip
\noindent
\small{
{\bf Annotation index} for Figure~\ref{fig:framework}:
\renewcommand\labelitemi{\tiny$^\bullet$}
\begin{itemize}
    \setlength{\itemsep}{0pt}
    \setlength{\parskip}{0pt}
    \setlength{\parsep}{0pt}  
\item interpretation of marking
\item reaction rates
\item semantics
\item execution type: animation/simulation type 
\item temporal logic type corresponding to the given semantics to describe expected properties
\end{itemize}
}

\normalsize
\subsection{How to use \Snoopy}

\Snoopy~\cite{HH+12} is written in C++, using the Standard Template Library and the cross-platform toolkit wxWidgets~\cite{wxWidgets:2021:Online}. 
The installation files, together with many other resources, are available at \url{https://www-dssz.informatik.tu-cottbus.de/DSSZ/Software/Snoopy}.
\Snoopy comes with a graphical user interface; thus all the following use cases start with {\it Open \Snoopy}.

\paragraph{(1) How to obtain the graphical representation for a CANDL file:}

\renewcommand\labelitemi{\tiny$^\bullet$}
\begin{myitemize}

\item 
$File \rightarrow Import$ x.candl

\item 
$Edit \rightarrow Layout \rightarrow$ chose layout algorithm

\item 
$View \rightarrow Show \; attributes
   \rightarrow$ for all net elements, select the attributes you want to see 
   
\item 
$File \rightarrow Save$ x (as colspn or colcpn)

\end{myitemize}

\paragraph{(2) How to obtain the unfolded net:} 

\renewcommand\labelitemi{\tiny$^\bullet$}
\begin{myitemize}

\item
$File \rightarrow Open$ x (colspn or colcpn)

\item
$File \rightarrow Export \; as$ andl file using dssd\_util // involves unfolding

\item
$File \rightarrow Import$ x.andl
    $\rightarrow$ select {\it new model}
    
\item
$Edit \rightarrow Layout \rightarrow$ chose layout algorithm

\item
$File \rightarrow Save$ x (as spn or cpn)

\end{myitemize}

{\it Remarks:}
\renewcommand\labelitemi{\tiny$^\bullet$}
\begin{myitemize}

\item
export of colspn (colcpn) to spn (cpn) also involves unfolding;

\item
alternatively, unfolding can be done via the command line tool \Spike:

\smallskip
\textsf{spike load -f=x.candl unfold save -f=x.andl}

\end{myitemize}

\paragraph{(3) How to convert \SPN $\Leftrightarrow$ \CPN  or \SPNC $\Leftrightarrow$ \CPNC:}

\renewcommand\labelitemi{\tiny$^\bullet$}
\begin{myitemize}

\item
$File \rightarrow Open$ x.spn

\item
$File \rightarrow Export \; as$ x.cpn

\item
$File \rightarrow Open$ x.cpn

\end{myitemize}

{\it Remarks:} or vice versa; likewise for \SPNC, \CPNC;

\paragraph{(4) How to obtain (C)ANDL files:}

\renewcommand\labelitemi{\tiny$^\bullet$}
\begin{myitemize}
\item $File \rightarrow Open$ x (spn/cpn)
\item $File \rightarrow Export$ to ANDL\\
or
\item $File \rightarrow Open$ x (colspn/colcpn)
\item $File \rightarrow Export$ to CANDL
\end{myitemize}

\paragraph{(5) How to obtain the ODEs of a given \CPN/\CPNC:}

\renewcommand\labelitemi{\tiny$^\bullet$}
\begin{myitemize}
\item $File \rightarrow Open$ x (cpn/colcpn)
\item $File \rightarrow Export \rightarrow$ {\it ODEs to LaTeX} or {\it Text, Octave, Matlab}
\end{myitemize}

{\it Remarks:} 
\renewcommand\labelitemi{\tiny$^\bullet$}
\begin{myitemize}
\item 
ODEs can also be exported as plain text by the simulation window's export options;
\item 
\CPNC have to be unfolded to \CPN, before generating the ODEs;
\end{myitemize}

\paragraph{(6) How to reuse colour definitions:}

\renewcommand\labelitemi{\tiny$^\bullet$}
\begin{myitemize}

\item
construct your (epidemic) model as \SPNC/\CPNC; 

alternatively, if you already have an uncoloured version x, do

\begin{myitemize}
\item $File \rightarrow Open$ x
\item $File \rightarrow export$ to corresponding coloured net class; \ie\\
 \SPN can be exported to \SPNC, and \CPN to \CPNC. 
\item $File \rightarrow Open$ the coloured version of x
\end{myitemize}

\item
$File \rightarrow Import$ y.candl // the file providing the required colour definitions
\renewcommand\labelitemi{\tiny$^\bullet$}
\begin{myitemize}
    \item chose {\it Selective import}
    \item chose {\it Select all} or individually select the subset of required definitions
    \item press OK
\end{myitemize}

\item
check if the colour definitions were imported as expected;    

\end{myitemize}

{\it Remark:}
Advanced users may alternatively edit CANDL files with their favourite text editor.

\paragraph{(7) How to simulate with \Snoopy:}

\renewcommand\labelitemi{\tiny$^\bullet$}
\begin{myitemize}
\item $File \rightarrow Open$ x (spn/colspn/cpn/colcpn)
\item $View \rightarrow Start \; Simulation \; Mode$ 
\item chose {\it Model Configuration}
\item chose {\it Simulator Configuration}
\item open some {\it Views} by double click
\item click {\it Start Simulation} 
\end{myitemize}

\noindent
{\it Remark:} For more details see Snoopy's Manuals~\cite{Liu:manual:2012,HLRH17.manual} and related textbook chapters~\cite{MRH12,BHM15}.

\subsection{How to use \Patty}

\Patty is a JavaScript to support Petri net animation (token flow) in a web browser; it does not require any installation on the user's site. \Patty reads (uncoloured) Petri nets in Snoopy's proprietary format. 

All uncoloured Petri nets given in this paper can be animated 
in a purely qualitative manner at \url{https://www-dssz.informatik.tu-cottbus.de/DSSZ/Research/ModellingEpidemics}.

\renewcommand\labelitemi{\tiny$^\bullet$}
\begin{myitemize}
\item A click on the picture opens a new window with a Petri net, directly executable in your web brower by help of Patty; nothing needs to be installed.
\item Then, the PN can be animated by either clicking on individual transitions (boxes) or
one of the smallish triangles in the control panel.
\item Clicking on a place adds a token.
\end{myitemize}

\subsection{How to use \Charlie}

\Charlie~\cite{HSW15} is written in Java; thus it requires a Java runtime environment.
The jar file is available at \url{https://www-dssz.informatik.tu-cottbus.de/DSSZ/Software/Charlie}.
Charlie comes with a graphical user interface; thus all the following use cases start with {\it Open \Charlie}.

\renewcommand\labelitemi{\tiny$^\bullet$}
\begin{myitemize}

\item $file \rightarrow open$ x (andl)

\item {\it net properties} --  summary vector of the results achieved so far; acronyms are explained by a tool tip feature.

\item {\it IM-based analysis} -- computation of P-/T-invariants, which can be written to a text file (see {\it options}) and are read by \Snoopy $\rightarrow Extras \rightarrow Load \; node \; set \;file$.

\item {\it siphon/trap computation} -- among others, bad siphons can be identified and written to a text file (click {\it export siphons}) and are read by \Snoopy $\rightarrow Extras \rightarrow Load \; node \; set \;file$.

\item {\it protocol} -- a more verbose summary of the analysis results achieved so far.

\item $file \rightarrow exit$ -- all analysis results can be found in a log file, written to the folder from where the ANDL file had been loaded.

\end{myitemize}

\noindent
{\it Remark:} For more details see~\cite{Fran09,HSW15} and \Charlie's webpage.

\subsection{How to use \Spike}
\Spike~\cite{CH19} is a command line tool written in C++, available at \url{https://www-dssz.informatik.tu-cottbus.de/DSSZ/Software/Spike}. Thus all commands have to be entered in a terminal.

\paragraph{(1) How to unfold:} Enter

\smallskip
\textsf{spike load -f=x.candl unfold save -f=x.andl}

\paragraph{(2) How to simulate:} For reproducible simulation experiments, \Spike builds on a configuration file (to be written in its own configuration language), where all details of how to configure the model and the simulator(s) to be used need to be specified. Assuming the configuration file has the name {\it config.spc}, enter

\smallskip
\textsf{spike exe -f=config.spc}

\bigskip
\noindent
{\it Remarks:} 
\renewcommand\labelitemi{\tiny$^\bullet$}
\begin{myitemize}
\item 
We provide configuration files to simulated P$_{10}$SIR with optimised kinetic rate constants and to produce Figure~\ref{figure:JC:lockunlock} illustrating dynamic rate constants.
\item For more details see~\cite{CH19} and \Spike's webpage. 
\end{myitemize}

\subsection{How to use {\it MC2}}

{\it MC2}~\cite{donaldson:cmsb:2008} is written in Java; thus it requires a Java runtime environment.
The jar file is available at \url{https://www-dssz.informatik.tu-cottbus.de/DSSZ/Software/MC2}.
{\em MC2}  is a command line tool and performs offline simulative model checking 
of Probabilistic Linear-time Temporal Logic (PLTL). It requires as
input a file of time series traces in CSV format, and a file of the PLTL properties to be checked.
The property file contains one property per line, in the form
$P_{\ge1}[\,G([x]=0\,)\,]$
where $x$ is an entity named (coloured/uncoloured place/transition/observer) in the traces file.
The call is:

\smallskip
\textsf{java -jar MC2v2.0beta2.jar det {\em TracesFile} {\em PropertiesFile}}
\smallskip

\noindent
and the output comprises, for each property checked, the property formula and on the next line the result ({\em true} or {\em false}).

We have compiled a library of useful properties which can be used directly or adapted as required.  These library properties employ {\em meta variables}
indicated by \$x in the properties, to be substituted one by one by all observables in the input time series csv file of behaviour traces, e.g.

\smallskip
\textsf{P>=1 [ G([\$x]>=0) ]}
\smallskip

We have developed a Unix script to automatically expand these meta variables to the observables in a given csv file so that there is a
version of the properties for each observable, and then to run the model
checker on the csv file and the expanded property file.  The property library and Unix script are provided.

\clearpage
\section{Exchange formats} 

\subsection{ANDL -- Abstract Net Description Language}

ANDL has been designed as a concise, but human-readable exchange format for (uncoloured) Petri nets, written as plain ASCII text.
For illustration of the ANDL syntax, we give a couple of examples; for formal definitions see~\cite{SRH16:marcie-manual}. 

The very first keyword (\node{spn}/\node{cpn}) determines the net class -- \SPN or \CPN. An ANDL specification consists basically of three lists: the constants, the places and the transitions. Constants can hold several value sets and can be organised in groups (here: {\it marking} and {\it parameter}, which are pre-defined groups, extended by the user-defined groups {\it par\_ratio}, {\it standard-incidence}). Models can be configured by choosing for each constant group a specific value set. For example, to apply the standard incidence, choose the {\it Main} value set ($N$ will be set to $I0+S0$), to apply Mass-action kinetics, choose {\it V\_Set\_0} ($N$ will be set to 1).

\paragraph{Example 1: SIR.spn - Figure~\ref{figure:SIR} in ANDL notation.}
\small{
\begin{verbatim}
spn [SIR]
{
constants:
  valuesets[Main:V_Set_0:V_Set_1:V_Set_2:V_Set_3:V_Set_4:V_Set_5:V_Set_6]
marking:
  int S0 = [100:1e3:1e4:1e5:1e6:1e7:1e8:];
  int I0 = [1:S0/100::::::];
standardIncidence:
  int N = [I0+S0:1::::::];
par_ratio:
  double ratio = [1:0.1:0.01:1.0e-3:2:3:4:5];
parameter:
  double k_infect = [1:0.1:0.01:1.0e-3:1.0e-4:1.0e-5:1.0e-6:];
  double k_recover = [k_infect*ratio:::::::];

places:
  Susceptible = S0;
  Infectious = I0;
  Recovered = 0;

transitions:
  Infect
    : 
    : [Infectious + 2] & [Susceptible - 1] & [Infectious - 1]
    : MassAction(k_infect/N)
    ;
  Recover
    : 
    : [Recovered + 1] & [Infectious - 1]
    : MassAction(k_recover)
    ;

}
\end{verbatim}
}
\normalsize

\paragraph{Example 2: SIR.cpn - Figure~\ref{figure:SIR} in ANDL notation.}   
In terms of ANDL syntax, the only difference between \SPN and \CPN (besides the very first keyword) is the keyword \node{continuous} (occurring after \node{places} and \node{transitions}) which reminds us that places and transitions of a \CPN are treated as continuous nodes.

\small{
\begin{verbatim}
cpn [SIR]
{
constants:
  valuesets[Main:V_Set_0:V_Set_1:V_Set_2:V_Set_3:V_Set_4:V_Set_5:V_Set_6]
marking:
  int S0 = [100:1e3:1e4:1e5:1e6:1e7:1e8:];
  int I0 = [1:S0/100::::::];
standardIncidence:
  int N = [I0+S0:1::::::];
par_ratio:
  double ratio = [1:0.1:0.01:1.0e-3:2:3:4:5];
parameter:
  double k_infect = [1:0.1:0.01:1.0e-3:1.0e-4:1.0e-5:1.0e-6:];
  double k_recover = [k_infect*ratio:::::::];

places:
continuous:
  Susceptible = S0;
  Infectious = I0;
  Recovered = 0;

transitions:
continuous:
  Infect
    : 
    : [Infectious + 2] & [Susceptible - 1] & [Infectious - 1]
    : MassAction(k_infect/N)
    ;
  Recover
    : 
    : [Recovered + 1] & [Infectious - 1]
    : MassAction(k_recover)
    ;

}
\end{verbatim}
}
\normalsize

\paragraph{Example 3: SIR-S2.spn - Figure~\ref{figure:SIRage}~(Top) in ANDL notation.} 
\small{
\begin{verbatim}
spn  [SIR-S2]
{
constants:
  valuesets[Main:V_Set_0:V_Set_1:V_Set_2:V_Set_3:V_Set_4:V_Set_5:V_Set_6]
marking:
  int SY0 = [50000:::::::];
  int SO0 = [50000:::::::];
  int IY0 = [1:::::::];
  int IO0 = [0:::::::];
standardIncidence:
  int N = [SY0+SO0+IY0+IO0:1::::::];
par_ratio:
  double ratio = [1:0.1:0.01:1.0e-3:2:3:4:5];
parameter:
  double k_infect = [1:0.1:0.01:1.0e-3:1.0e-4:1.0e-5:1.0e-6:];
  double k_recover = [k_infect*ratio:::::::];
  double k_crossInfect = [k_infect/0.1:::::::];

places:
  Susceptible_Young = SY0;
  Susceptible_Old = SO0;
  Infectious_Young = IY0;
  Infectious_Old = IO0;
  Recovered_Young = 0;
  Recovered_Old = 0;

transitions:
  InfectYoung
    : 
    : [Infectious_Young + 2] & [Susceptible_Young - 1] & [Infectious_Young - 1]
    : MassAction(k_infect/N)
    ;
  RecoverYoung
    : 
    : [Recovered_Young + 1] & [Infectious_Young - 1]
    : MassAction(k_recover)
    ;
  RecoverOld
    : 
    : [Recovered_Old + 1] & [Infectious_Old - 1]
    : MassAction(k_recover)
    ;
  InfectOld
    : 
    : [Infectious_Old + 2] & [Infectious_Old - 1] & [Susceptible_Old - 1]
    : MassAction(k_infect/N)
    ;
  OldInfectYoung
    : 
    : [Infectious_Young + 1] & [Infectious_Old + 1] & 
      [Susceptible_Young - 1] & [Infectious_Old - 1]
    : MassAction(k_crossInfect/N)
    ;
  YoungInfectOld
    : 
    : [Infectious_Young + 1] & [Infectious_Old + 1] & 
      [Susceptible_Old - 1] & [Infectious_Young - 1]
    : MassAction(k_crossInfect/N)
    ;

}
\end{verbatim}
}
\normalsize

\subsection{CANDL -- Coloured ANDL}
 
CANDL is an extension of ANDL;
it has been designed as a concise, but human-readable exchange format for coloured Petri nets, written as plain ASCII text. For illustration of the CANDL syntax, we give two examples; for formal definitions see~\cite{CANDL-manual:2021}.

Example 4 shows the coloured version of Example 3, \ie unfolding the CANDL code in Example 4 gives the ANDL code in Example 3. The naming convention for unfolded places is exactly as shown here (name of the coloured place followed by one of its colours, separated by an underscore), the naming of unfolded transitions slightly differs.

The rates of the transition \node{Infect} are colour-dependent; the appropriate colours are determined by a Boolean expression, forming a guard, which in turn is given in square brackets.

\paragraph{Example 4 - SIR-S\_enum.colspn - Figure~\ref{figure:SIRage}~(Middle) in CANDL notation.} 
\small{
\begin{verbatim}
colspn  [SIR-S2_enum]
{
constants:
    valuesets[Main:V_Set_0:V_Set_1:V_Set_2:V_Set_3:V_Set_4:V_Set_5:V_Set_6]
marking:
  int SY0 = [50000:::::::];
  int SO0 = [50000:::::::];
  int IY0 = [1:::::::];
  int IO0 = [0:::::::];
standardIncidence:
  int N = [SY0+SO0+IY0+IO0:1::::::];
par_ratio:
  double ratio = [1:0.1:0.01:1.0e-3:2:3:4:5];
parameter:
  double k_infect = [1:0.1:0.01:1.0e-3:1.0e-4:1.0e-5:1.0e-6:];
  double k_crossInfect = [k_infect/0.1:::::::];
  double k_recover = [k_infect*ratio:::::::];

colorsets:
  Dot = {dot};
  enum Strata = {Young,Old};

variables:
  Strata : x;
  Strata : y;

places:
discrete:
  Strata Infectious = IY0`Young++IO0`Old;
  Strata Recovered = 0`Young++0`Old;
  Strata Susceptible = SY0`Young++SO0`Old;

transitions:
  Infect
    : 
    : [Infectious + {x++y}] & [Susceptible - {x}] & [Infectious - {y}]
    : [(x=Young&y=Young)|(x=Old&y=Old)] MassAction(k_infect/N) ++ 
      [(x=Young&y=Old)|(x=Old&y=Young)] MassAction(k_crossInfect/N)
    ;
  Recover
    : 
    : [Recovered + {x}] & [Infectious - {x}]
    : MassAction(k_recover)
    ;

}
\end{verbatim}
}
\normalsize

Example 5 is structurally equivalent to Example 4, but uses a different colour set to encode stratification. Example 4 defines an enumeration set, which can be extended by adding further keywords. In contrast, Example 5 defines an integer subset, ranging from 1 to an upper bound given by the constant \node{StrataNum}, which simplifies scaling to an arbitrary number of strata. In terms of implementation, there isn't much of a difference as enumeration types are internally mapped to integer types. Thus the keywords of an enumeration type (here: \node{Young}, \node{Old}) are merely syntactic sugar, which may improve readability. For further details of the supported colour sets and related colouring principles, see~\cite{Liu:manual:2012}.

\paragraph{Example 5 - SIR-S\_int.colspn - Figure~\ref{figure:SIRage}~(Middle) in CANDL notation.} 
\small{
\begin{verbatim}
colspn  [SIR-S2_int]
{
constants:
  valuesets[Main:V_Set_0:V_Set_1:V_Set_2:V_Set_3:V_Set_4:V_Set_5:V_Set_6]
coloring:
  int StrataNum = [2:::::::];
marking:
  int SY0 = [50000:::::::];
  int SO0 = [50000:::::::];
  int IY0 = [1:::::::];
  int IO0 = [0:::::::];
standardIncidence:
  int N = [SY0+SO0+IY0+IO0:1::::::];
par_ratio:
  double ratio = [1:0.1:0.01:1.0e-3:2:3:4:5];
parameter:
  double k_infect = [1:0.1:0.01:1.0e-3:1.0e-4:1.0e-5:1.0e-6:];
  double k_crossInfect = [k_infect/0.1:::::::];
  double k_recover = [k_infect*ratio:::::::];

colorsets:
  Dot = {dot};
  Strata = {1..StrataNum};

variables:
  Strata : x;
  Strata : y;

places:
discrete:
  Strata Infectious = IY0`1++IO0`2;
  Strata Recovered = 0`1++0`2;
  Strata Susceptible = SY0`1++SO0`2;

transitions:
  Infect
    : 
    : [Infectious + {x++y}] & [Susceptible - {x}] & [Infectious - {y}]
    : [(x=1&y=1)|(x=2&y=2)] MassAction(k_infect/N) ++ 
      [(x=1&y=2)|(x=2&y=1)] MassAction(k_crossInfect/N)
    ;
  Recover
    : 
    : [Recovered + {x}] & [Infectious - {x}]
    : MassAction(k_recover)
    ;

}
\end{verbatim}
}
\normalsize

\subsection{ODEs generated}

For illustration we provide here for selected \CPN/\CPNC the ODEs which are automatically generated when it comes to simulating a model; for \CPNC this involves unfolding. To feed the ODEs to another simulator tool, consider the option to export to Matlab, Octave, or ERODE~\cite{cardelli2017erode}; otherwise you may
start from the plain text representation to obtain your required representation style;
all of these exports are supported by \Snoopy. 


\paragraph{(1) ODEs - SIQR - Figure~\ref{figure:SIQR-SIAR}~(Top).}

\begin{displaymath}\begin{array}{rll}

\frac{d\mathrm{Susceptible}}{d\mathrm{t}} & = &
 \mathrm{-(k\_\mathrm{infect/N*Susceptible*Infectious)}}\\ 
\\
\frac{d\mathrm{Recovered}}{d\mathrm{t}} & = &
 \mathrm{+(k\_\mathrm{recover*Infectious)+(k\_\mathrm{recover*Quarantine)}}}\\ 
\\
\frac{d\mathrm{Infectious}}{d\mathrm{t}} & = &
    +(k\_\mathrm{infect*Susceptible*Infectious)}\\
& & -(k\_\mathrm{recover*Infectious)-(k\_\mathrm{quar*Infectious)}}\\ 
\\
\frac{d\mathrm{Quarantine}}{d\mathrm{t}} & = &
 \mathrm{+(k\_\mathrm{quar*Infectious)-(k\_\mathrm{recover*Quarantine)}}}\\ 
\\
\end{array}\end{displaymath}

\paragraph{(2) ODEs - SIAR - Figure~\ref{figure:SIQR-SIAR}~(Bottom).}

\begin{displaymath}\begin{array}{rll}

\frac{d\mathrm{InfectiousSymptomatic}}{d\mathrm{t}} & = &
 
 + (k\_\mathrm{infect/N*Susceptible*InfectiousSymptomatic)}\\
 & & +(k\_\mathrm{infect/N*Susceptible*InfectiousAsymptomatic)}\\

& & +(k\_\mathrm{symptoms*InfectiousAsymptomatic)}\\

& & -(k\_\mathrm{recover*InfectiousSymptomatic)}\\

\\
\frac{d\mathrm{RecoveredSymptomatic}}{d\mathrm{t}} & = &
 \mathrm{(k\_\mathrm{recover*InfectiousSymptomatic)}}\\ 
\\
\frac{d\mathrm{Susceptible}}{d\mathrm{t}} & = &
 \mathrm{-2(k\_\mathrm{infect/N*Susceptible*InfectiousSymptomatic)}}\\
& & +(k\_\mathrm{infect/N*Susceptible*InfectiousAsymptomatic)}\\
\\

\frac{d\mathrm{InfectiousAsymptomatic}}{d\mathrm{t}} & = &
 \mathrm{+(k\_\mathrm{infect/N*Susceptible*InfectiousSymptomatic)}}\\
 & & +(k\_\mathrm{infect/N*Susceptible*InfectiousAsymptomatic)}\\
& & -(k\_\mathrm{symptoms*InfectiousAsymptomatic)}\\
& & -(k\_\mathrm{recover*InfectiousAsymptomatic)}\\

\\
\frac{d\mathrm{RecoveredAsymptomatic}}{d\mathrm{t}} & = &
 \mathrm{(k\_\mathrm{recover*InfectiousAsymptomatic)}}\\ 
\\
\end{array}\end{displaymath}

\paragraph{(3) ODEs - SIR-S2 - Figure~\ref{figure:SIRage}.}

\begin{displaymath}\begin{array}{rll}

\frac{d\mathrm{Infectious_\mathrm{Young}}}{d\mathrm{t}} & = &
 +(k\_\mathrm{crossInfect/N*Susceptible\_\mathrm{Young*Infectious\_\mathrm{Old)}}}\\
& & +(k\_\mathrm{infect/N*Susceptible\_\mathrm{Young*Infectious\_\mathrm{Young)}}}\\
& & -(k\_\mathrm{recover*Infectious\_\mathrm{Young)}}\\
\\
\frac{d\mathrm{Recovered_\mathrm{Young}}}{d\mathrm{t}} & = &
 \mathrm{+(k\_\mathrm{recover*Infectious\_\mathrm{Young)}}}\\ 
\\
\frac{d\mathrm{Susceptible_\mathrm{Young}}}{d\mathrm{t}} & = &
 \mathrm{-(k\_\mathrm{infect/N*Susceptible\_\mathrm{Young*Infectious\_\mathrm{Young)}}}}\\
& & -(k\_\mathrm{crossInfect/N*Susceptible\_\mathrm{Young*Infectious\_\mathrm{Old)}}}\\ 
\\
\frac{d\mathrm{Susceptible_\mathrm{Old}}}{d\mathrm{t}} & = &
 \mathrm{-(k\_\mathrm{infect/N*Infectious\_\mathrm{Old*Susceptible\_\mathrm{Old)}}}}\\
 & &-(k\_\mathrm{crossInfect/N*Susceptible\_\mathrm{Old*Infectious\_\mathrm{Young)}}}\\ 
\\
\frac{d\mathrm{Recovered_\mathrm{Old}}}{d\mathrm{t}} & = &
 \mathrm{(k\_\mathrm{recover*Infectious\_\mathrm{Old)}}}\\ 
\\
\frac{d\mathrm{Infectious_\mathrm{Old}}}{d\mathrm{t}} & = &
 \mathrm{+(k\_\mathrm{crossInfect/N*Susceptible\_\mathrm{Old*Infectious\_\mathrm{Young)}}}}\\
& & +(k\_\mathrm{infect/N*Infectious\_\mathrm{Old*Susceptible\_\mathrm{Old)}}}\\
& & -(k\_\mathrm{recover*Infectious\_\mathrm{Old)}}\\

\\
\end{array}\end{displaymath}

\paragraph{(4) ODEs - P$_2$SIR - Figure~\ref{figure:SIRtravel2}.}

\begin{displaymath}\begin{array}{rll}

\frac{d\mathrm{S_\mathrm{DE}}}{d\mathrm{t}} & = &
 -(k\_\mathrm{infect/N*I\_\mathrm{DE*S\_\mathrm{DE)+\mathrm{(k\_\mathrm{travel*S\_\mathrm{FR)-(k\_\mathrm{travel*S\_\mathrm{DE)}}}}}}}}\\ 
\\
\frac{d\mathrm{R_\mathrm{DE}}}{d\mathrm{t}} & = &
 \mathrm{+(k\_\mathrm{recover*I\_\mathrm{DE)+(k\_\mathrm{travel*R\_\mathrm{FR)-(k\_\mathrm{travel*R\_\mathrm{DE)}}}}}}}\\ 
\\
\frac{d\mathrm{I_\mathrm{DE}}}{d\mathrm{t}} & = &
+(k\_\mathrm{infect/N*I\_\mathrm{DE*S\_\mathrm{DE)}}}\\
& & \mathrm{+(k\_\mathrm{travel*I\_\mathrm{FR)}}}\\
& & -(k\_\mathrm{recover*I\_\mathrm{DE)}}\\
& & -(k\_\mathrm{travel*I\_\mathrm{DE)}}\\ 
\\
\frac{d\mathrm{S_\mathrm{FR}}}{d\mathrm{t}} & = &
 -(k\_\mathrm{infect/N*I\_\mathrm{FR*S\_\mathrm{FR)+\mathrm{(k\_\mathrm{travel*S\_\mathrm{DE)-(k\_\mathrm{travel*S\_\mathrm{FR)}}}}}}}}\\ 
\\
\frac{d\mathrm{R_\mathrm{FR}}}{d\mathrm{t}} & = &
 \mathrm{+(k\_\mathrm{recover*I\_\mathrm{FR)+(k\_\mathrm{travel*R\_\mathrm{DE)-(k\_\mathrm{travel*R\_\mathrm{FR)}}}}}}}\\ 
\\
\frac{d\mathrm{I_\mathrm{FR}}}{d\mathrm{t}} & = &
\mathrm{+(k\_\mathrm{infect/N*I\_\mathrm{FR*S\_\mathrm{FR)}}}}\\

& & +(k\_\mathrm{travel*I\_\mathrm{DE)}}\\
& & -(k\_\mathrm{recover*I\_\mathrm{FR)}}\\
& & -(k\_\mathrm{travel*I\_\mathrm{FR)}}\\ 
\\
\end{array}\end{displaymath}


\clearpage
\section{Additional Figures and Tables}
The following subsection titles correspond to the ones in the main paper.

\subsection{Extended epidemic models}

\begin{figure}[h!tb]
            \centering
            \includegraphics[width=0.80\textwidth]{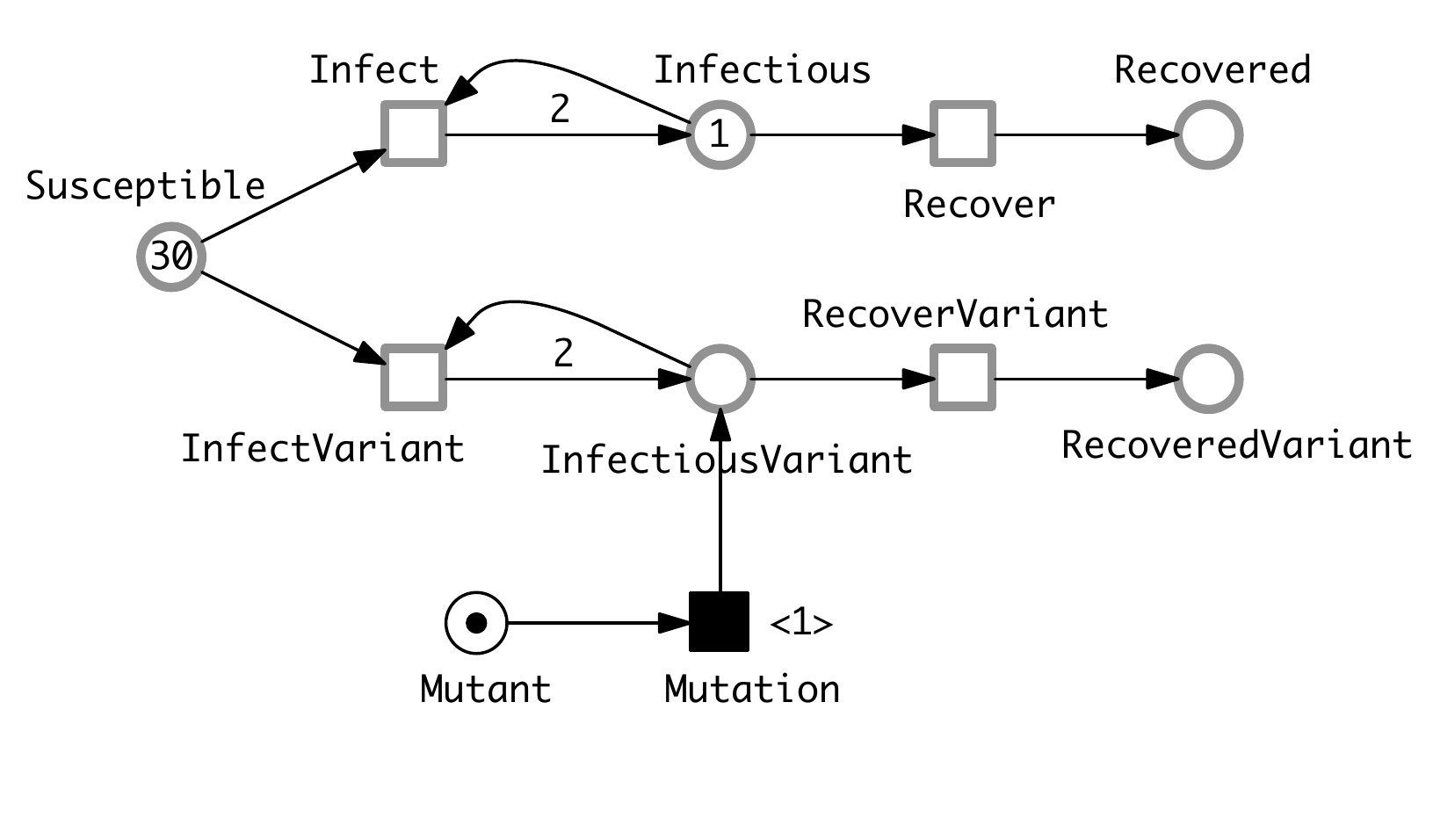}
            \includegraphics[width=0.95\textwidth]{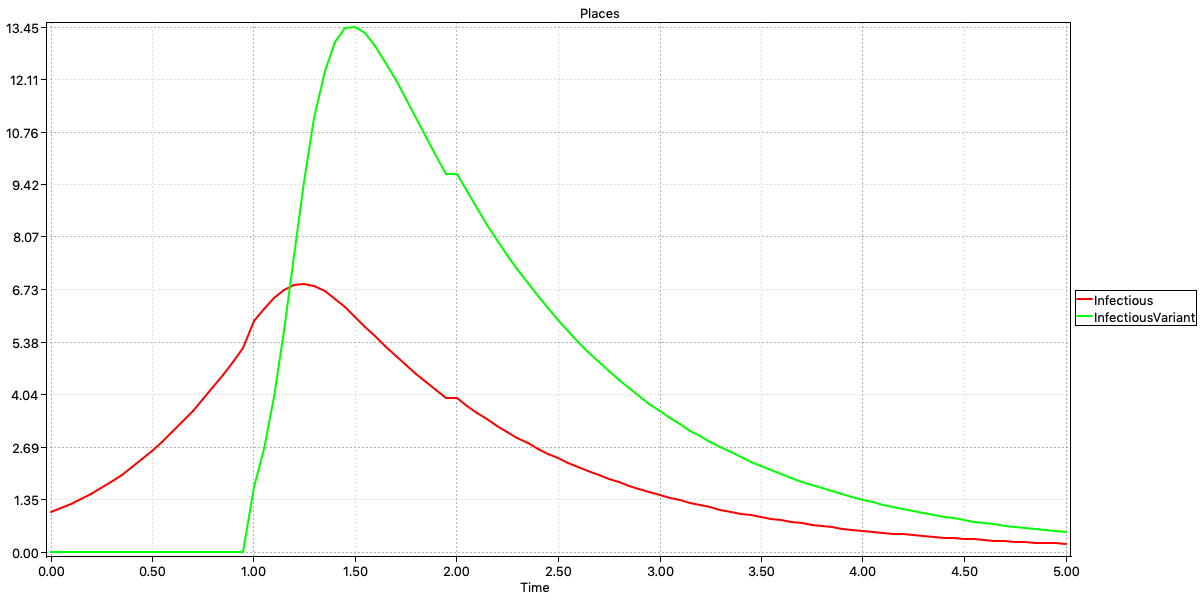}
\caption[SIVR]{{\bf SIVR model.} {\bf (Top)} \HPN describing the introduction of a  virus variant, using a deterministic transition \node{Mutation} 
which fires with a delay of 1 time unit after simulation start.  Kinetic parameters: ratio between \node{Infect} and \node{Recover}, \node{RecoverVariant} is 1:10; ratio
between \node{Infect} and \node{InfectVariant} is 1:5.
{\bf (Bottom)} Hybrid simulation trace for \node{Infectious} and \node{InfectiousVariant}.}
\label{figure:SIVR}
\end{figure}

Figure~\ref{figure:SIVR} shows SIVR -- an SIR model extended by a virus variant represented as \HPN,  and a hybrid simulation trace.

\subsection{Pandemic models}

\begin{figure}[h!tb]
            \centering
            \includegraphics[width=0.80\textwidth]{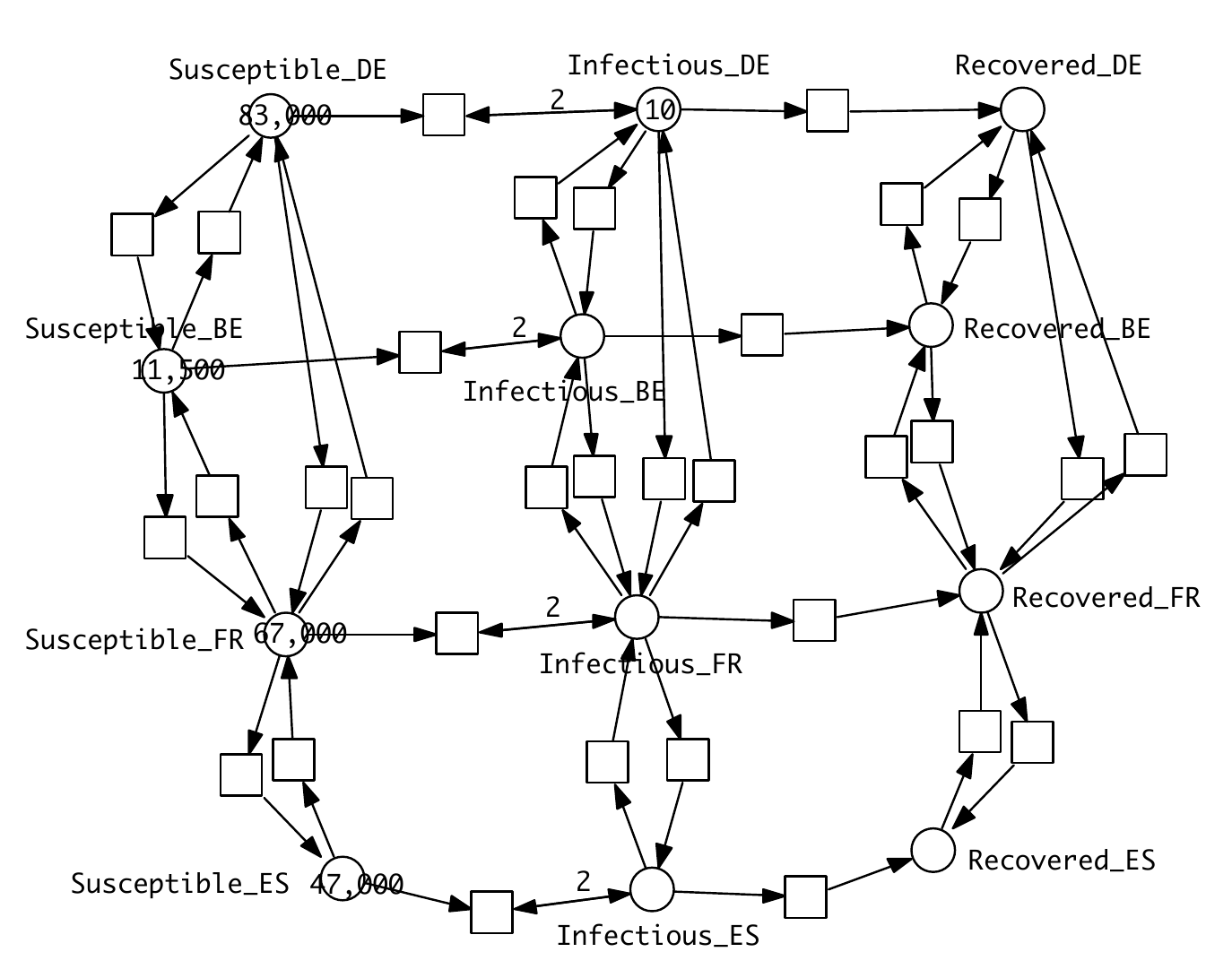}
   
\caption[P4-SIR, unfolded]{{\bf P$_4$SIR model, unfolded} obtained by unfolding P$_4$SIR, shown in Figure~\ref{figure:Countries4} (Middle), layout automatically generated.}
\label{figure:P4-SIR-unfolded}
\end{figure}

P$_4$SIR model, unfolded is given in Figure~\ref{figure:P4-SIR-unfolded}.
\hspace{3em}

\clearpage
\subsection{Combined models}

\begin{figure}[h!tb]
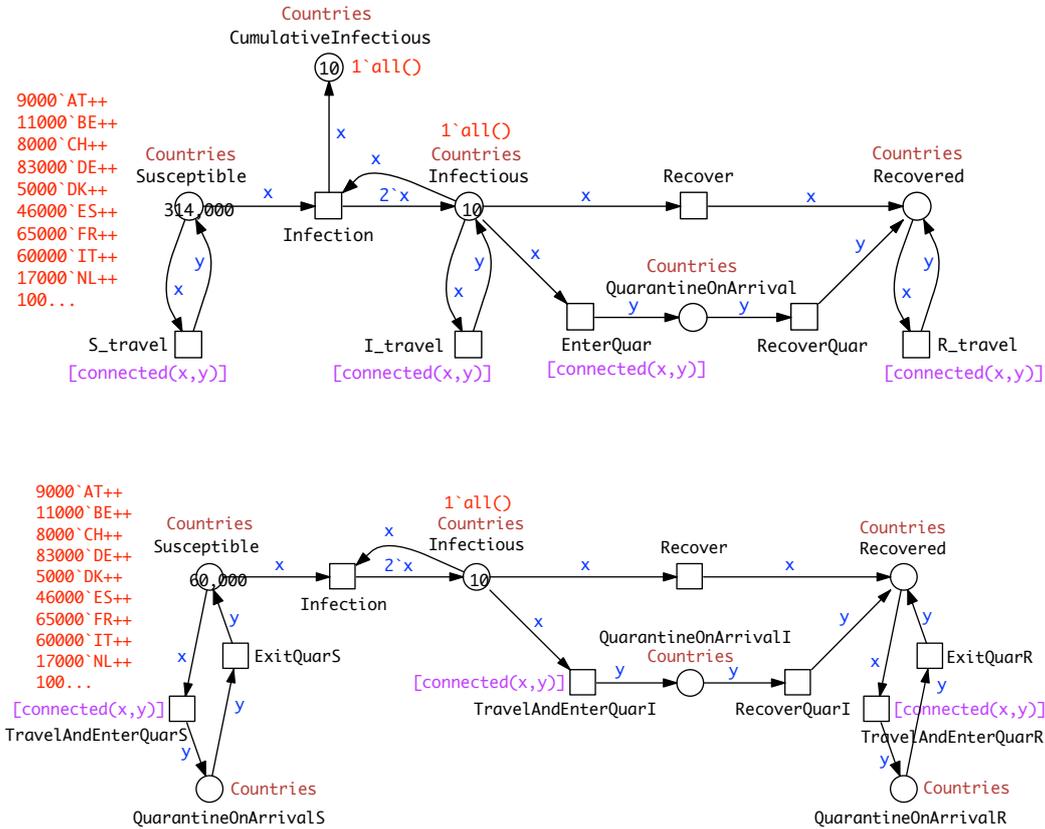

            \centering
            \includegraphics[width=0.95\textwidth]{Figures/P10_SIQR-may-quarantine-on-arrival-colspn.pdf}
            \includegraphics[width=0.95\textwidth]{Figures/P10_SIQR-colspn.pdf}
\caption[P$_{10}$QSIR model]{{\bf P$_\textrm{10}$QSIR model} in two versions for West Europe (10 countries).
{\bf(Top)} P$_{10}$Q$_\textrm{I}$SIR. Arrival by \node{Infectious} may possibly involve quarantine.
There is an additional place \node{CumulativeInfectious} to keep a record of total infections over time. 
Note: this additional place destroys the conservativeness and induces a second P-invariant, comprising (all uncoloured places of) \node{Susceptible} and \node{CumulativeInfectious}.
A related analysis is given in Figure~\ref{figure:SC:Figure30}.
{\bf(Bottom)} P$_\textrm{10}$QSIR. All arrivals will involve quarantine.
}
\label{figure:SC:model7}
        \end{figure}

P$_{10}$SIQR model in two versions -- see Figure~\ref{figure:SC:model7}.

\clearpage

\subsection{Parameter fitting}

\begin{figure}[h!tb]
            \centering
            \includegraphics[width=0.75\textwidth]{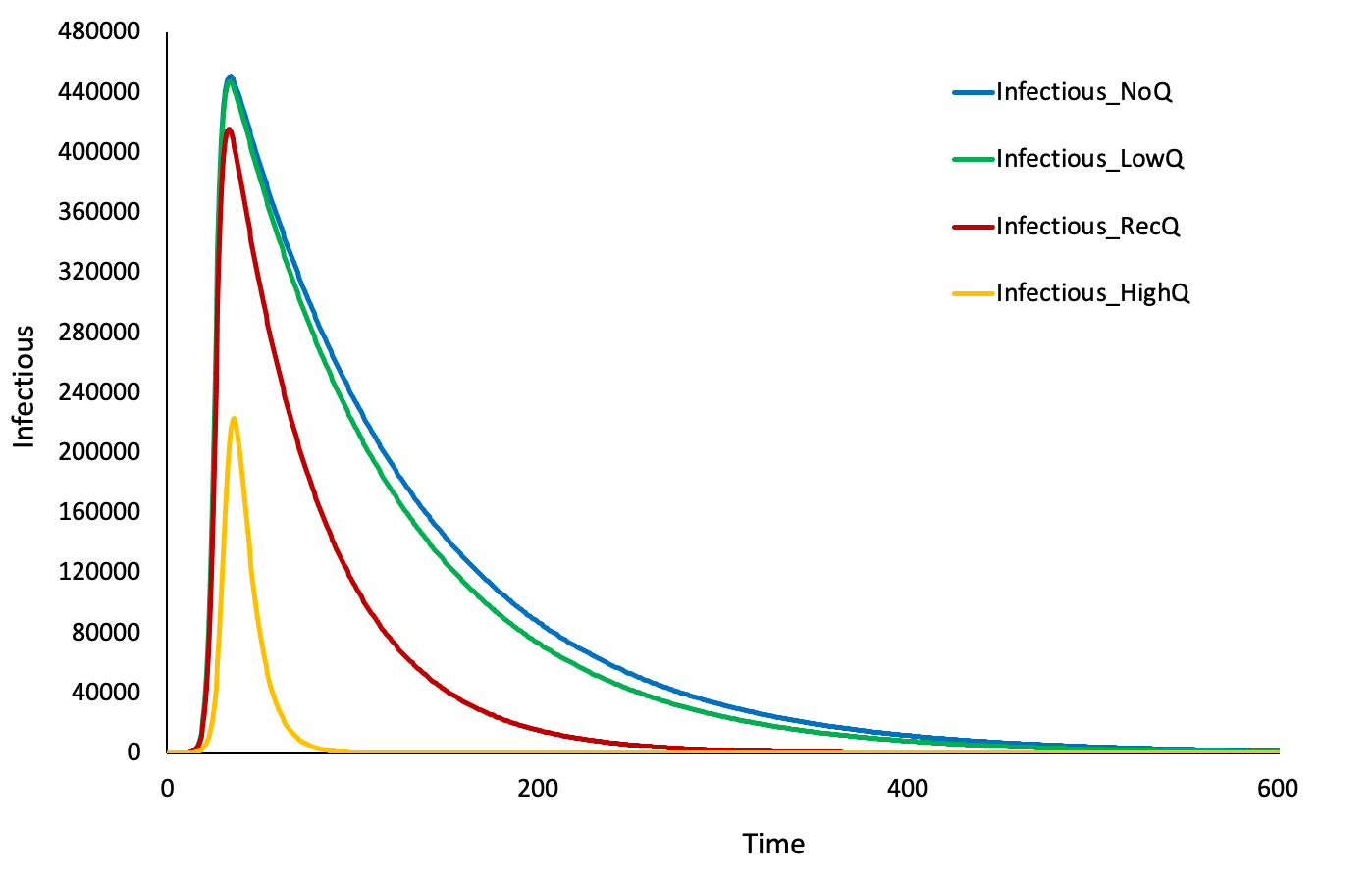}
            \includegraphics[width=0.75\textwidth]{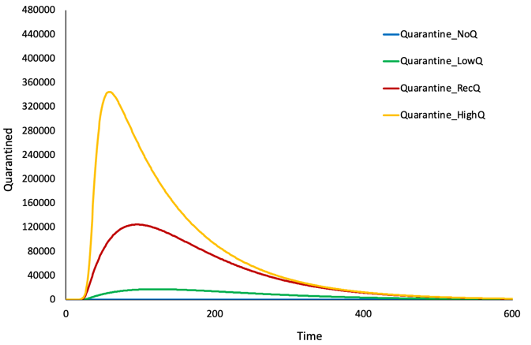}
\caption[Varying Quarantine Rates]{{\bf SIQR model with variable quarantine rates}.
Four simulations of model SIQR (Figure~\ref{figure:SIQR-SIAR} (Top)) with different Quarantine rates. NoQ = 0, LowQ =
10x less than Recover, RecQ = Recover, HighQ = 10x greater than Recover.
{\bf(Top)} Infectious compartment, {\bf(Bottom)} Quarantined compartment.
}
\label{figure:SC:Figure27+28}
        \end{figure}

\begin{figure}[h!tb]
            \centering
            \includegraphics[width=0.95\textwidth]{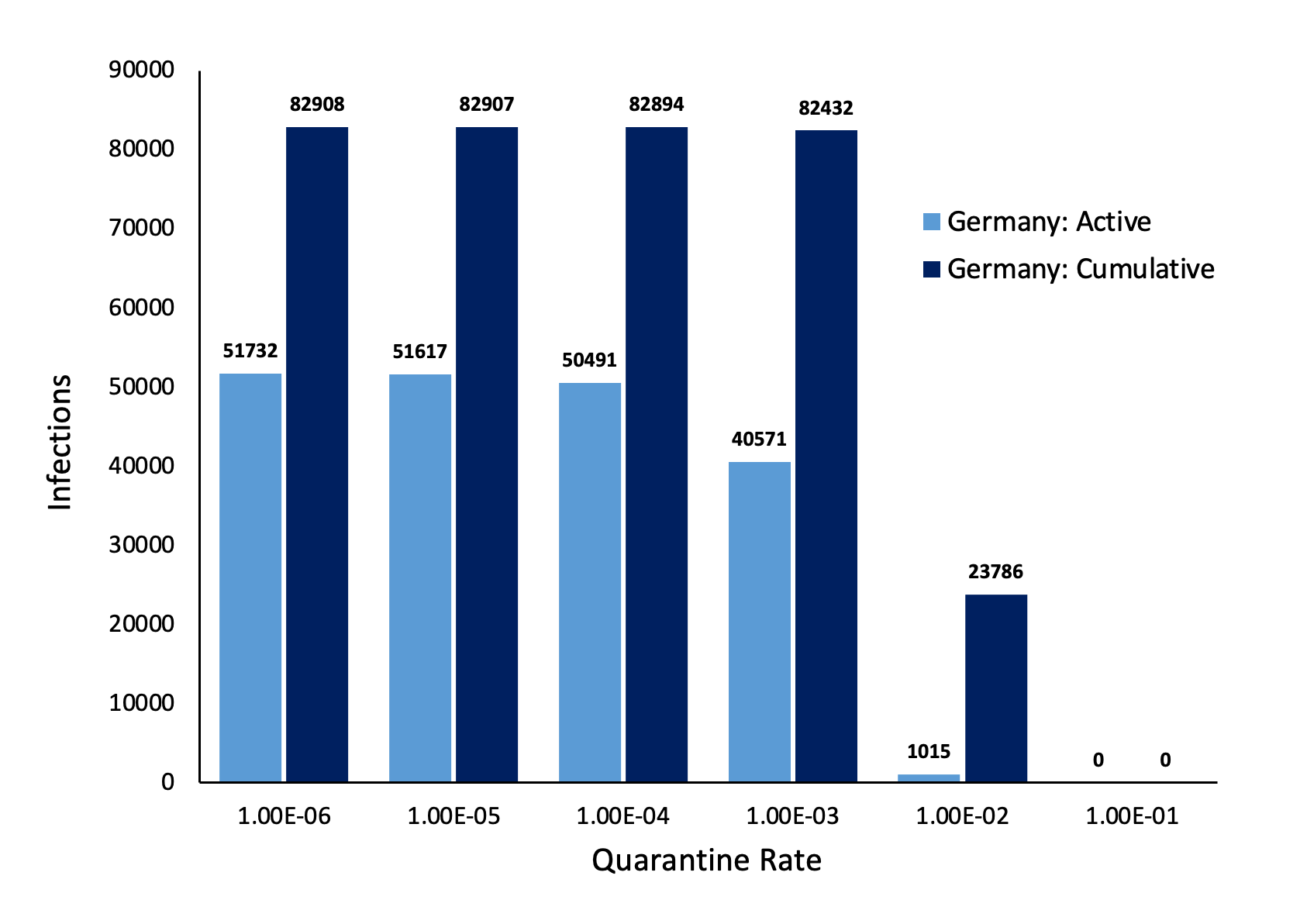}
\caption[Quarantine Rates]{{\bf P$_{10}$Q$_\textrm{I}$SIR model -- effect of modifying quarantine rates}. 
Active and CumulativeInfectious for Germany with varying rates of Quarantine. Output from
model in Figure~\ref{figure:SC:model7} (Top).  X axis: rates for how many people go into quarantine.  Y axis: Data for the peaks
(active infectious at peak, cumulative at peak).  Interpretation: up to a certain point
quarantine only suppresses the active cases but not cumulative, lengthening out the curve,
until quarantine is high.
}
\label{figure:SC:Figure30}
        \end{figure}

\paragraph{Parameter scanning.} Examples for varying rate parameters, see
 Figure~\ref{figure:SC:Figure27+28} for SIQR,
and Figure~\ref{figure:SC:Figure30} for P$_{10}$Q$_\textrm{I}$SIR.

\clearpage

\begin{figure}[h!tb]
            \centering
            \includegraphics[width=0.95\textwidth]{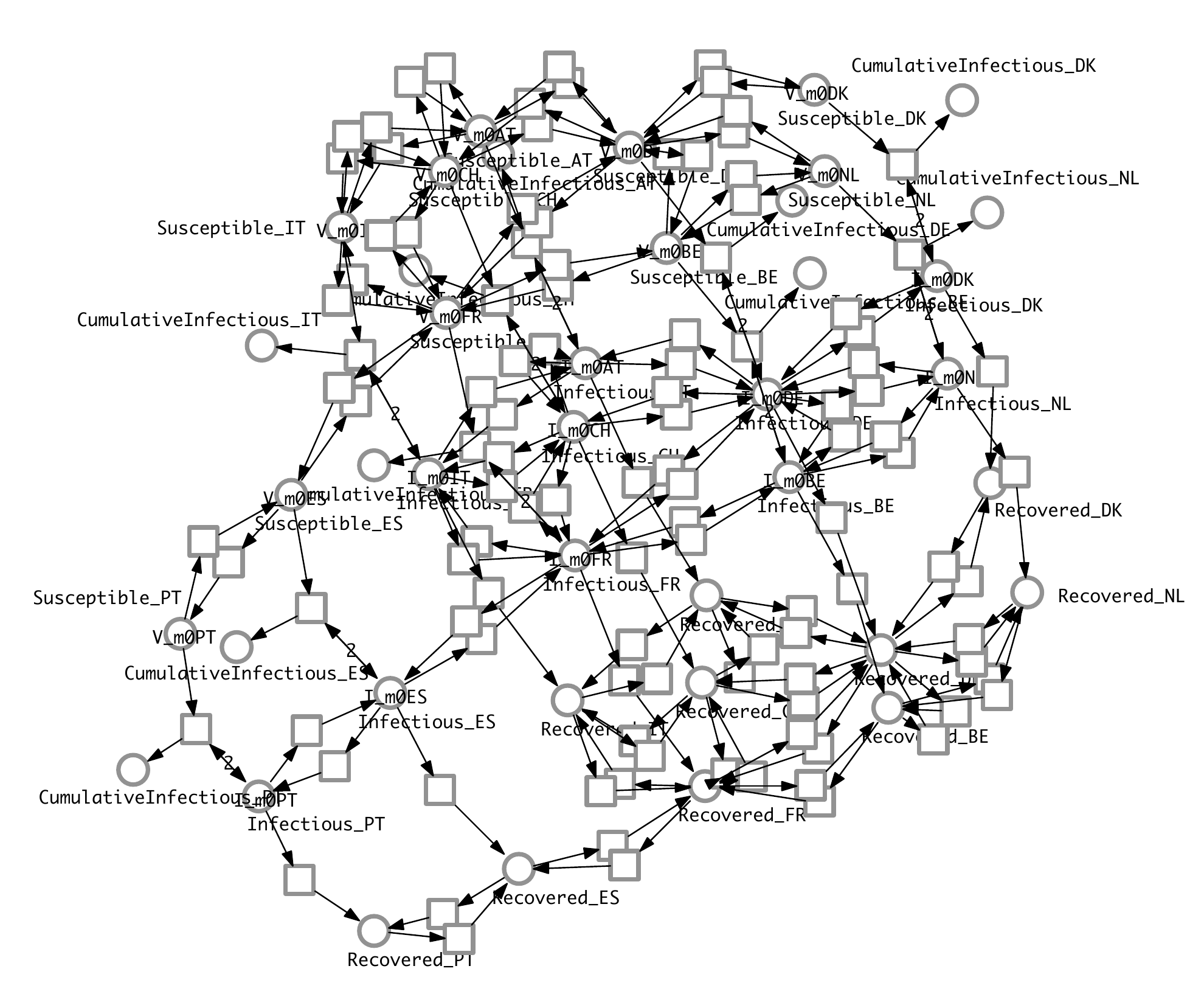}
\caption[P$_{10}$SIR model]{{\bf  P$_{10}$SIR model}. 
\CPN for SIR, West Europe, 
with a \node{CumulativeInfectious} place to keep a record of total infections over time; obtained by unfolding and automatic layout. 
}
\label{figure:SC:model6}
        \end{figure}

\paragraph{Parameter optimisation.}
The model fitted was P$_{10}$SIR in the unfolded version
(see Figure~\ref{figure:SC:model6}, and supplementary files P10SIR.candl, P10SIR\_Separate\_Rates.andl and P10SIR\_Separate\_Rates.cpn)
employing a Random Restart Hill Climbing Algorithm (RRHCA) using real world data
given in Table~\ref{table:sc:table6}, with results for \node{Infection} rates given in
Table~\ref{table:sc:table9} (Objective~1),
Table~\ref{table:sc:table10} (Objective~2), and
Table~\ref{table:sc:table11} (Objective~3). 
The \node{Connection} rates for Objective~3 are given in Table~\ref{table:sc:fittedconnectionrates}.

\begin{figure}[h!tb]
            \centering
            \includegraphics[width=0.95\textwidth]{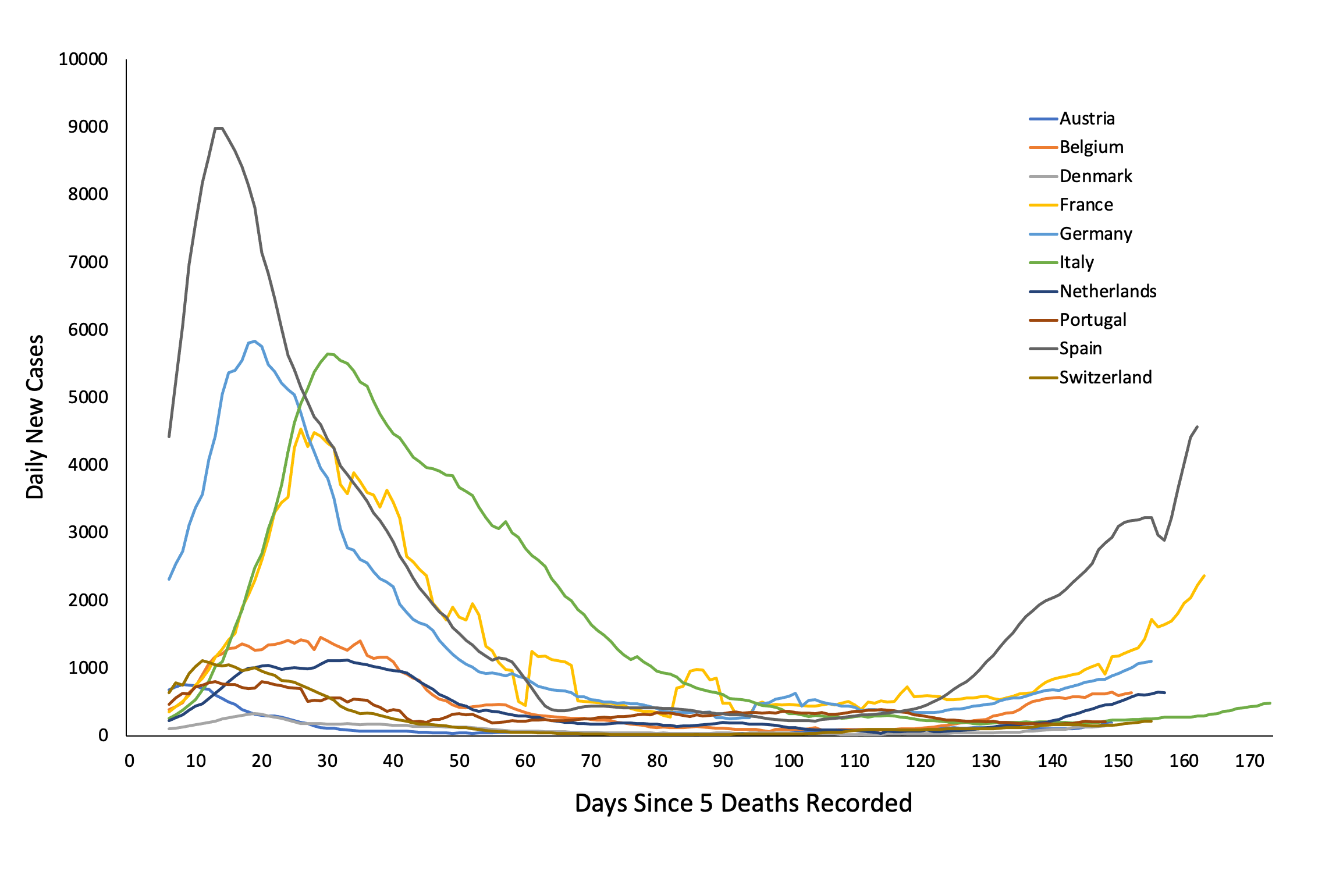}
\caption[Data used to fit P$_{10}$SIR]{{\bf Data used to fit P$_{10}$SIR }.
Real world data: daily new cases for 10 Western Europe countries,
source data from WorldOmeters
(https://www.worldometers.info/coronavirus/) for 19th January to 16th August 2020.
Data smoothed by using a 7-day rolling average.
Time: day 0 is after 5 or more cumulative deaths.  Peak is the raw number. 
The figure shows clearer peaks for countries which reported a higher number of cases.
Data used to fit the model given in Figure~\ref{figure:SC:model6}.
} 
\label{figure:SC:Figure17}
        \end{figure}

\begin{table}[h!]
\caption[Real world data]{Real world data: peak magnitude and timing of daily new cases for 10 Western Europe countries,
source data from WorldOmeters
(https://www.worldometers.info/coronavirus/) for 19th January to 16th August 2020.}
\vspace{1em}
\begin{tabular}{lcc}
\hline
Country & Time & Peak \\
\hline
Austria & 8 & 754 \\
Switzerland & 11 & 1112 \\
Portugal & 13 & 803 \\
Spain & 14 & 8987 \\
Denmark & 19 & 328 \\
Germany & 19 & 5837 \\
France & 26 & 4537 \\
Belgium & 29 & 1453 \\
Italy & 30 & 5646 \\
Netherlands & 33 & 1120 \\
\hline
\end{tabular}
\label{table:sc:table6}
\end{table}

\begin{table}[h!]
\caption[Objective~1 results]{Objective~1: Summary statistics for Random Restarts achieving the best Levenshtein Distance of 2 for
model given in Figure~\ref{figure:SC:model6}.
RR = Random Restart and n refers to the number of solutions producing the summary statistics.}
\vspace{1em}
\tiny
\begin{tabular}{lccc|ccc|ccc}
\hline
\multicolumn{1}{c}{} & \multicolumn{3}{c}{Infection Rates RR0 (n=202)} & \multicolumn{3}{|c|}{Infection Rates RR5 (n=186)} & \multicolumn{3}{c}{Infection Rates RR9 (n=315)}\\
\hline
{\bf Country} & {\bf Mean} & {\bf Min} & {\bf Max} & {\bf Mean} & {\bf Min} & {\bf Max} & {\bf Mean} & {\bf Min}
& {\bf Max} \\ \hline
Austria & 8.34e$^{-8}$ & 3.24e$^{-8}$ & 1.69e$^{-7}$ & 4.05e$^{-8}$ & 2.33e$^{-8}$ & 6.72e$^{-8}$ & 5.92e$^{-7}$ & 1.69e$^{-7}$ & 1.03e$^{-6}$ \\
Belgium & 2.82e$^{-6}$ & 2.07e$^{-6}$ & 4.72e$^{-6}$ & 1.98e$^{-6}$ & 1.24e$^{-6}$ & 3.41e$^{-6}$ & 5.40e$^{-7}$ & 6.68e$^{-8}$ & 1.06e$^{-6}$ \\
Switzerland & 6.72e$^{-6}$ & 4.09e$^{-6}$ & 1.20e$^{-5}$ & 5.16e$^{-7}$ & 2.64e$^{-7}$ & 1.10e$^{-6}$ & 1.41e$^{-7}$ & 3.86e$^{-9}$ & 6.89e$^{-7}$ \\
Germany & 5.55e$^{-7}$ & 3.92e$^{-7}$ & 7.41e$^{-7}$ & 1.98e$^{-5}$ & 5.38e$^{-6}$ & 3.59e$^{-5}$ & 1.61e$^{-6}$ & 4.88e$^{-7}$ & 4.58e$^{-6}$ \\
Denmark & 6.19e$^{-7}$ & 1.45e$^{-7}$ & 1.54e$^{-6}$ & 2.93e$^{-6}$ & 4.90e$^{-7}$ & 9.79e$^{-6}$ & 1.05e$^{-7}$ & 3.51e$^{-8}$ & 2.46e$^{-7}$ \\
Spain & 2.53e$^{-6}$ & 9.18e$^{-7}$ & 1.00e$^{-5}$ & 1.49e$^{-6}$ & 8.47e$^{-7}$ & 2.52e$^{-6}$ & 3.83e$^{-6}$ & 1.01e$^{-6}$ & 9.52e$^{-6}$ \\
France & 6.11e$^{-7}$ & 4.80e$^{-7}$ & 8.72e$^{-7}$ & 4.83e$^{-7}$ & 2.46e$^{-7}$ & 7.26e$^{-7}$ & 7.47e$^{-7}$ & 3.44e$^{-7}$ & 1.49e$^{-6}$ \\
Italy & 4.72e$^{-7}$ & 2.98e$^{-7}$ & 7.00e$^{-7}$ & 2.99e$^{-7}$ & 2.23e$^{-7}$ & 6.02e$^{-7}$ & 3.45e$^{-7}$ & 1.83e$^{-7}$ & 5.68e$^{-7}$ \\
Netherlands & 5.71e$^{-7}$ & 3.11e$^{-7}$ & 1.28e$^{-6}$ & 8.31e$^{-7}$ & 6.16e$^{-7}$ & 1.40e$^{-6}$ & 7.81e$^{-7}$ & 6.10e$^{-7}$ & 1.14e$^{-6}$ \\
Portugal & 1.51e$^{-7}$ & 5.13e$^{-8}$ & 4.50e$^{-7}$ & 3.35e$^{-5}$ & 5.24e$^{-6}$ & 8.36e$^{-5}$ & 8.82e$^{-5}$ & 2.63e$^{-5}$ & 2.21e$^{-4}$ \\
\hline
\end{tabular}
\label{table:sc:table9}
\end{table}

\begin{table}[h!]
\caption[Objective~2 results]{Objective~2: Infection rates output from the best solution from the RRHCA for model given in
Figure~\ref{figure:SC:model6}. (left) Sorted alphabetically,
(right) sorted by rate.}
\vspace{1em}
\begin{tabular}{lc}
\hline
{\bf Country}	& {\bf Infection Rate} \\
\hline
Austria & 3.64e$^{-5}$ \\
Belgium & 4.31e$^{-5}$ \\
Switzerland & 3.46e$^{-6}$ \\
Germany & 3.48e$^{-6}$ \\
Denmark & 6.94e$^{-6}$ \\
Spain & 1.80e$^{-5}$ \\
France & 4.44e$^{-6}$ \\
Italy & 6.57e$^{-6}$ \\
Netherlands & 2.36e$^{-6}$ \\
Portugal & 3.22e$^{-5}$ \\
\hline
\end{tabular}
\hspace{3em}
\begin{tabular}{lc}
\hline
{\bf Country}   & {\bf Infection Rate} \\
\hline
Netherlands & 2.36e$^{-6}$ \\
Switzerland & 3.46e$^{-6}$ \\
Germany & 3.48e$^{-6}$ \\
France & 4.44e$^{-6}$ \\
Italy & 6.57e$^{-6}$ \\
Denmark & 6.94e$^{-6}$ \\
Spain & 1.80e$^{-5}$ \\
Portugal & 3.22e$^{-5}$ \\
Austria & 3.64e$^{-5}$ \\
Belgium & 4.31e$^{-5}$ \\
\hline
\end{tabular}
\label{table:sc:table10}
\end{table}

\begin{table}[h!]
\caption[Objective~3 results for infection rates]{Objective~3: Infection rates output from the best solution from the RRHCA for model given in
Figure~\ref{figure:SC:model6}.(left) Sorted alphabetically, (right) sorted by rate.}
\vspace{1em}
\begin{tabular}{lc}
\hline
{\bf Country} & {\bf Infection~Rate} \\
\hline 
Austria & 4.05e$^{-5}$ \\
Belgium & 8.03e$^{-6}$ \\
Switzerland & 3.49e$^{-5}$ \\
Germany & 2.89e$^{-6}$ \\
Denmark & 8.69e$^{-6}$ \\
Spain & 7.58e$^{-6}$ \\
France & 3.75e$^{-6}$ \\
Italy & 6.20e$^{-6}$ \\
Netherlands & 1.11e$^{-6}$ \\
Portugal & 3.49e$^{-5}$ \\
\hline
\end{tabular}
\hspace{3em}
\begin{tabular}{lc}
\hline
{\bf Country} & {\bf Infection~Rate} \\
\hline
Netherlands & 1.11e$^{-6}$ \\
Germany & 2.89e$^{-6}$ \\
France & 3.75e$^{-6}$ \\
Italy & 6.20e$^{-6}$ \\
Spain & 7.58e$^{-6}$ \\
Belgium & 8.03e$^{-6}$ \\
Denmark & 8.69e$^{-6}$ \\
Portugal & 3.49e$^{-5}$ \\
Switzerland & 3.49e$^{-5}$ \\
Austria & 4.05e$^{-5}$ \\
\hline
\end{tabular}
\label{table:sc:table11}
\end{table}

\begin{table}[h!]
\caption[Objective~3 results for travel rates]{Objective~3: Travel rates output from the best solution from the RRHCA for model given in
Figure~\ref{figure:SC:model6} (Infection rates in Table~\ref{table:sc:table11}).  Columns: Connection from;
Rows: Connection to.}
\vspace{1em}
\tiny
\begin{tabular}{c|cccccccccc}
\hline
 & AT & BE & CH & DE & DK & ES & FR & IT & NL & PT \\ \hline
AT & & & 3.553e$^{-6}$ & 1.471e$^{-6}$ & & & & 1.086e$^{-5}$ & & \\
BE & & & & 1.115e$^{-6}$ & & & 5.525e$^{-6}$ & & 5.408e$^{-7}$ & \\
CH & 3.786e$^{-6}$ & & & 1.56e$^{-6}$ & & & 3.389e$^{-7}$ & 6.361e$^{-6}$ & & \\
DE & 3.390e$^{-6}$ & 7.833e$^{-6}$ & 3.443e$^{-6}$ & & 2.125e$^{-6}$ & & 4.267e$^{-6}$ & & 3.187e$^{-6}$ & \\
DK & & & & 1.960e$^{-6}$ & & & & & & \\
ES & & & & & & & 1.462e$^{-6}$ & & & 1.565e$^{-6}$ \\
FR & & 1.291e$^{-6}$ & 1.425e$^{-6}$ & 1.082e$^{-5}$ & & 2.630e$^{-6}$ & & 7.151e$^{-6}$ & & \\
IT & 2.535e$^{-6}$ & & 2.529e$^{-6}$ & & & & 1.75e$^{-6}$ & & & \\
NL & & 9.549e$^{-6}$ & & 2.342e$^{-6}$ & & & & & & \\
PT & & & & & & 2.071e$^{-5}$ & & & & \\
\end{tabular}
\label{table:sc:fittedconnectionrates}
\end{table}

\clearpage
\subsection{Analysis}

\begin{figure}[h!tb]
            \centering
            \includegraphics[width=0.95\textwidth]{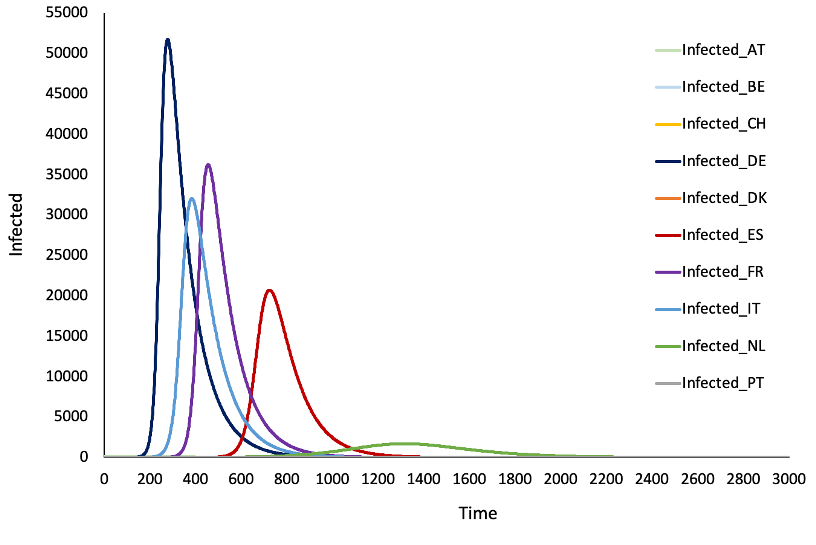}
\caption[P$_{10}$SIR behaviour]{{\bf P$_{10}$SIR behaviour.}
Modelled infections for 10 Western European countries connected by road. AT = Austria, BE = Belgium, CH =
Switzerland, DE = Germany, DK = Denmark, ES = Spain, FR = France, IT = Italy, NL = Netherlands and PT =
Portugal. Plots for countries under 100 Infected are not visible due to the scale of the graph.}
\label{figure:SC:Figure20}
        \end{figure}

\paragraph{Visual analysis.} For an example see Figure~\ref{figure:SC:Figure20}.
\vspace{3em}

\begin{figure}[h!tb]
            \centering
            \includegraphics[width=0.40\textwidth]{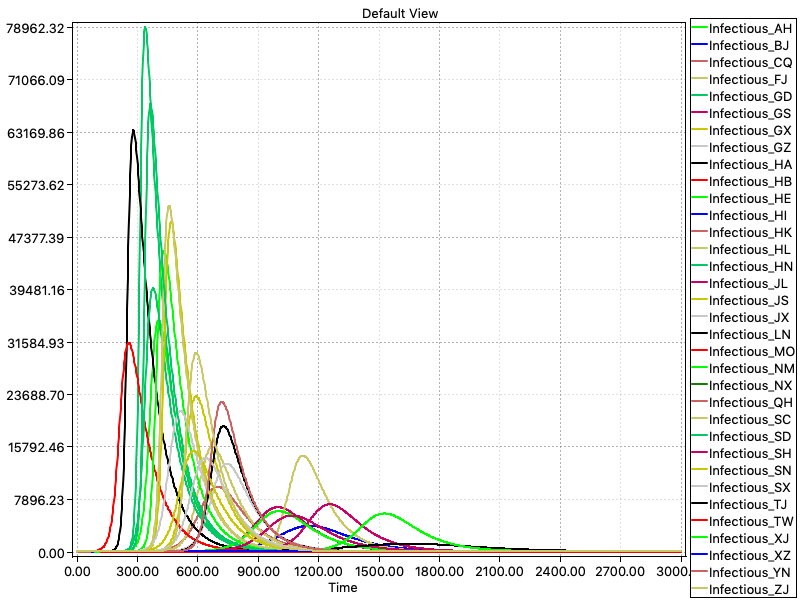}
            \includegraphics[width=0.40\textwidth]{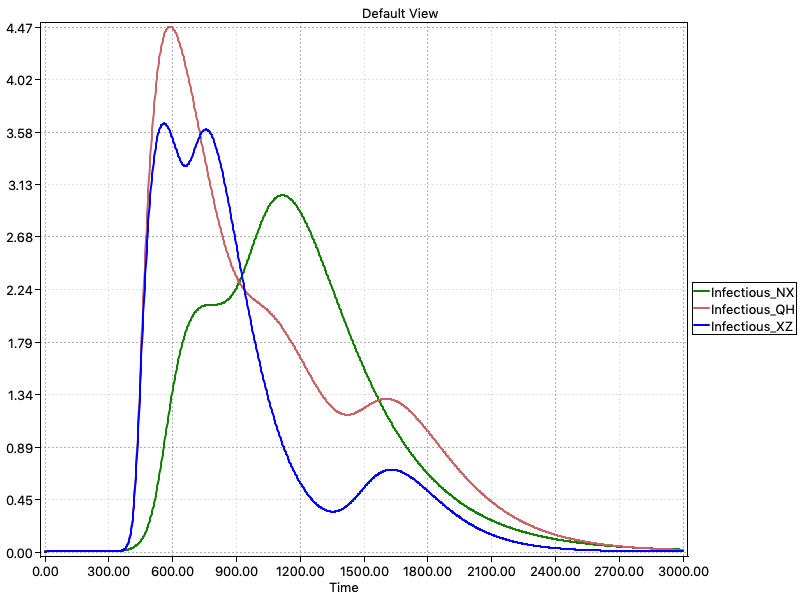}
            \includegraphics[width=0.70\textwidth]{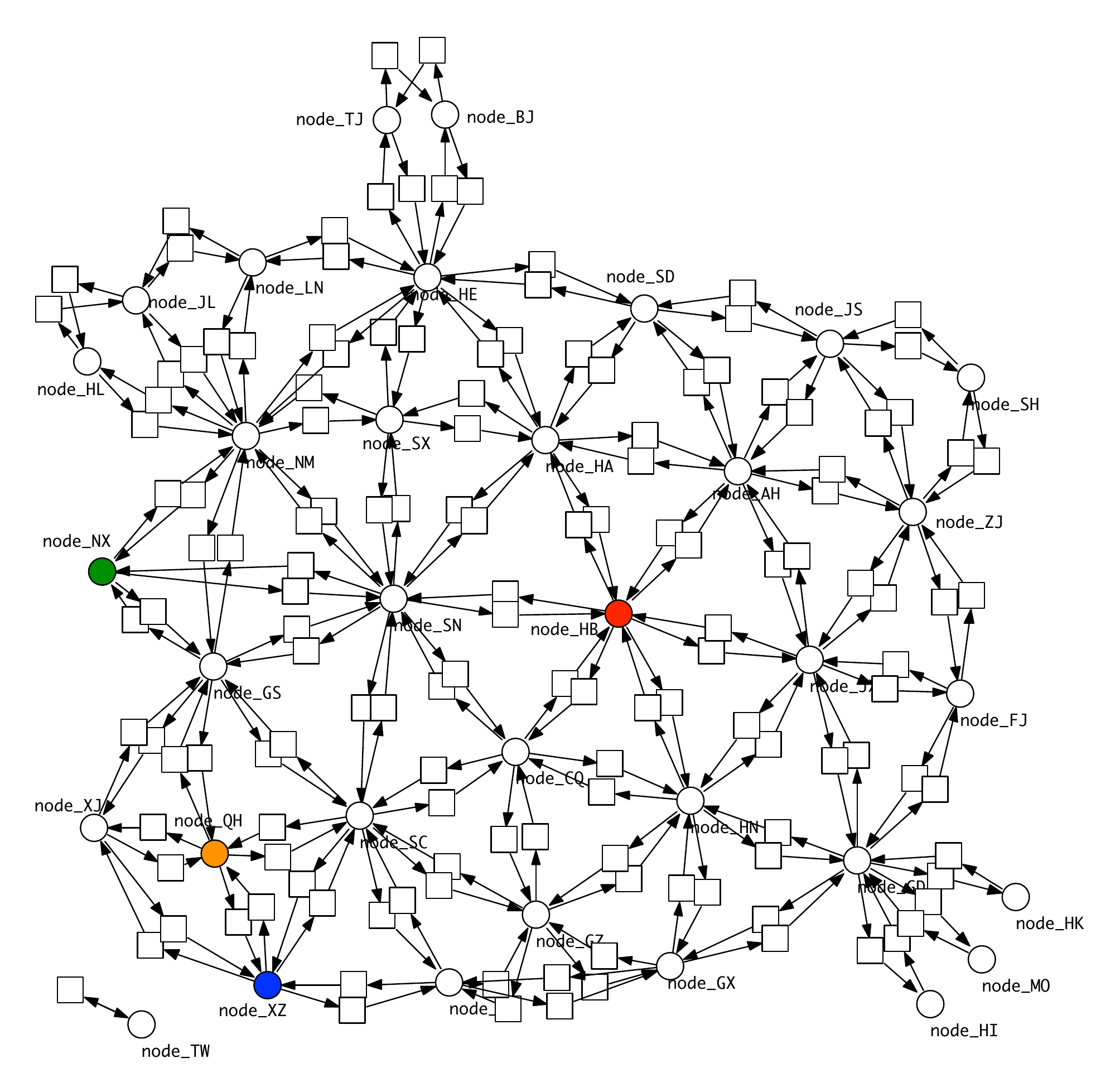}
\caption[P$_\textrm{China}$SIR model checking]{{\bf P$_\textrm{China}$SIR model checking.}
{\bf(Top, left)} Time series traces of \node{Infectious} for the 34 provinces, with infection starting in Hubei.
The ratio between rate constants for \node{Infect} and \node{Recover} is $1:10^4$
($k_{infect}=1.0e^{-6}$, $k_{recover}=1.0e^{-2}$.)
{\bf(Top, right)} Model checking reveals three provinces with more than one peak.
{\bf(Bottom)} Network of China highlighted to indicate the geographical context of the three provinces: HB - red, NX - green, QH - brown, XZ - blue.}
\label{figure:MCpeaks}
\end{figure}

\clearpage
\paragraph{Model checking}
For illustration we include one experiment involving model checking, see Figure~\ref{figure:MCpeaks}.
\vspace{1em}

\clearpage

\begin{figure}[h!tb]
            \centering
            \includegraphics[width=0.95\textwidth]{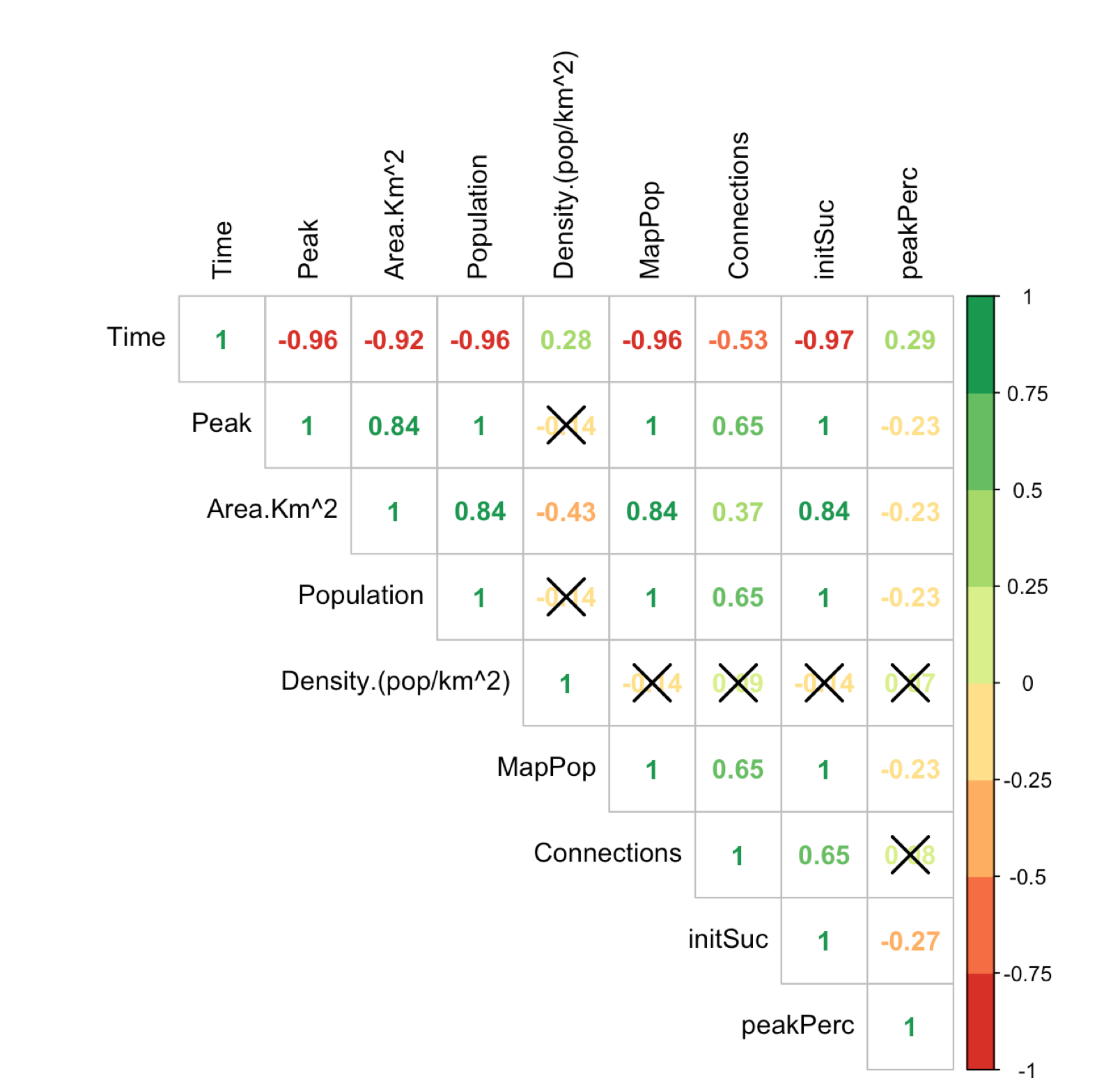}
\caption[Correlation matrix for P10SIR, not fitted]{{\bf Correlation matrix for P$_{10}$SIR, not fitted.}
Green indicates a strong positive correlation and red indicates a strong
negative correlation. Black crosses represent a non-significant correlation.  
Infection rates and travel rates all identical at 1.0e$^{-6}$; recovery rates 1.0e$^{-2}$.}
\label{figure:SC:Figure23}
        \end{figure}

\begin{figure}[h!tb]
            \centering
            \includegraphics[width=0.95\textwidth]{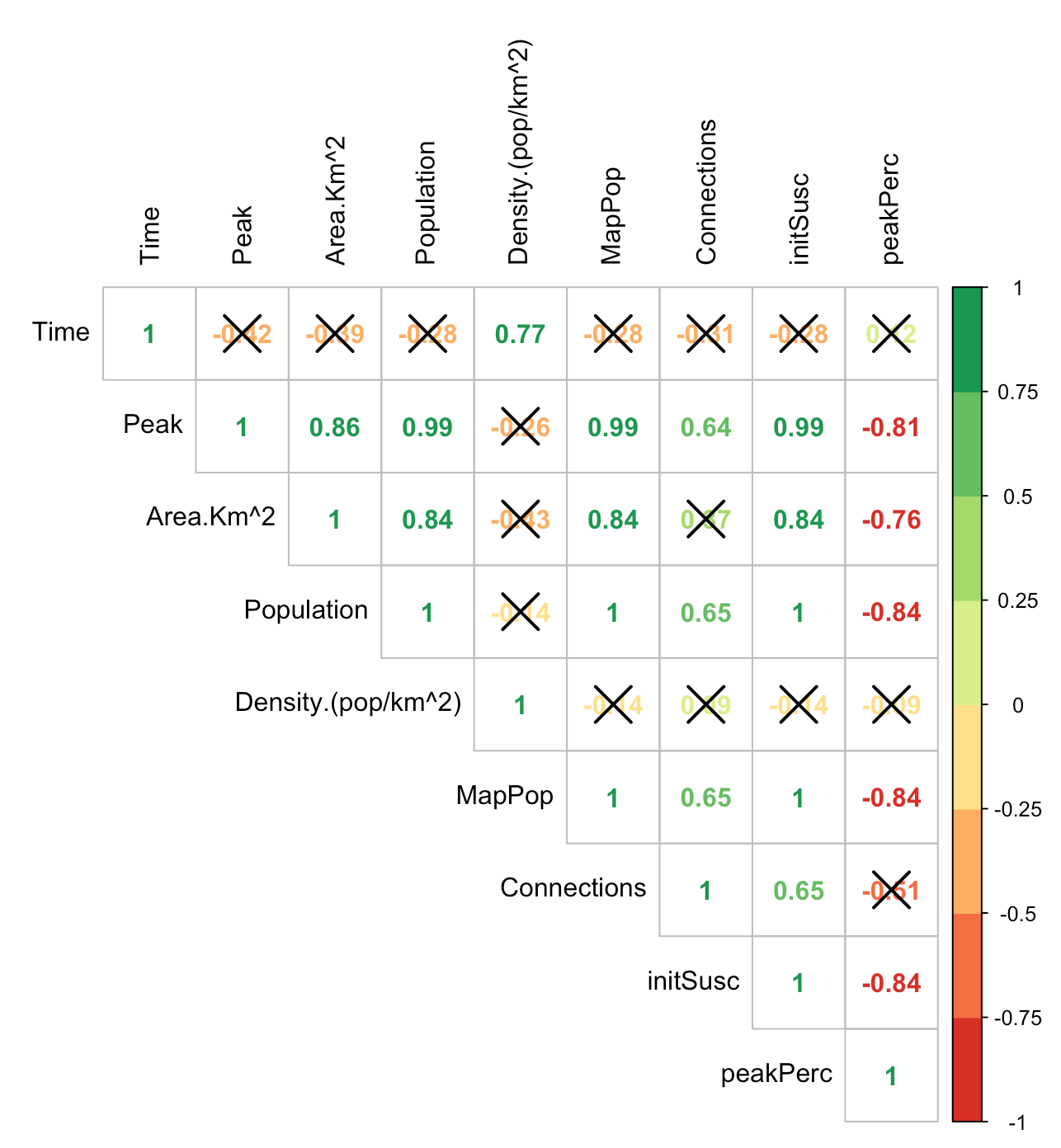}
\caption[Correlation matrix for West Europe10, fitted]{{\bf Correlation matrix for P$_{10}$SIR, fitted.}
Green indicates a strong positive correlation and red indicates a strong
negative correlation. Black crosses represent a non-significant correlation.}
\label{figure:SC:Europe10fitted}
        \end{figure}

\begin{figure}[h!tb]
            \centering
            \includegraphics[width=0.95\textwidth]{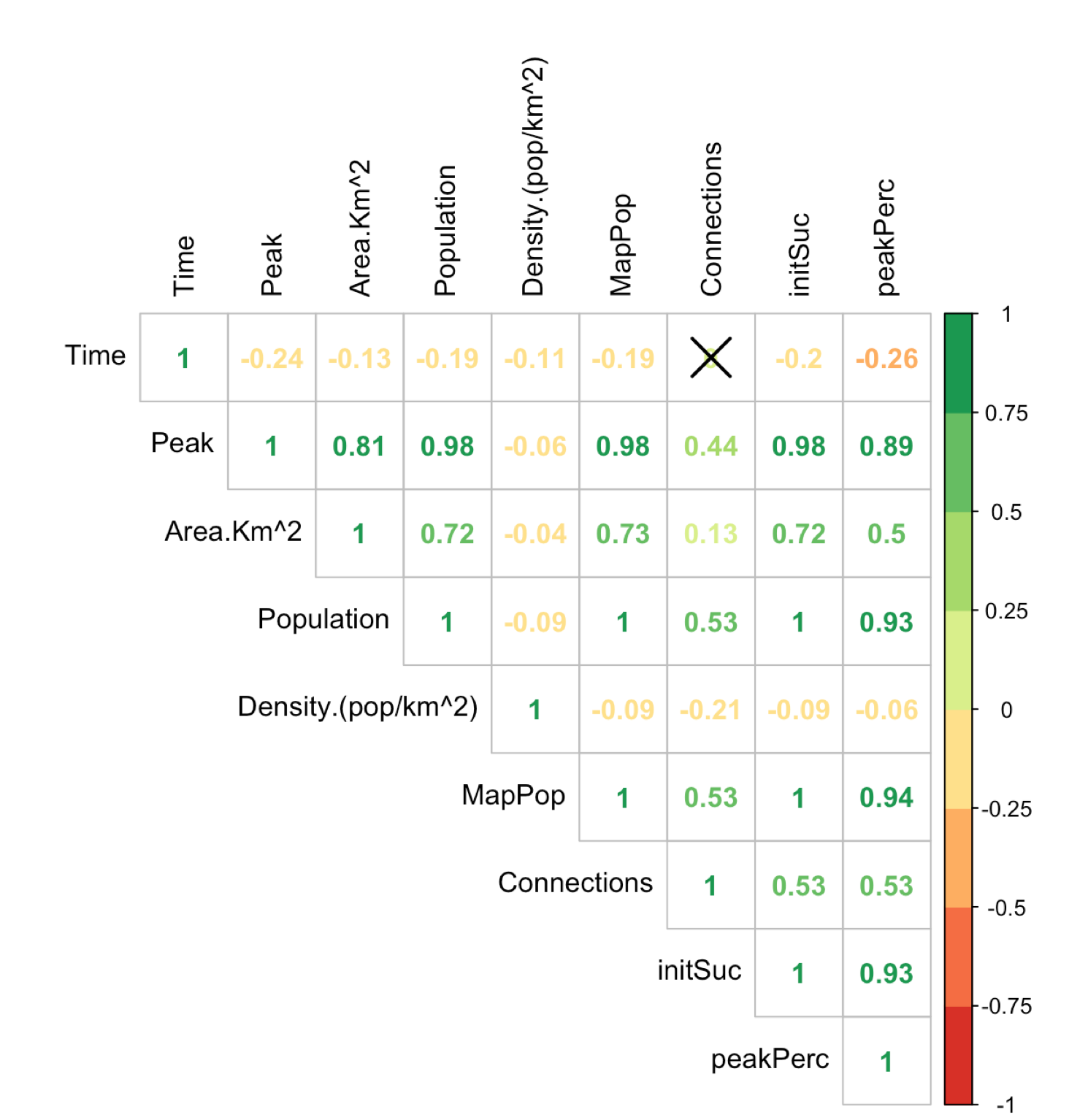}
\caption[Correlation matrix for Europe48, not fitted]{{\bf Correlation matrix for P$_{48}$SIR, not fitted.}
Green indicates a strong positive correlation and red indicates a strong
negative correlation. Black crosses represent a non-significant correlation.}
\label{figure:SC:Figure24}
        \end{figure}

\paragraph{Correlation analysis.}  Table~\ref{table:sc:table7} records the variables, correlation matrices are
given in
Figure~\ref{figure:SC:Figure23},~\ref{figure:SC:Europe10fitted},~\ref{figure:SC:Figure24}.  
The correlation matrices for the unfitted models indicate that the
percent of population infected at peak is related to the population size for
P$_{48}$SIR, but not for the smaller model P$_{10}$SIR, perhaps because there are more smaller countries in 
the smaller model.  The number of
connections is related to the peak and  peak pecentage in the larger model, but only to the peak in
the smaller model.
Area is a partial proxy for the whole population and is strongly related to
peak and time (negatively - i.e. earlier) but not to peak percentage in P$_{10}$SIR.  However in P$_{48}$SIR
area is related to peak and peak percentage but not to time.

However, the correlation over the fitted data indicates that the percentage of the population at the peak is
less related to the population size, compared with the unfitted model.  Also the size of the peak was less
related to length of time from the start of infections in the fitted model compared to the unfitted model.
Finally, the time of the peak of infections was not correlated with the number of inter-country connections in
the fitted model, as opposed to being negatively correlated in the unfitted model (more connections implying
earlier peaks), indicating that the
intervention policies that governments have taken to control the epidemic in their countries have overriden the
effect of geographical connections (i.e. closing down inter-country travel).

\begin{table}[h!]
\caption[Real world variables]{Variables used to investigate road connected models further. 
Categories: (1) real world data external to model,
(2) derived measure connecting external (black) to blue,
(3) real world data incorporated in model, 
(4) data generated by model.  }
\vspace{1em}
\footnotesize
\begin{tabular}{lll}
\hline
{\bf Category} & {\bf Variable} & {\bf Description}\\
\hline
1 & Area.Km2 & Area in km$^2$ of a given country. \\
2 & Density & Population density of a given country (population/km2). \\
3 & Connections & Number of connections a country has in land borders. \\
 & initSusc & Susceptible + Infected at time 0.  \\
 & Population & The real-world country populations. \\
 & MapPop & The population used in the model (real world population / 1000).  \\
4 & peak & The peak of Infected, this represents active cases. \\
 & peakPerc & The percentage of initSusc that were infected at the peak.  \\
 & time & The number of steps into the simulation the peak occurred.  \\
\hline
\end{tabular}
\label{table:sc:table7}
\end{table}

\clearpage

\begin{figure}[h!tb]
            \centering
            \includegraphics[width=0.85\textwidth]{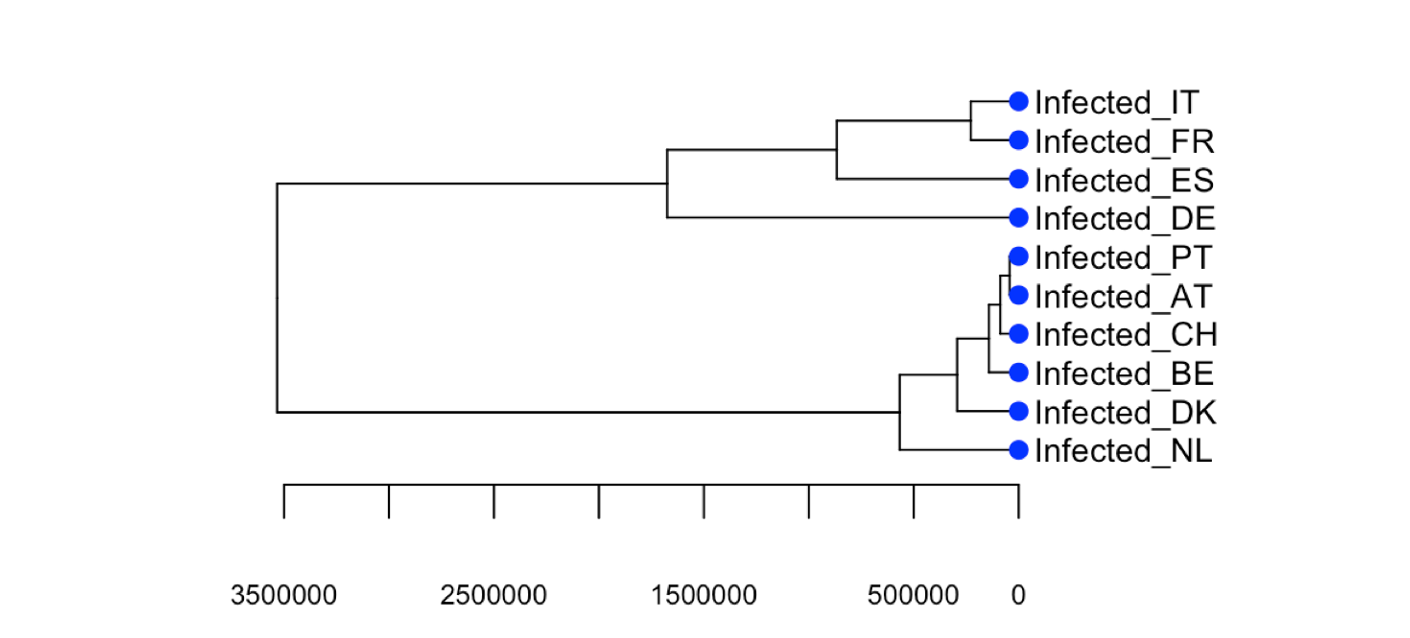}
\caption[Dendrogram P$_{10}$SIR ]{{\bf Dendrogram of P$_{10}$SIR} for West Europe,
hierarchical clustering. Height is on the x-axis. Data not normalised}
\label{figure:SC:Figure25}
        \end{figure}

\begin{figure}[h!tb]
            \centering
            \includegraphics[width=0.85\textwidth]{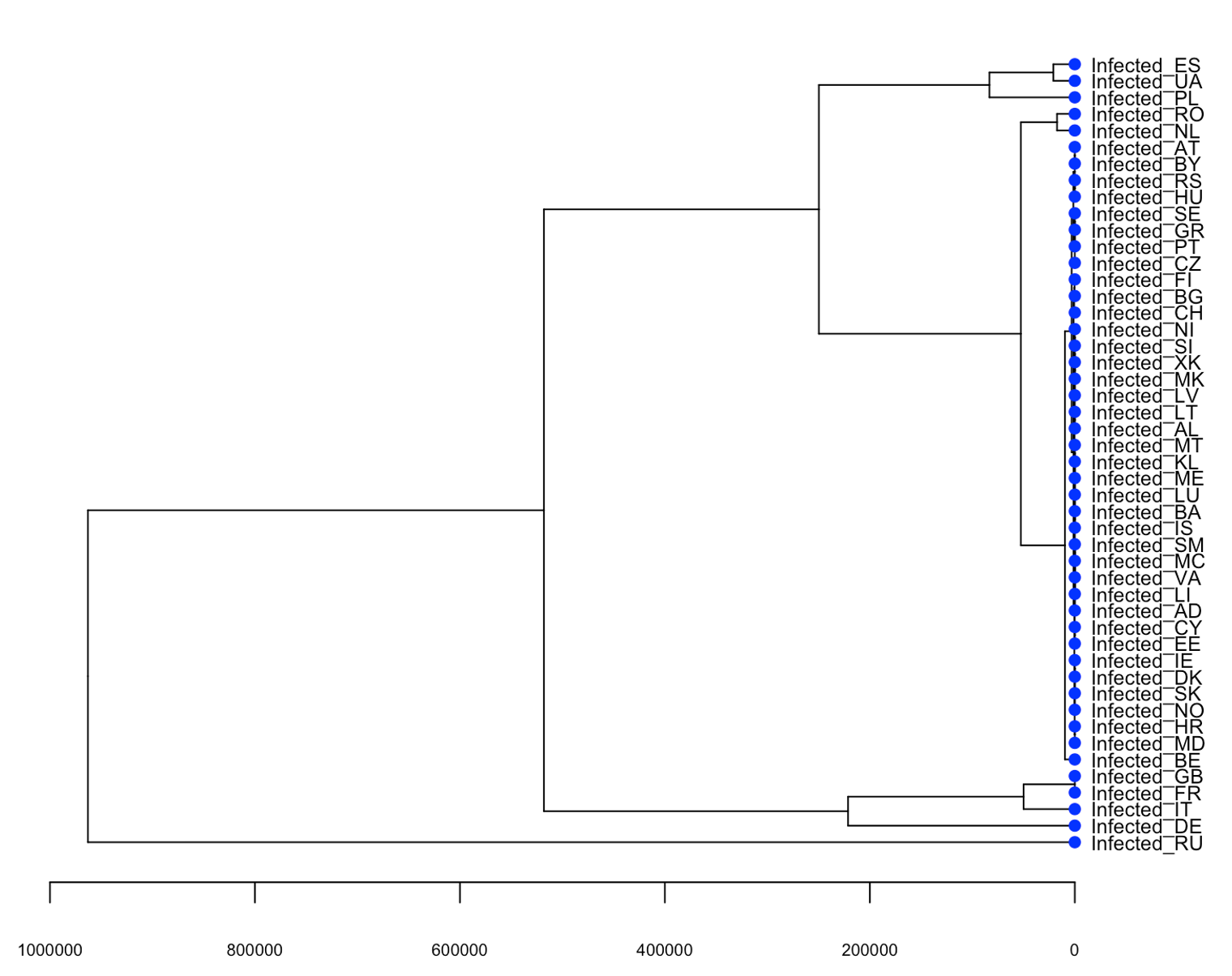}
\caption[Dendrogram P$_{48}$SIR Europe]{{\bf Dendrogram of P$_{48}$SIR} for Europe,
hierarchical clustering. Height is on the x-axis. Data not normalised.}
\label{figure:SC:Figure26}
        \end{figure}

\paragraph{Dendrograms.} Examples are given in Figures~\ref{figure:SC:Figure25},~\ref{figure:SC:Figure26}.


\clearpage

\subsection{Discussion.}  

\begin{figure}[h!tb]
            \centering
            \includegraphics[width=0.70\textwidth]{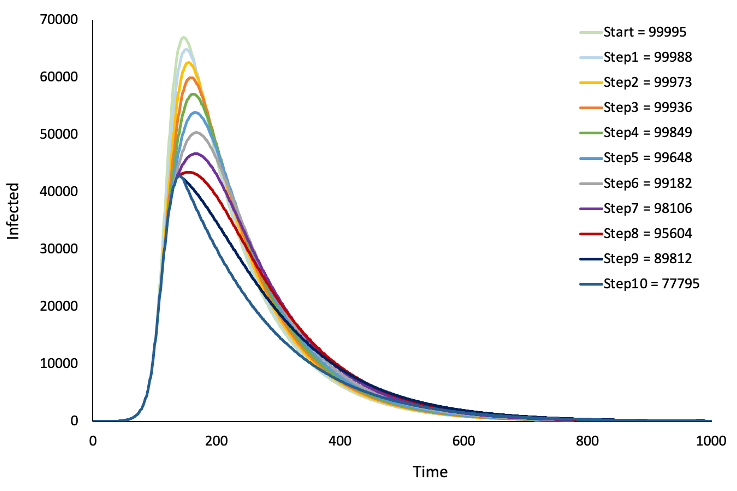}
\caption[Dynamic change of infection rates: lockdown]{{\bf SIR model - Dynamic change of infection rates:
lockdown.}
Output from the SIR model in Figure~\ref{figure:SIR} where infection rates were stepwise
decremented during simulation. Figure legend shows
cumulative infections for different number of decrements.}
\label{figure:SC:Figure29}
        \end{figure}

\begin{figure}[h!tb]
            \centering
            \includegraphics[width=0.70\textwidth]{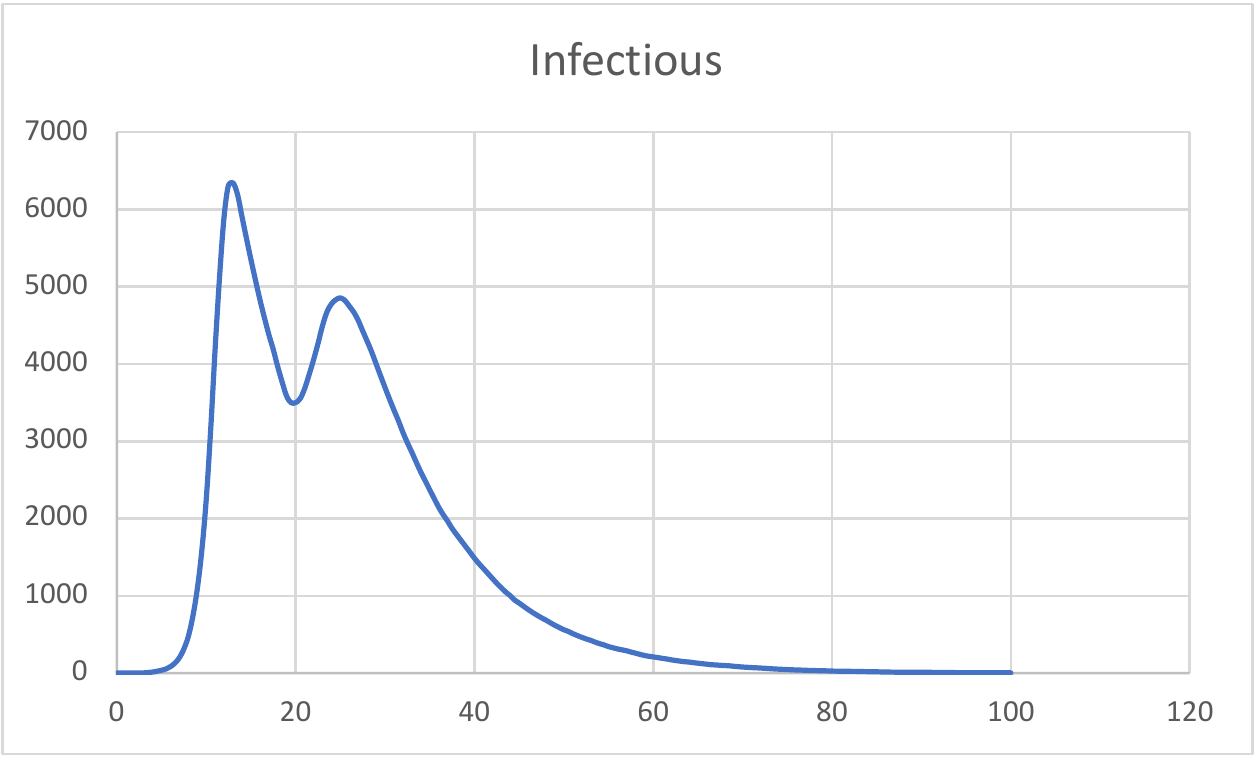}
\caption[Dynamic change of infection rate]{{\bf SIR model, dynamic change of infection rate}
increased back to pre-intervention level at time step 20.  }
\label{figure:JC:lockunlock}
        \end{figure}

Results of simulating lockdown and unlock by dynamic change of infection rates in the SIR model; 
Figure~\ref{figure:SC:Figure29} - lockdown only, 
Figure~\ref{figure:JC:lockunlock} - lockdown followed by unlock.

\clearpage
\section{List of Files Provided}

\paragraph{All models given in figures (provided as \Snoopy files):}


\begin{myitemize}
\renewcommand\labelitemi{\tiny$^\bullet$}

\item {\bf SIR.spn} - standard SIR, see Figure~\ref{figure:SIR} 
\item {\bf SIQR.spn} - SIR extended by quarantine, see Figure~\ref{figure:SIQR-SIAR} (Top)  
\item {\bf SIAR.spn} - SIR extend by symptomatic/asymptomatic compartments, see Figure~\ref{figure:SIQR-SIAR} (Bottom) 
\item {\bf SIR-S2.spn} - SIR-S$_2^{age}$, uncoloured, see Figure~\ref{figure:SIRage} (Top)
\item {\bf SIR-S2\_enum.colspn}, {\bf SIR-S2\_int.colspn} - SIR-S$_2^{age}$, coloured, see Figure~\ref{figure:SIRage} (Middle)
\item {\bf SIR-S20.spn} - SIR-S$_{20}$, see Figure~\ref{figure:SIRage} (Bottom)
\item {\bf P2-SIR.spn} - P$_2$SIR, uncoloured, see Figure~\ref{figure:SIRtravel2} (Top)  
\item {\bf P2-SIR.colspn} - P$_2$SIR, coloured, see Figure~\ref{figure:SIRtravel2} (Bottom) 
\item {\bf network-Europe4.spn} - connectivity graph for Europe, 4 countries, uncoloured, see Figure~\ref{figure:Countries4} (Top middle) 
\item {\bf network-Europe4.colspn} - connectivity graph for Europe, 4 countries, coloured, see Figure~\ref{figure:Countries4} (Top right) 
\item {\bf P4-SIR.colspn} - P$_4$SIR, coloured, see Figure~\ref{figure:Countries4} (Middle) 
\item {\bf P4-SIR.spn} - P$_4$SIR, uncoloured, see Figure~\ref{figure:P4-SIR-unfolded}
\item {\bf P48-SIR-S10.colspn} - P$_{48}$SIR-S$_{10}$, see Figure~\ref{figure:Countries4} (Bottom) 
\item {\bf P48-SIR-S10.andl.spn} - P$_{48}$SIR-S$_{10}$ unfolded, see Figure~\ref{figure:P48SIR} (Bottom)
\item {\bf SIVR.hpn} - SIR with variant virus, see Figure\ref{figure:SIVR}
\item {\bf P10-QI-SIR.colspn} - P$_{10}$Q$\_\textrm{I}$SIR, see Figure~\ref{figure:SC:model7} (Top)
\item {\bf P10-QSIR.colspn} - P$_{10}$QSIR, see Figure~\ref{figure:SC:model7} (Bottom)
\item {\bf P10-SIR\_Separate\_Rates.cpn} - \CPN used for parameter optimisation - unfolded version of P$_{10}$SIR, given in Figure~\ref{figure:SC:model6}.

\end{myitemize}

\paragraph{Connectivity networks (provided as CANDL files):}

\begin{myitemize}
\renewcommand\labelitemi{\tiny$^\bullet$}

\item {\bf network-Europe04.candl}\\ - unfolding: |P|=4, |T|=8, |A|=16   
\item {\bf network-Europe05.candl}\\ - unfolding: |P|=5, |T|=16, |A|=32
\item {\bf network-Europe10.candl}\\ - unfolding: |P|=10, |T|=30, |A|=60 
\item {\bf network-Europe48.candl}\\
     - unfolding: |P|=47, |T|=170, |A|=340\\
    {\it - note:} IS isolated, https://countrycode.org\\
    - unfolding time: 1.302sec
\item {\bf network-China.candl}\\  
    - unfolding: P|=33, |T|=142, |A|=286\\
    {\it - note:} 34 states; TW isolated\\
    - unfolding time: 0.283sec
\item {\bf network-USA.candl} \\
     - unfolding: |P|=48, |T|=210, |A|=420\\
    {\it - note:} 50 states; AK, HI isolated\\ 
    - unfolding time: 1.643sec
\end{myitemize}  


\paragraph{P$_n$SIR models (provided as CANDL files):}

\begin{myitemize}
\renewcommand\labelitemi{\tiny$^\bullet$}
\item P-Europe02-SIR - Europe fragment, 2 countries
\item P-Europe04-SIR - Europe fragment, 4 countries
\item P-Europe10-SIR - Western Europe, 10 countries
\item P-Europe48-SIR - All Europe, 48 countries
\item P-China-SIR    - China, 34 provinces
\item P-USA-SIR      - USA, 50 states
\end{myitemize}

\paragraph{Python programs:}

\begin{myitemize}
\renewcommand\labelitemi{\tiny$^\bullet$}
\item 
RRHCA (Random Restart Hill Climbing) 
\item
Python Web Scraper, used to to collect COVID19 Data. 
\end{myitemize}

\paragraph{R code:}

\begin{myitemize}
\renewcommand\labelitemi{\tiny$^\bullet$}

\item 
Europe.Rmd, Western\_Europe.Rmd: Both these files take in csv outputs from \Snoopy, \Spike to produce correlation matrices and regression analysis. The Time and \node{Infectious} traces are extracted from the csv files and joined to real world data (e.g. population, density, country size). Correlation matrices are then produced. Regression analysis is also conducted along with checking the assumptions of regression. 

\item
Clustering.Rmd:
This code takes in one csv file (output from \Snoopy, \Spike) and performs agglomerative hierarchical clustering using Euclidean distance and complete-linkage. This code selects Time and Infected traces to perform the clustering. The code plots a dendrogram of the clustering results for visualisation.

\end{myitemize}

\paragraph{\Spike:}

\begin{myitemize}
\renewcommand\labelitemi{\tiny$^\bullet$}

\item
config file + andl/candl file

\item
all source files to produce Figure~\ref{figure:JC:lockunlock}.

\end{myitemize}

\paragraph{MC2:}

\begin{myitemize}
\renewcommand\labelitemi{\tiny$^\bullet$}

\item
property library and Unix script to produce Figure~\ref{figure:MCpeaks}.

\end{myitemize}

\end{document}